\documentclass[aps,prd,showpacs,notitlepage,nofootinbib,floatfix,showkeys]{revtex4-2}

\usepackage{blindtext}
\usepackage{hyperref}
\usepackage{amsmath,amssymb}
\usepackage{float}
\usepackage{microtype}

\usepackage{graphicx}
\usepackage{bm}
\usepackage{latexsym}
\usepackage{epsfig}
\usepackage{psfrag}
\usepackage{color}
\usepackage[dvipsnames]{xcolor}
\usepackage{subfigure}
\usepackage{graphicx}

\usepackage{amssymb}
\usepackage{amsmath}
\usepackage{bm}
\usepackage{latexsym}
\usepackage{epsfig}
\usepackage{psfrag}
\usepackage[normalem]{ulem}
\usepackage{textcomp}
\usepackage{color}
\usepackage{pstricks}
\usepackage[utf8]{inputenc}

\def\nn{\nonumber} 
\def\pa{{\partial}}
\def\f{\frac}
\def\l{\left}
\def\r{\right}
\def\d{{\rm d}}
\def\Mpl{M_{_{\mathrm{Pl}}}}
\def\Mp{M_{_{\mathrm{Pl}}}}

\def\ps{\mathcal{P}_{_{\mathrm{S}}}}
\def\pc{\mathcal{P}_{_{\mathrm{C}}}}
\def\pt{\mathcal{P}_{_{\mathrm{T}}}}
\def\ph{\mathcal{P}_{h}}
\def\ns{n_{_{\mathrm{S}}}}

\def\cB{{\mathcal B}}
\def\cG{{\mathcal G}}
\def\cI{{\mathcal I}}
\def\cP{{\mathcal P}}
\def\cR{{\mathcal R}}
\def\cS{{\mathcal S}}
\def\cT{{\mathcal T}}

\def\ei{\eta_{\rm i}}

\def\ee{\eta_{\rm e}}

\def\fpbh{f_{_{\mathrm{PBH}}}}
\def\ogw{\Omega_{_{\mathrm{GW}}}}
\def\fnl{f_{_{\rm NL}}}

\def\vka{{\bm k}_{1}}
\def\vkb{{\bm k}_{2}}
\def\vkc{{\bm k}_{3}}

\def\vx{{\bm{x}}}
\def\vk{{\bm{k}}}
\def\vp{{\bm{p}}}

\def\d{{\mathrm{d}}}

\allowdisplaybreaks

\begin{document}
\title{PBHs and secondary GWs from ultra slow roll and punctuated inflation}
\author{H.~V.~Ragavendra$^1$}
\email{E-mail: ragavendra@physics.iitm.ac.in} 
\affiliation{$^1$Department of Physics, Indian Institute of Technology Madras, 
Chennai~600036, India}
\author{Pankaj Saha$^1$}
\email{E-mail: pankaj@physics.iitm.ac.in}
\affiliation{$^1$Department of Physics, Indian Institute of Technology Madras, 
Chennai~600036, India}
\author{L.~Sriramkumar$^1$}
\email{E-mail: sriram@physics.iitm.ac.in}
\affiliation{$^1$Department of Physics, Indian Institute of Technology Madras, 
Chennai~600036, India}
\author{Joseph Silk$^{2,3,4,5}$}
\email{E-mail: silk@iap.fr}
\affiliation{$^2$Institut d'Astrophysique de Paris, UMR 7095, CNRS/UPMC Universit\'e 
Paris~6, Sorbonne Universit\'es, 98 bis boulevard Arago, F-75014 Paris, France}
\affiliation{$^3$Institut Lagrange de Paris, Sorbonne Universit\'es, 
98 bis Boulevard Arago, 75014 Paris, France}
\affiliation{$^4$Department of Physics and Astronomy, The Johns Hopkins University, 
3400 N.~Charles Street, Baltimore, MD 21218, U.S.A.}
\affiliation{$^5$Beecroft Institute for Cosmology and Particle Astrophysics,
University of Oxford, Keble Road, Oxford OX1 3RH, U.K.}

\begin{abstract}
The primordial scalar power spectrum is well constrained by the cosmological 
data on large scales, primarily from the observations of the anisotropies in 
the cosmic microwave background (CMB).
Over the last few years, it has been recognized that a sharp rise in power on 
small scales will lead to enhanced formation of primordial black holes (PBHs) 
and also generate secondary gravitational waves (GWs) of higher and, possibly,
detectable amplitudes.
It is well understood that scalar power spectra with COBE normalized amplitude 
on the CMB scales and enhanced amplitudes on smaller scales can be generated due 
to deviations from slow roll in single, canonical scalar field models of 
inflation.
In fact, an epoch of so-called ultra slow roll inflation can lead to the
desired amplification.
We find that scenarios that lead to ultra slow roll can be broadly classified
into two types, one wherein there is a brief departure from inflation (a scenario
referred to as punctuated inflation) and another wherein such a departure does 
not arise.
In this work, we consider a set of single field inflationary models involving
the canonical scalar field that lead to ultra slow roll and punctuated inflation
and examine the formation of PBHs as well as the generation of secondary GWs in
these models.
Apart from considering specific models, we reconstruct potentials from 
certain functional choices of the first slow roll parameter leading to
ultra slow roll and punctuated inflation and investigate their observational
signatures.
In addition to the secondary tensor power spectrum, we calculate the secondary 
tensor bispectrum in the equilateral limit in these scenarios.
Moreover, we calculate the inflationary scalar bispectrum that arises in all the 
cases and discuss the imprints of the scalar non-Gaussianities on the extent of 
PBHs formed and the amplitude of the secondary GWs generated. 
We conclude with a discussion on the wider implications of our results.
\end{abstract}

\maketitle


\section{Introduction}

With the recent observations of gravitational waves (GWs) from merging 
binary black holes involving a few to tens of solar 
masses~\cite{TheLIGOScientific:2016agk,TheLIGOScientific:2016qqj,
TheLIGOScientific:2016wfe,Abbott:2016blz,Abbott:2016nmj,Abbott:2017vtc,
Abbott:2017gyy,Abbott:2017oio,TheLIGOScientific:2017qsa,LIGOScientific:2020stg,
Abbott:2020uma,Abbott:2020khf}, there has been a considerable
interest in examining whether such black holes could have a primordial 
origin~\cite{DeLuca:2020qqa,Jedamzik:2020ypm,Jedamzik:2020omx}.
The most popular mechanism to generate primordial black holes (PBHs) is the 
inflationary scenario (for earlier discussions, see, for example,
refs.~\cite{Carr:1975qj,Carr:2009jm}; also see the recent 
reviews~\cite{Carr:2016drx,Carr:2018rid,Sasaki:2018dmp,Carr:2020xqk}).
PBHs are formed when the curvature perturbations generated during inflation
reenter the Hubble radius during the radiation and matter dominated epochs.
However, most inflationary models permit only slow roll inflation and, in 
such cases, the extent of PBHs produced proves to be considerably smaller
than required for any astrophysical implications (see, for example,
ref.~\cite{Chongchitnan:2006wx}).
Recall that, on large scales, the primordial scalar power spectrum is strongly
constrained by the increasingly precise observations of the anisotropies in 
the cosmic microwave background (CMB) (for recent constraints from Planck,
see refs.~\cite{Ade:2015lrj,Akrami:2018odb}).
In order to lead to a significant amount of PBHs, the scalar power spectrum 
on small scales should be considerably enhanced from the COBE normalized 
values over the CMB scales (for an early discussion in this context, see, 
for instance, ref.~\cite{Chongchitnan:2006wx}).
In inflation, this is possible only when there are strong departures from
slow roll.
It boils down to identifying inflationary potentials that permit slow roll
initially and then violating it for a certain period of time, before restoring 
it again until close to the termination of inflation.

\par

In models of inflation driven by a single, canonical scalar field, the 
so-called ultra slow scenario has turned out to be the most popular 
mechanism in the literature to enhance scalar power on small scales.
This scenario involves a period during inflation wherein the first slow 
roll parameter turns very small (for the initial discussions, see 
refs.~\cite{Garcia-Bellido:2017mdw,Ballesteros:2017fsr,Germani:2017bcs}; 
in this context, also see, for instance,
refs.~\cite{Dalianis:2018frf,Bhaumik:2019tvl}).
In fact, one finds that the scenario can be further divided into two types, 
those which admit a brief period of departure from inflation and another 
wherein no such departure arises.
The scenario wherein inflation is interrupted briefly is referred to as 
punctuated inflation (for the original discussions, see 
refs.~\cite{Roberts:1994ap,Leach:2000yw,Leach:2001zf};
for later and recent efforts, see 
refs.~\cite{Jain:2007au,Jain:2008dw,Jain:2009pm,Ragavendra:2020old}; for a 
discussion in the context of PBHs, see refs.~\cite{Kannike:2017bxn,Dalianis:2018frf}).
Interestingly, in such scenarios, the interruption of inflation is inevitably 
followed by an epoch of ultra slow roll which aids in boosting the power on 
small scales.
While, in the case of punctuated inflation, all the slow roll parameters 
(including the first) turn large briefly, in ultra slow roll inflation, the 
first slow parameter remains small until the very end of inflation and slow 
roll is said to be violated due to the large values achieved by the second and 
higher slow roll parameters.

\par

Often, the above-mentioned scenarios are achieved with the aid of potentials 
which contain a point of inflection~\cite{Jain:2008dw,Jain:2009pm,
Garcia-Bellido:2017mdw,Ballesteros:2017fsr,Germani:2017bcs,Bhaumik:2019tvl}.
The inflection point seems to play a crucial role in these scenarios in inducing
a period of ultra slow roll after the short epoch of deviation from slow roll.
The two stages of slow roll and ultra slow roll lead to either a step or a 
bump-like feature in the resulting inflationary scalar power spectrum, depending 
on the details of the intermediate departure from slow roll.
The lower level of the step is associated with the large scale modes that leave 
the Hubble radius during the first epoch of slow roll and the power is enhanced 
on small scales corresponding to modes that leave the Hubble radius during the 
later epoch of ultra slow roll.
We should mention here that the punctuated inflationary scenario has been 
considered to explain the lower power observed at the small multipoles in the 
CMB data.
If one chooses the drop in power to occur at scales roughly corresponding 
to the Hubble radius today, one finds that the resulting power spectrum
can improve the fit to the CMB data to a certain extent (for an earlier
analysis, see ref.~\cite{Jain:2008dw}; for a recent discussion, see 
ref.~\cite{Ragavendra:2020old}).

\par

We mentioned above that both ultra slow roll inflation and punctuated inflation 
can lead to a sharp rise in power on small scales. 
Evidently, if one chooses the rise to occur at suitable scales, one can utilize 
these power spectra to lead to enhanced formation of PBHs.
As has been established, such an enhanced amplitude for the scalar power spectrum
can induce secondary GWs when these modes reenter the Hubble radius at later 
times during the radiation dominated epoch (for the original discussions, see,
for example, refs.~\cite{Ananda:2006af,Baumann:2007zm,Saito:2008jc,Saito:2009jt};
for recent discussions 
in this context, see refs.~\cite{Kohri:2018awv,Espinosa:2018eve,Pi:2020otn}).
These secondary GWs with boosted amplitudes can, in principle, be detected 
by current and forthcoming observatories such as 
LIGO/Virgo~\cite{TheLIGOScientific:2016dpb}, Pulsar Timing 
Arrays (PTA)~\cite{Sazhin:1977tq,Detweiler:1979wn,Arzoumanian:2018saf},
the Laser Interferometer Space Antenna 
(LISA)~\cite{Audley:2017drz,Barausse:2020rsu},
the Big Bang Observer (BBO)~\cite{Crowder:2005nr,Corbin:2005ny,Baker:2019pnp},
the Deci-hertz Interferometer Gravitational wave 
Observatory~(DECIGO)~\cite{Kawamura:2011zz,Kawamura:2019jqt} and
the Einstein Telescope~(ET)~\cite{Punturo:2010zz,Sathyaprakash:2012jk}.
Moreover, the deviations from slow roll inflation, even as they boost the scalar 
power spectrum on small scales, also lead to larger levels of scalar 
non-Gaussianities on these scales (in this context, see, for example,
refs.~\cite{Chen:2008wn,Martin:2011sn,Hazra:2012yn}).
These non-Gaussianities can, in principle, further increase the extent of PBH
formation (for early discussions, see, for example,
refs.~\cite{Chongchitnan:2006wx,Seery:2006wk,Hidalgo:2007vk}; for recent 
discussions, see refs.~\cite{Motohashi:2017kbs,Atal:2018neu,Franciolini:2018vbk,
Kehagias:2019eil,Atal:2019erb,DeLuca:2019qsy,Passaglia:2018ixg,Ezquiaga:2019ftu})
as well as the strength of the secondary GWs 
(see refs.~\cite{Cai:2018dig,Unal:2018yaa,Cai:2019elf}; for a
very recent discussion, also see ref.~\cite{Yuan:2020iwf}).
In this work, we examine the enhanced formation of PBHs and the generation of 
secondary GWs in ultra slow roll and punctuated inflation.
We also numerically evaluate the inflationary scalar bispectrum generated on 
small scales in these scenarios and utilize the results to discuss the 
corresponding imprints on the extent of PBHs formed and the amplitude of 
secondary GWs.
In addition to considering specific potentials that lead to the scenarios of 
our interest, we choose functional forms for the first slow roll parameter 
leading to ultra slow roll and punctuated inflation, reverse engineer potentials
and examine the observational implications (for other efforts in these 
directions, see, for instance, Refs.~\cite{Chongchitnan:2006wx,
Hertzberg:2017dkh,Byrnes:2018txb,Motohashi:2019rhu}).
Interestingly, such an exercise also confirms the understanding that, in models 
of inflation involving a single, canonical scalar field, a point of inflection in 
the potential seems essential to lead to ultra slow roll or punctuated inflation.

\par

This paper is organized as follows.
In the following section, we shall introduce the different models of our interest
which lead to ultra slow roll and punctuated inflation.
In section~\ref{sec:ps}, we shall discuss the power spectra that arise in these
models and illustrate how the intrinsic entropy perturbation associated with the 
scalar field proves to be responsible for enhancing the amplitude of the curvature
perturbations.
In this section, we shall also highlight some of the challenges that one encounters
in constructing viable models of ultra slow roll and punctuated inflation.
In section~\ref{sec:re}, we shall consider specific forms for the first slow roll
parameter leading to ultra slow roll and punctuated inflation, and reverse 
engineer the potentials that lead to such scenarios.
We shall also discuss the power spectra that arise in these cases.
In sections~\ref{sec:pbhs} and~\ref{sec:sgws}, we shall discuss extent of PBHs 
formed and calculate the dimensionless parameters characterizing the power as 
well as bispectra of secondary GWs generated in the models and scenarios of 
interest.
We shall also compare our results with the constraints from observations.
In section~\ref{sec:sgws-dng}, we shall calculate the dimensionless 
non-Gaussianity parameter~$\fnl$ associated with the scalar bispectrum in 
all the different cases.
We shall highlight some of the properties of the non-Gaussianity parameter~$\fnl$ 
and then go on to discuss the imprints of the scalar non-Gaussianities on the
formation of PBHs and the generation of secondary GWs.
In section~\ref{sec:c}, we shall conclude with a summary of the main results.  
We shall relegate some of the related discussions to six appendices.

\par

A few remarks on our conventions and notations are in order at this stage 
of our discussion. 
We shall work with natural units such that $\hbar=c=1$ and set the reduced 
Planck mass to be $\Mpl=\l(8\,\pi\, G\r)^{-1/2}$.
We shall adopt the signature of the metric to be~$(-,+,+,+)$.
Note that Latin indices shall represent the spatial coordinates, except 
for~$k$ which shall be reserved for denoting the wave number. 
We shall assume the background to be the spatially flat
Friedmann-Lema\^itre-Robertson-Walker~(FLRW) 
line element described by the scale factor~$a$ and the Hubble parameter~$H$.
Also, an overdot and an overprime shall denote differentiation with 
respect to the cosmic time~$t$ and the conformal time~$\eta$, respectively.
Moreover, $N$ shall denote the number of e-folds.


\section{Models of ultra slow roll and punctuated inflation}\label{sec:mi}

In this section, we shall briefly describe the specific models of interest 
that lead to ultra slow roll and punctuated inflation.
We should mention that all the five models that we shall discuss in the following
two subsections contain a point of inflection.
Recall that, the first slow roll parameter is defined as $\epsilon_1=-\dot{H}/H^2$. 
The higher order slow roll parameters are defined in terms of the first slow
roll parameter $\epsilon_1$ through the relations 
\begin{equation}
\epsilon_{n+1}=\f{\d\, \mathrm{ln}\,\epsilon_n}{\d N}   
\end{equation}
for $n \geq 1$.
As it is the first three slow roll parameters, viz. $\epsilon_1$,  $\epsilon_2$, 
and $\epsilon_3$, that determine the amplitude and shape of the power spectrum
as well as the bispectrum, we shall illustrate the behavior of these slow roll 
parameters in the models of interest.


\subsection{Potentials leading to ultra slow roll inflation}

We shall consider two specific models that permit ultra slow roll inflation.
The first potential we shall consider which leads to a period of ultra slow 
roll inflation is often written in the following form (see, for instance,
ref.~\cite{Garcia-Bellido:2017mdw}):
\begin{equation}
V(\phi) = V_0\;
\f{6\,x^2 - 4\,\alpha\,x^3 + 3\,x^4}{(1 + \beta\,x^2)^2},\label{eq:phi4}
\end{equation}
where $x = \phi/v$, with $v$ being a constant rescaling factor.
We shall work with the following choices of the parameters involved:~$V_0/\Mpl^4 
= 4\times10^{-10}$, $v/\Mpl = \sqrt{0.108}$, $\alpha = 1$ and $\beta = 1.4349$.
For these choices of parameters, the inflection point, say, $\phi_0$, is located 
at $0.39\,\Mpl$.
We find that, if we choose the initial value of the field to be 
$\phi_\mathrm{i}=3.614\,\Mpl$, then inflation lasts for about $63$~e-folds 
in the model.
For convenience, we shall hereafter refer to the potential~\eqref{eq:phi4}, 
along with the above-mentioned set of parameters, as USR1.

\par

The second potential that we shall consider is given by~\cite{Dalianis:2018frf}
\begin{equation}
V(\phi) = V_0\,\l\{\mathrm{tanh}\l(\f{\phi}{\sqrt{6}\,\Mpl}\r) 
+ A\,\sin\l[\f{\mathrm{tanh}\l[\phi/\l(\sqrt{6}\,\Mpl\r)\r]}{f_\phi}\r]\r\}^2,
\label{eq:D1}
\end{equation}
and we shall work with the following values of the parameters involved: $V_0/\Mpl^4 
= 2\times10^{-10}$, $A = 0.130383$ and $f_\phi = 0.129576$.
We find that, for these values of the parameters, the inflection point occurs at 
$\phi_0 = 1.05\,\Mpl$.
For the initial value of the field $\phi_\mathrm{i}=6.1\,\Mpl$, we obtain about
$66$~e-folds of inflation in the model.
We shall refer to the potential~\eqref{eq:D1} and the above set of parameters 
as~USR2.

\par

\begin{figure}[!t]
\begin{center}
\includegraphics[width=7.50cm]{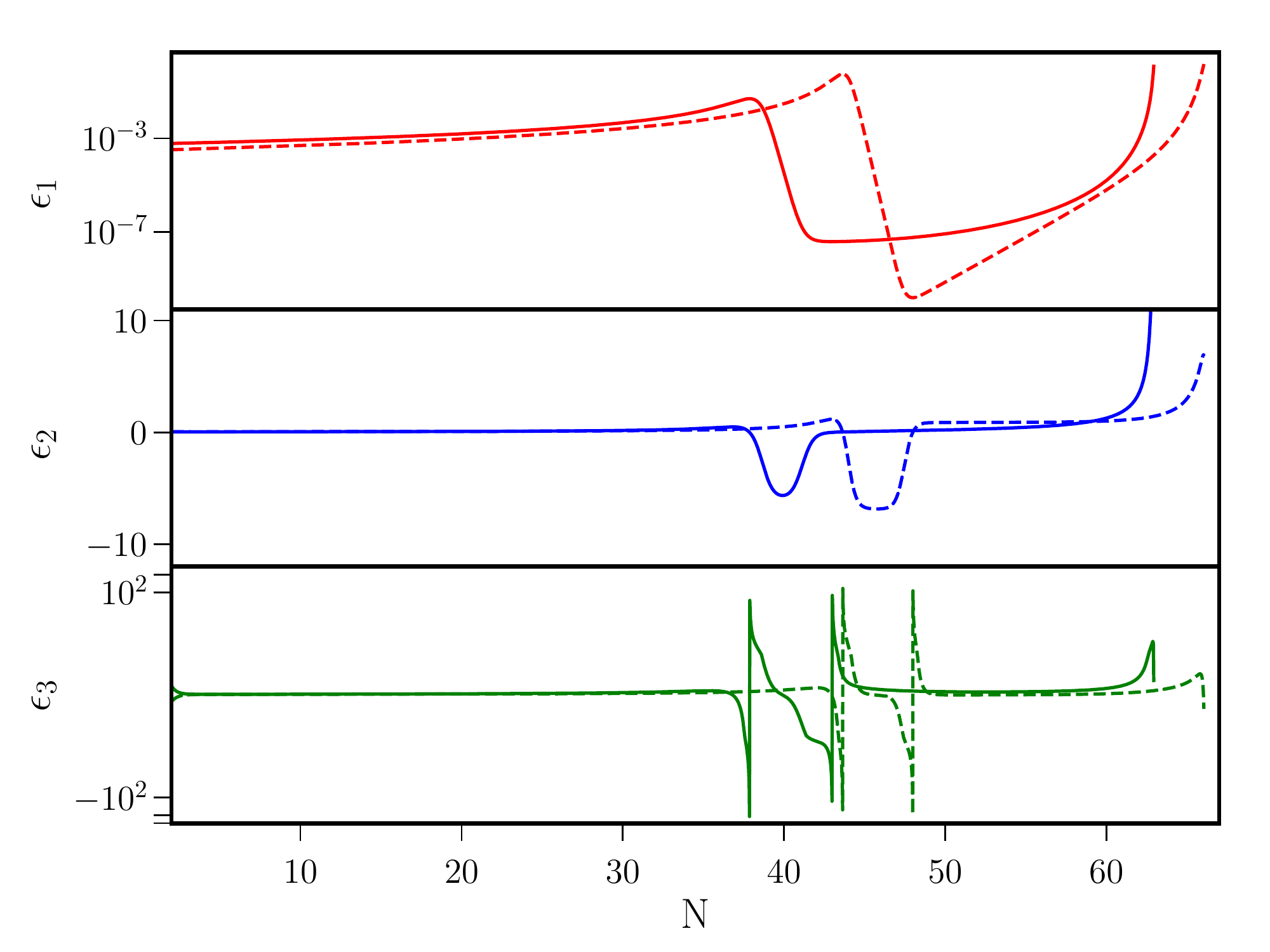}
\hskip 5pt
\includegraphics[width=7.50cm]{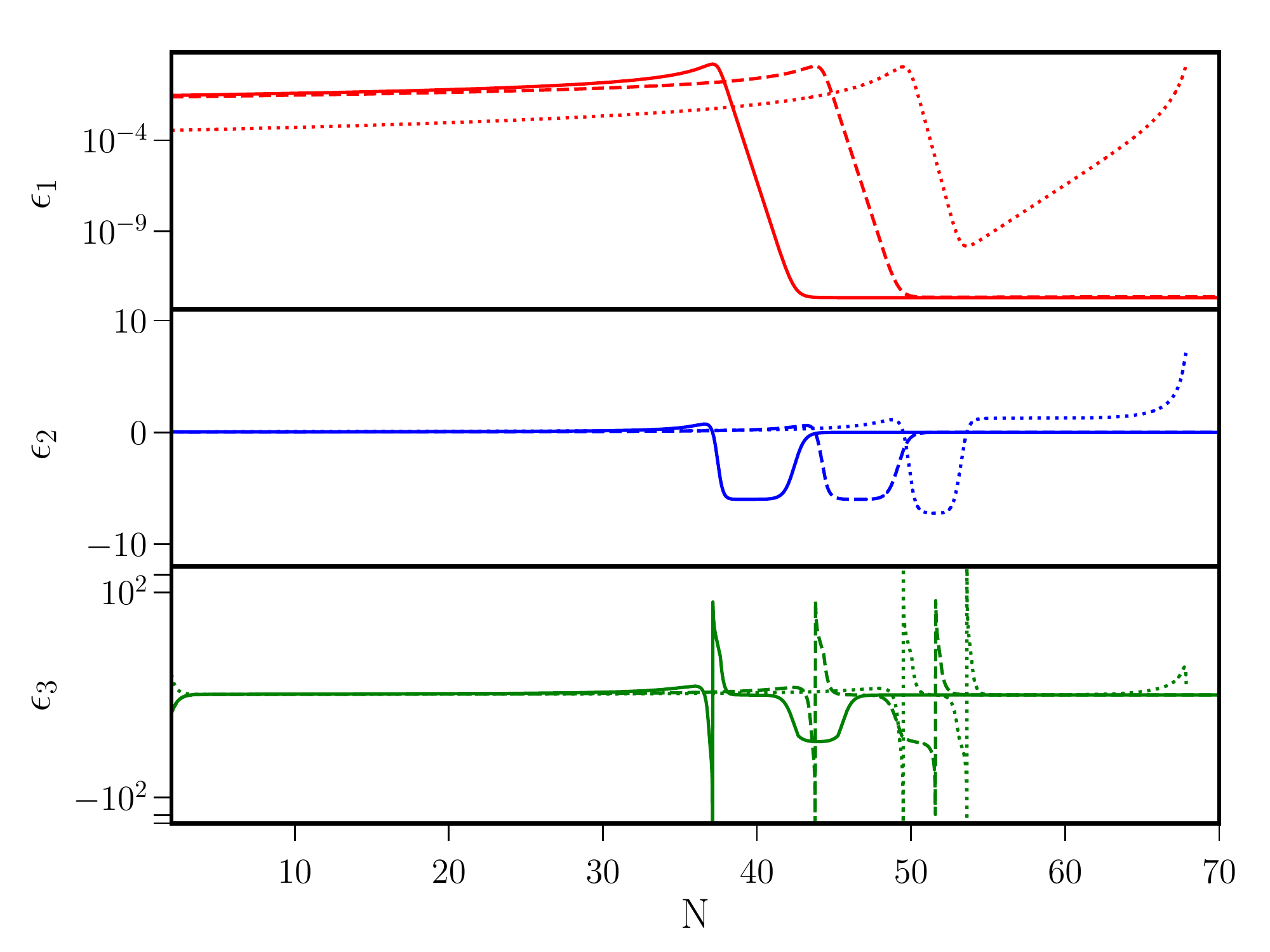}
\end{center}
\vskip -15pt
\caption{The behaviors of the first three slow roll parameters $\epsilon_1$ (on
top), $\epsilon_2$ (in the middle) and $\epsilon_3$ (at the bottom) have been 
plotted in the models of interest which lead to ultra slow roll and punctuated 
inflation.
We have plotted the behaviors for all the five models we have discussed,
viz. USR1 and USR2 (as solid and dashed curves, on the left) as well as 
PI1, PI2 and PI3 (as solid, dashed and dotted curves, on the right).
Note that all the models consist of two distinct regimes of slow roll and
ultra slow roll inflation, while the punctuated inflationary models also
contain a short period of departure from inflation.}\label{fig:epss-usr-pi}
\end{figure}

As we mentioned, the background dynamics driven by these potentials can be 
well captured by the behavior of the first three slow roll parameters
$\epsilon_1$, $\epsilon_2$ and $\epsilon_3$.
We have plotted the evolution of these quantities as a function of e-folds~$N$
in figure~\ref{fig:epss-usr-pi}. 
It is clear from the behavior of $\epsilon_1$ that these models permit two 
different regimes of slow roll, separated by a short phase of departure 
from slow roll. 
Note that the value of $\epsilon_1$ during the second regime of slow roll is
a few orders of magnitude smaller than its value during the initial regime,
thereby leading to the nomenclature of ultra slow roll inflation.
We should point out that there is no deviation from inflation in these models, 
as the first slow roll parameter always remains smaller than unity until the 
very end of inflation. 
The transition from slow roll to ultra slow roll is rather rapid and this 
aspect is reflected by the sharp rise and fall in the amplitude of the 
second and third slow roll parameters within a short period. 
It should also be highlighted that the second slow roll parameter~$\epsilon_2$ 
is large and negative (about $-6$ and $-7$ in USR1 and USR2) during the ultra 
slow phase when the first slow roll parameter~$\epsilon_1$ is rapidly decreasing. 
The parameter~$\epsilon_2$ changes sign when $\epsilon_1$ begins to rise 
as the field crosses the point of inflection and rolls down towards the 
minimum of the potential.
But, $\epsilon_2$ continues to remain relatively large (it is about $0.2$ 
and $0.9$ in the cases of USR1 and USR2) even during this latter phase, 
when compared to the typical slow roll values encountered, say, at early 
times before the transition to the epoch of ultra slow roll.

\par 

\begin{figure}[!t]
\begin{center}
\includegraphics[width=7.5cm]{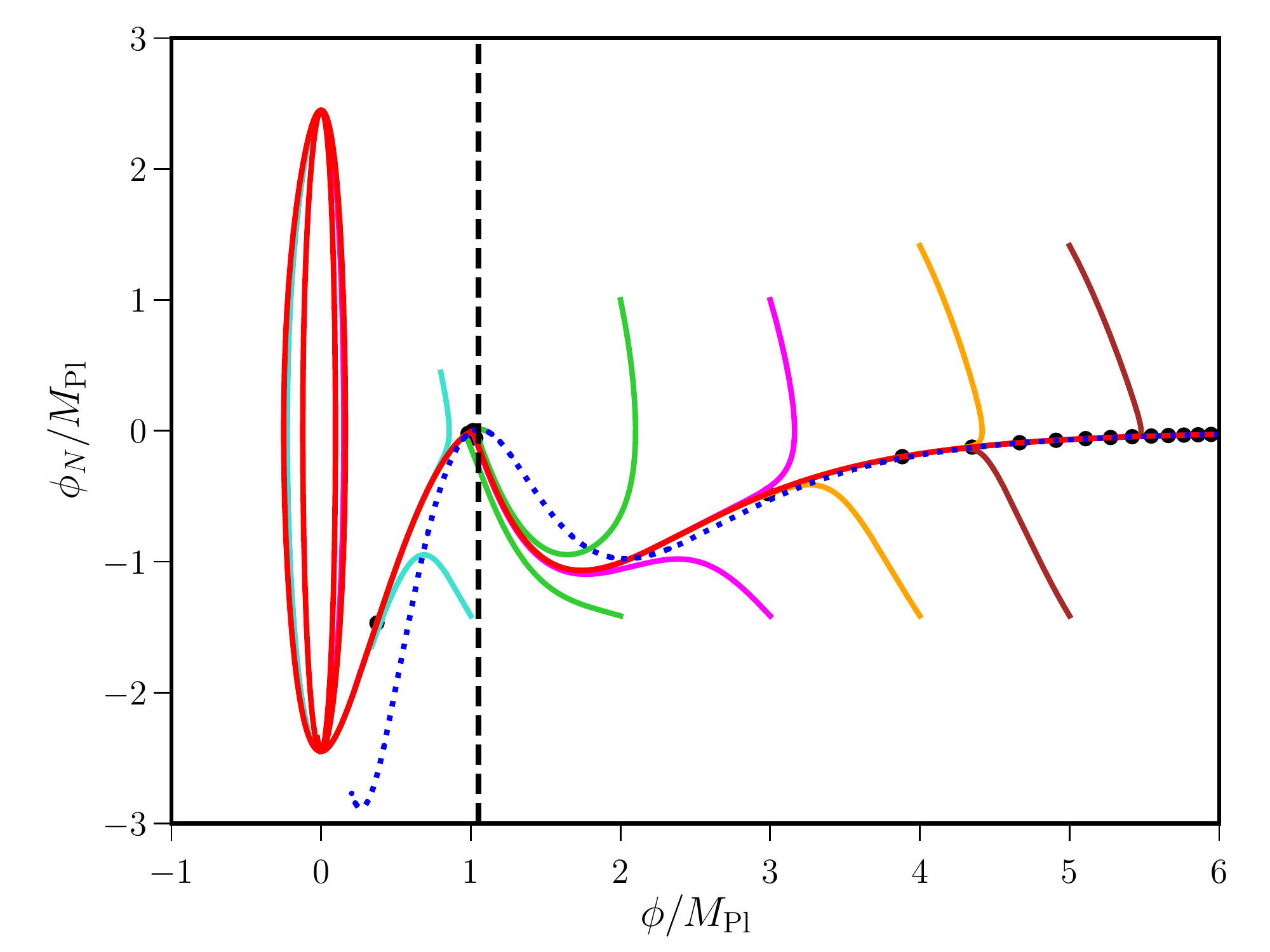}
\includegraphics[width=7.5cm]{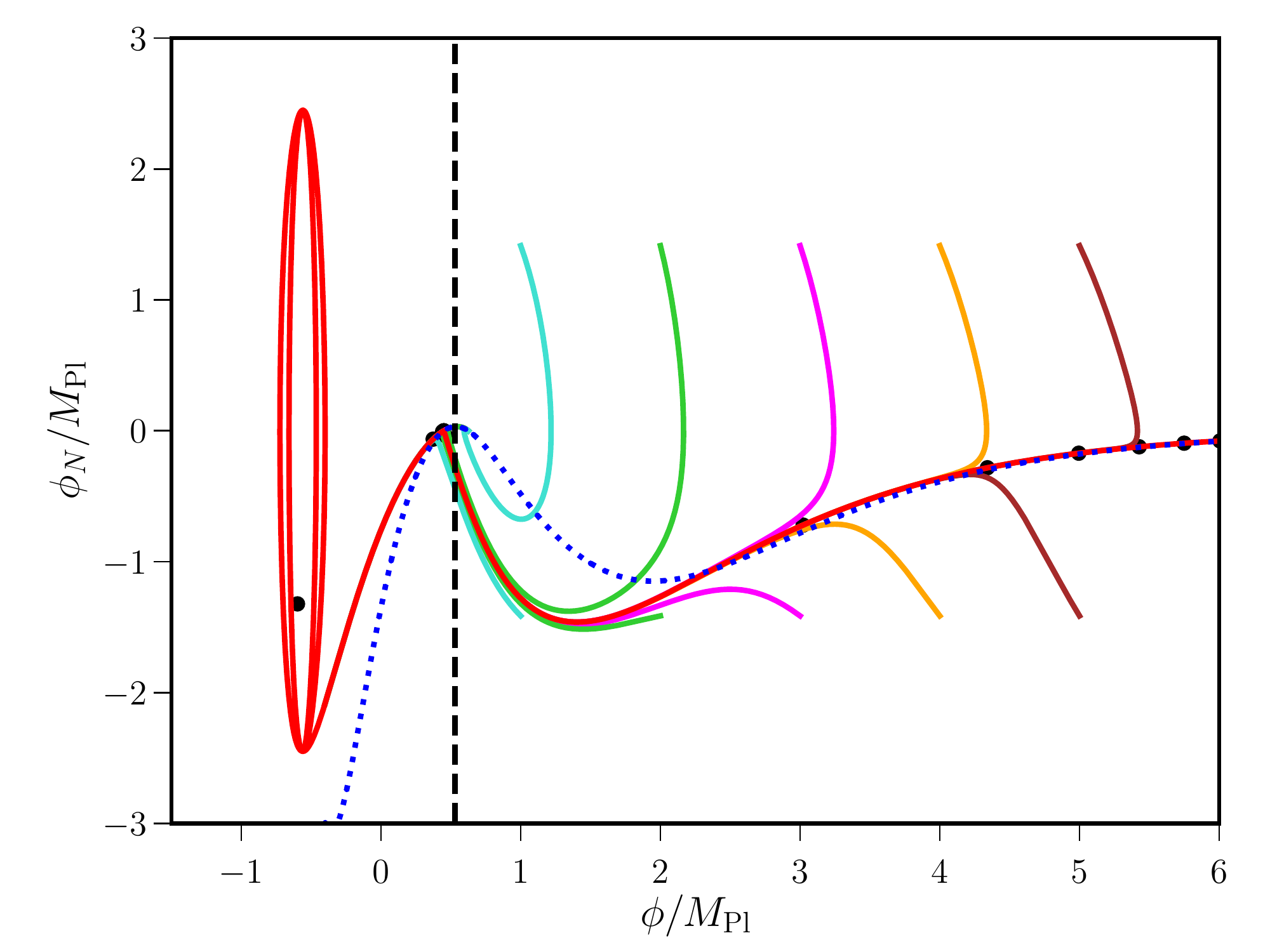}
\end{center}
\vskip -15pt
\caption{The dynamics of the scalar field in the phase space $\phi$-$\phi_N$,
where $\phi_N=\d \phi/\d N$, has been illustrated for the models USR2 (on the
left) and PI3 (on the right).
Apart from the trajectory for the specific initial conditions we shall be 
working with (plotted in red), we have also plotted the evolution for a few
other initial conditions (as solid curves in different colors). 
Moreover, in the case of the primary trajectory, we have indicated the lapse
in time every $3$~e-folds (as black dots on the red curves).
Further, we have highlighted the evolution arrived at using the standard slow 
roll approximation (as dotted blue curves).
Note that the vertical lines (in dashed black) identify the point of inflection.}
\label{fig:phase}
\end{figure}

To gain a better understanding of the dynamics involved, in figure~\ref{fig:phase}, 
we have also plotted the evolution of the scalar field in phase space for the 
case of USR2.
Evidently, trajectories from different initial conditions eventually merge 
with the primary trajectory of interest.
The transition to the ultra slow roll regime corresponds to the sharp upward 
turn in the phase space trajectory when the velocity of the field decreases
as it nears the point of inflection.
It is interesting to note that the solution obtained in the slow roll
approximation closely follows the primary trajectory even during the ultra 
slow roll regime. 
The field crosses the point of inflection, eventually emerging from the ultra 
slow regime, and inflation ends as the field approaches the minimum of the
potential.


\subsection{Potentials permitting punctuated inflation}

As we have discussed, punctuated inflation corresponds to a scenario wherein 
a short period of departure from inflation is sandwiched between two epochs
of slow roll.
With the help of specific examples, we shall illustrate that the period of 
departure from inflation is inevitably followed by an epoch of ultra slow roll 
inflation.

\par

A simple model that has been examined in the early literature which permits 
interrupted inflation is described by the potential (see ref.~\cite{Roberts:1994ap};
also see refs.~\cite{Leach:2000yw,Leach:2001zf})
\begin{equation}
V(\phi) = V_0\,\l(1 + B\,\phi^4\r).\label{eq:pi-phi4}
\end{equation}
It should be evident that the inflection point for this model is located
at $\phi = 0$.
For $B/\Mpl^4 = 0.5520$, one finds that the model leads to two epochs of 
inflation separated by a brief interruption of inflation.
In fact, around the interruption, the first slow roll parameter rises above 
unity and quickly falls to very small values, resulting in a period of 
ultra slow roll.
It is easy to argue that such a behavior arises due to the constant 
term~$V_0$ in the potential~\cite{Roberts:1994ap}.
But, the presence of the constant term simultaneously leads to an 
important drawback of the model.
Once inflation is restored after the interruption, it is found that 
the eventual slow roll regime lasts forever.
There is no conventional termination of inflation as the constant 
term~$V_0$ sustains slow roll evolution even when the field has 
reached the bottom of the potential.
So, one is either forced to terminate inflation by hand or invoke an 
additional source to end inflation.
Despite these drawbacks, we shall nevertheless briefly discuss the model 
due to its simplicity.
We shall work with the above-mentioned value for the parameter~$B$ and 
choose $V_0/\Mpl^4 = 8 \times 10^{-13}$.
We shall set the initial value of the field to be $\phi_\mathrm{i}= 17\,
\Mpl$, and we shall assume that inflation ends after~$70$ e-folds.
We shall hereafter refer to this model as~PI1.

\par

The second potential that we shall consider can be expressed as (see, for 
instance, refs.~\cite{Allahverdi:2006we,Jain:2008dw,Jain:2009pm})
\begin{equation}
V(\phi) = \frac{m^2}{2}\,\phi^2
- \l(\f{\sqrt{2\,\lambda\,(n-1)}\,m}{n}\r)\,\phi^n 
+ \f{\lambda}{4}\,\phi^{2\,(n-1)},\label{eq:pi-pg}
\end{equation}
where~$n$ is an integer.
These potentials contain a point of inflection at  
\begin{equation}
\phi_0=\l[\f{2\,m^2}{\lambda\,(n-1)}\r]^{1/[2\,(n-2)]}.
\end{equation}
We shall focus on the case $n=3$, wherein the potential above reduces to
\begin{equation}
V(\phi) = \f{m^2}{2}\,\phi^2 - \f{2\,m^2}{3\,\phi_0}\, \phi^3 
+ \f{m^2}{4\,\phi_0^2}\,\phi^4,\label{eq:pi-p}
\end{equation}
and we shall work with the following values of the parameters: $m/\Mpl= 1.8 
\times 10^{-6}$ and $\phi_0/\Mpl = 1.9777$.
As we shall soon discuss, these choice of parameters indeed admit 
punctuated inflation.
However, one finds, as in the case of PI1, the above potential (for 
the parameters mentioned) does not naturally result in an end of 
inflation.
Despite this limitation, we shall discuss the model, since, it should
be clear that, modulo the denominator, the potential describing 
USR1 [cf. eq.~\eqref{eq:phi4}] is essentially the same as 
the potential~\eqref{eq:pi-pg}.
We shall choose the initial value of the field to be $\phi_\mathrm{i}
= 20\,\Mpl$, and we shall again assume that inflation ends after~$70$ e-folds.
We shall refer to this model as PI2.

\par

Another model we shall consider that permits punctuated inflation is
motivated by supergravity.
It is described by the potential (see ref.~\cite{Dalianis:2018frf}; 
for a very recent discussion, also see ref.~\cite{Dalianis:2020cla})
\begin{equation}
V(\phi) = V_0\,\l[c_0 + c_1\,\tanh\, \l(\f{\phi}{\sqrt{6\,\alpha}}\r) 
+ c_2\,\tanh^2\l(\f{\phi}{\sqrt{6\,\alpha}}\r)
+ c_3\,\tanh^3{\l(\f{\phi}{\sqrt{6\,\alpha}}\r)}\r]^2,\label{eq:pi-tanh}
\end{equation}
and we shall work with the following values for the parameters involved: 
$V_0/\Mpl^4=2.1 \times 10^{-10}$, $c_0=0.16401$, $c_1=0.3$, $c_2=-1.426$,
$c_3=2.20313$ and $\alpha=1$.
This model too contains a point of inflection and, for the above values for
the parameters, the inflection point is located at $\phi_0 = 0.53\,\Mpl$.
If we choose the initial value of the field to be $\phi_\mathrm{i} = 7.4\,\Mpl$, 
we find that inflation ends after about $68$ e-folds.
We shall refer to this model as PI3. 
For the above choice of the parameters, apart from a plateau for large field 
values, the potential admits a second plateau at smaller values of the field.
As we shall see soon, it is these aspects of the potential that permits 
punctuated inflation and thereby aids in boosting the scalar power spectrum 
at small scales.

\par

As in the case of the ultra slow roll models we had discussed in the previous
sub-section, we have plotted the first three slow roll parameters $\epsilon_1$,
$\epsilon_2$ and $\epsilon_3$ for the models PI1, PI2 and PI3 in 
figure~\ref{fig:epss-usr-pi}.
It is easy to see from the plots that the behavior of the three slow roll 
parameters are very similar across the models and they differ only in their 
location of the departures from slow roll.
Evidently, after an initial slow roll regime, a brief departure from inflation 
occurs with $\epsilon_1$ growing above unity.
The interruption of inflation is immediately followed by a period of ultra slow
roll with $\epsilon_1$ falling to a value that is considerably smaller than its
value during the initial slow roll regime. 
Moreover, other than PI3, the models have no definite end of inflation since
$\epsilon_1$ does not rise to unity once the ultra slow roll regime has 
begun. 
Further, note that, when the epoch of ultra slow roll sets in, as in USR1 and 
USR2, the second slow roll parameter $\epsilon_2$ turns large and negative in 
all the cases of PI1, PI2 and PI3.   
The parameter $\epsilon_2$ eventually approaches zero in the cases of PI1 and 
PI2, since the first slow roll parameter never rises from its very low values
in these models. 
However, in PI3, since $\epsilon_1$ rises ultimately leading to the end of
inflation, the second slow roll parameter~$\epsilon_2$ eventually turns 
positive (from nearly $-7$) and attains a large value (around $1.2$), in 
very much the same manner it had in USR2.
As with USR2, we have plotted the behavior of the field in phase space for the
case of PI3 in figure~\ref{fig:phase}.
It should be clear from the figure that the velocity of the field reaches 
larger values in the case of PI3 than in the case of USR2 prior to entering
the ultra slow roll regime.
Evidently, it is this behavior that is responsible for the brief interruption
of inflation. 


\section{Evolution of the curvature perturbation and power spectra}\label{sec:ps}

In this section, we shall discuss the scalar and tensor power spectra that 
arise in the models permitting ultra slow roll and punctuated inflation
we had introduced in the previous section.
However, before we go on to discuss the power spectra, we shall illustrate
the behavior of the curvature perturbations during the period of deviation 
from slow roll.
Specifically, we shall highlight the role played by the intrinsic entropy
perturbations in the enhancement of the amplitude of the curvature 
perturbations over wave numbers that leave the Hubble radius either immediately
prior to or during the departure from slow roll.


\subsection{Scalar and tensor modes, and power spectra}\label{subsec:ps}

Let $\cR$ and $\gamma_{ij}$ denote the curvature and the tensor perturbations 
at the first order, respectively.
Also, let~$\cR_\vk$ and~$\gamma_{ij}^\vk$ denote the Fourier modes associated
with these perturbations. 
Recall that the modes~$\cR_\vk$ and~$\gamma_{ij}^\vk$ satisfy the differential
equations
\begin{subequations}\label{eq:de-p}
\begin{eqnarray}
\cR_\vk''+2\, \f{z'}{z}\, \cR_\vk' + k^2\, \cR_\vk &=& 0,\label{eq:de-fk}\\
{\gamma_{ij}^\vk}''+2\, \f{a'}{a}\, {\gamma_{ij}^\vk}' + k^2\, \gamma_{ij}^\vk 
&=& 0,\label{eq:de-gk}
\end{eqnarray}
\end{subequations}
where $z=\sqrt{2\,\epsilon_1}\,\Mpl\,a$, with $\epsilon_1$ being the first 
slow roll parameter.
Moreover, note that, if~$\hat{\cR}_\vk$ and~$\hat{\gamma}_{ij}^\vk$ denote the 
operators associated with the scalar and tensor modes on quantization, the 
scalar and tensor power spectra $\ps(k)$ and $\pt(k)$ are defined in terms of
these operators through the relations
\begin{subequations}
\begin{eqnarray}
\label{eq:twopointfourier}
\langle\hat{\cR}_{\vk}(\ee)\, 
{\hat \cR}_{\vk'}(\ee)\rangle 
&=&\f{2\, \pi^2}{k^3}\; {\cal P}_{_{\rm S}}(k)\;
\delta^{(3)}\l(\vk+\vk'\r),\label{eq:sps-d}\\
\langle\, {\hat \gamma}_{ij}^{\vk}(\ee)\,
\hat{\gamma}^{ij}_{\vk'}(\ee)\,\rangle
&=&\f{2\,\pi^2}{k^3}\,{\mathcal P}_{_{\rm T}}(k)\;
\delta^{(3)}(\vk+\vk'),\label{eq:tps-d}
\end{eqnarray}
\end{subequations}
where $\ee$~is the conformal time at late times, close to the end of 
inflation.
We should mention that, in the above expressions, the expectation values 
on the left hand side are to be evaluated in the specified initial quantum 
state, which we shall assume to be the Bunch-Davies vacuum.
Let $f_k$ and $g_k$ denote the positive frequency modes (associated with
the Bunch-Davies vacuum) in terms of which the operators~$\hat{\cR}_\vk$
and~$\hat{\gamma}_{ij}^\vk$ are decomposed. 
Then, in terms of the quantities $f_k$ and $g_k$, the power spectra $\ps(k)$ 
and $\pt(k)$ can be expressed as
\begin{subequations}
\begin{eqnarray}
\ps(k)
&=&\f{k^3}{2\, \pi^2}\, \vert f_k(\ee)\vert^2,\label{eq:sps}\\
\pt(k)
&=& 8\,\f{k^3}{2\, \pi^2}\, \vert g_k(\ee)\vert^2.\label{eq:tps}
\end{eqnarray}
\end{subequations}


\subsection{Role of the intrinsic entropy perturbation}

Often the evolution of the curvature perturbations in 
non-trivial scenarios involving departures from slow roll inflation are 
examined in terms of the behavior of the quantity~$z$ (see, for instance,
Refs.~\cite{Dalianis:2018frf,Ozsoy:2018flq,Ozsoy:2019lyy}).
We find that it proves to be instructive to understand this 
aspect from the behavior of the intrinsic entropy 
perturbations~\cite{Leach:2000yw,Jain:2007au}. 
It is well known that, in contrast to perfect fluids, scalar fields, 
in general, possess non-vanishing non-adiabatic pressure perturbation 
$\delta p_{_{\mathrm{NA}}}$ or, equivalently, the intrinsic entropy
perturbation~$\cS$, which are related through the expression (in this 
context, see, for example, refs.~\cite{Gordon:2000hv,Unnikrishnan:2010ag})
\begin{equation}
\delta p_{_{\mathrm{NA}}}=\f{p'}{\mathcal H}\,\cS,
\end{equation}
where $p$ denotes the pressure associated with the background and $\mathcal{H}
=a\,H$ is the conformal Hubble parameter.
In the case of inflation driven by a single, canonical scalar field, one can
show that the intrinsic entropy perturbation~$\cS_k$ associated with a given
mode of the field can be expressed in terms of the corresponding curvature
perturbation, say, $\cR_k$, as follows~\cite{Leach:2000yw,Jain:2007au}:
\begin{equation}
\cR_k'=-\l[\f{2\,a^2\,p'}{\Mpl^2\, (\mathcal{H}'-\mathcal{H}^2)}\r]
\,\l(\f{1}{1-c_{_{\mathrm{A}}}^2}\r)\, \cS_k,
\end{equation}
where $c_{_{\mathrm{A}}}=\sqrt{p'/\rho'}$ is adiabatic speed of the scalar 
perturbations, with $\rho$ being the background energy density.
It is easy to show using the equation of motion~\eqref{eq:de-fk} describing 
the curvature perturbation that, in the super Hubble limit, the intrinsic 
entropy perturbation~$\cS_k$ decays as~$\mathrm{e}^{-2\,N}$.
However, it is found that, during deviations from slow roll, for modes which 
are either about to leave or have just left the Hubble radius, the amplitude
of the intrinsic entropy perturbation briefly increases, sourcing the
curvature perturbation~\cite{Leach:2001zf,Jain:2007au}.
This, in turn, alters the amplitude of the curvature perturbation for modes 
which cross the Hubble radius just before or during the departure from slow 
roll.

\par

\begin{figure}[!t]
\begin{center}
\includegraphics[width=7.50cm]{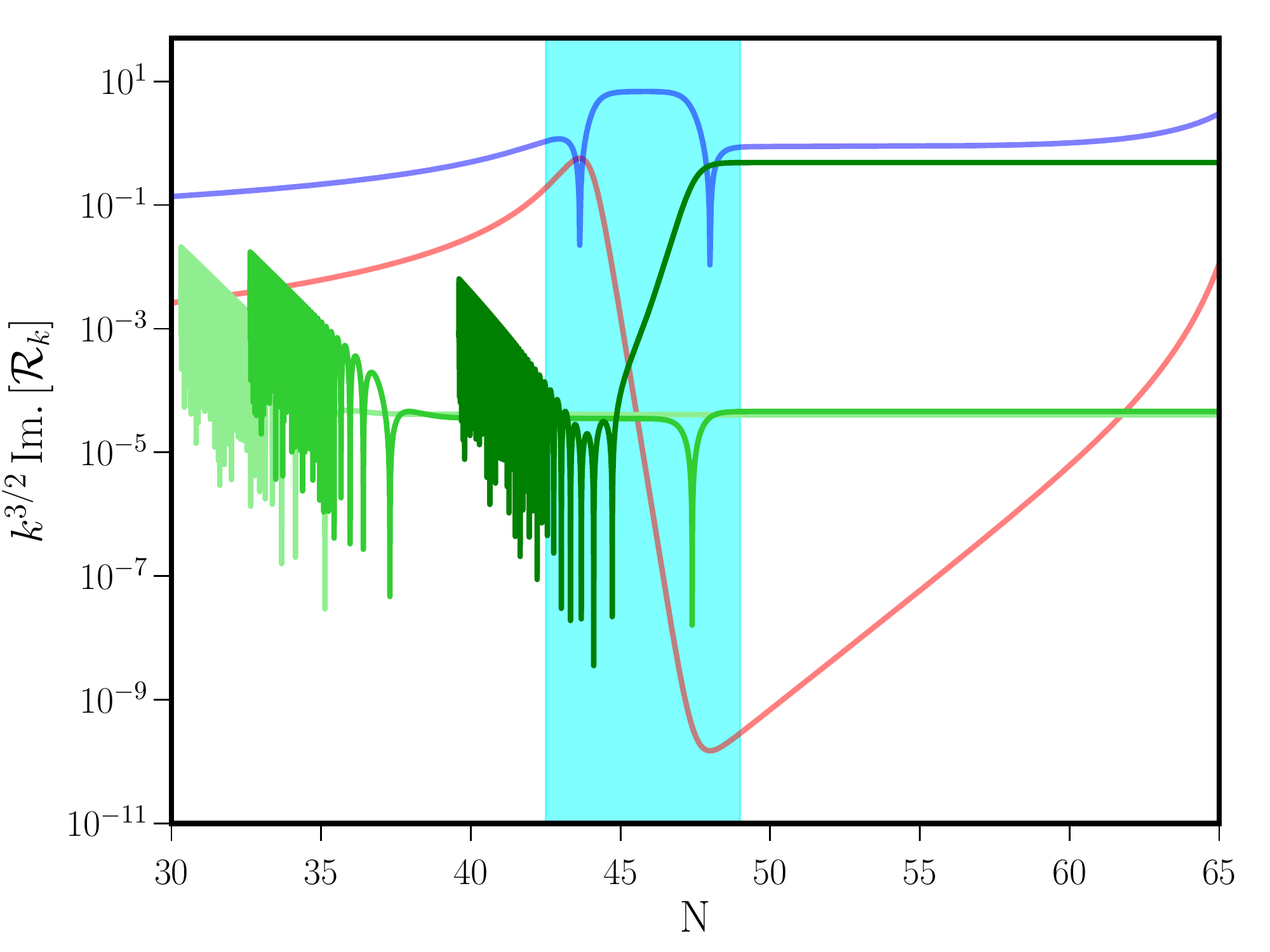}
\includegraphics[width=7.50cm]{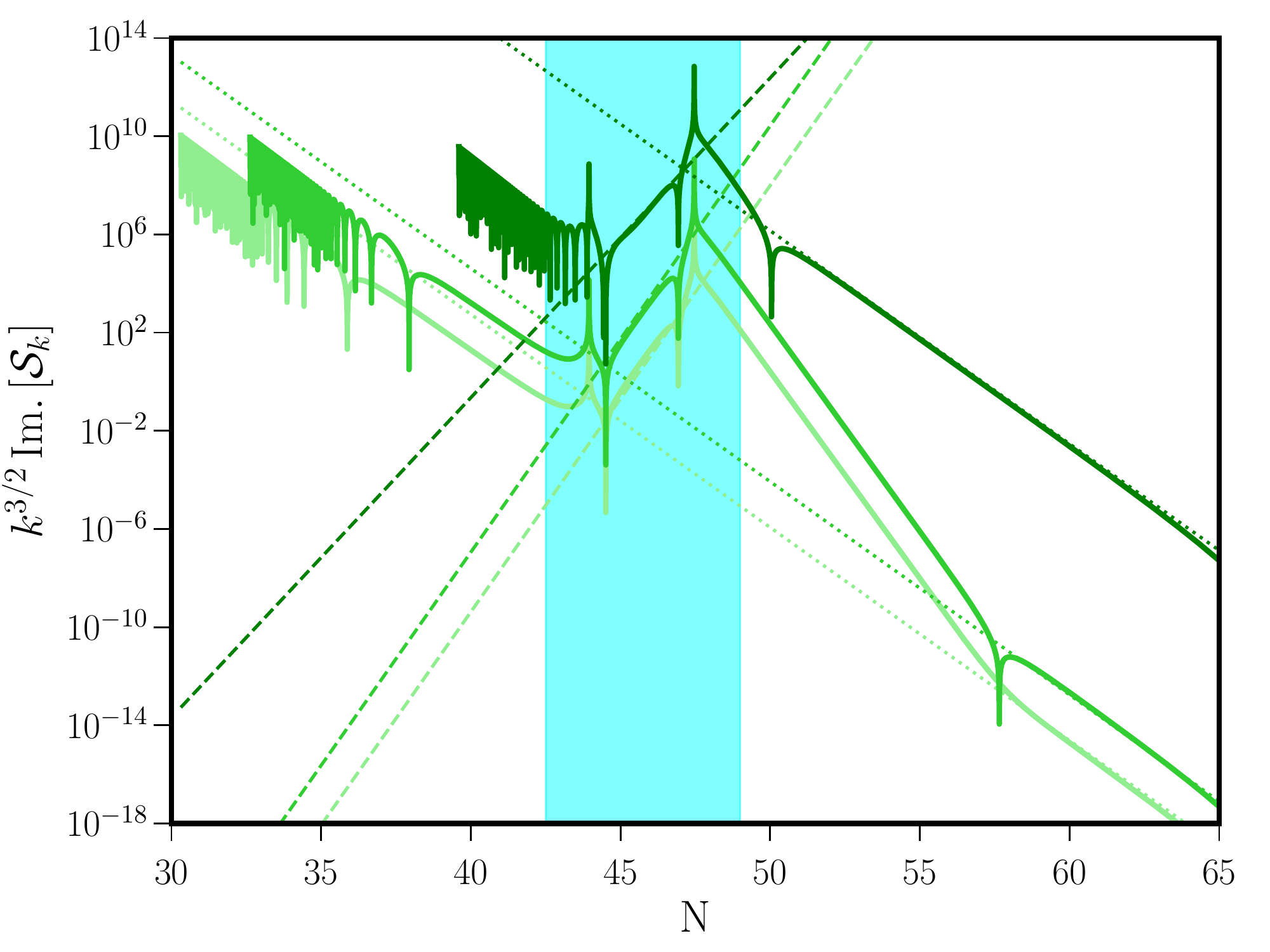}
\vskip 10pt
\includegraphics[width=7.50cm]{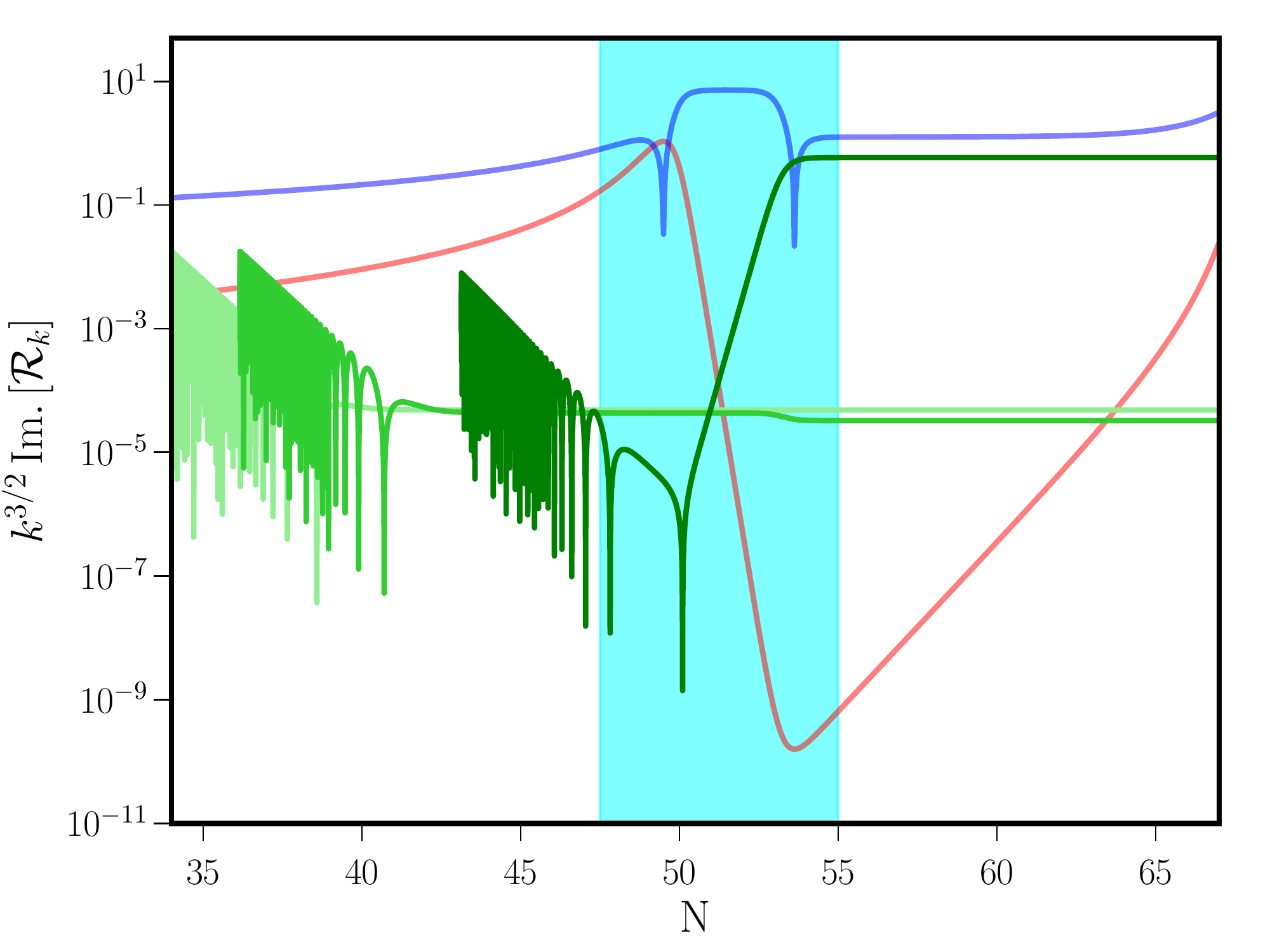}
\includegraphics[width=7.50cm]{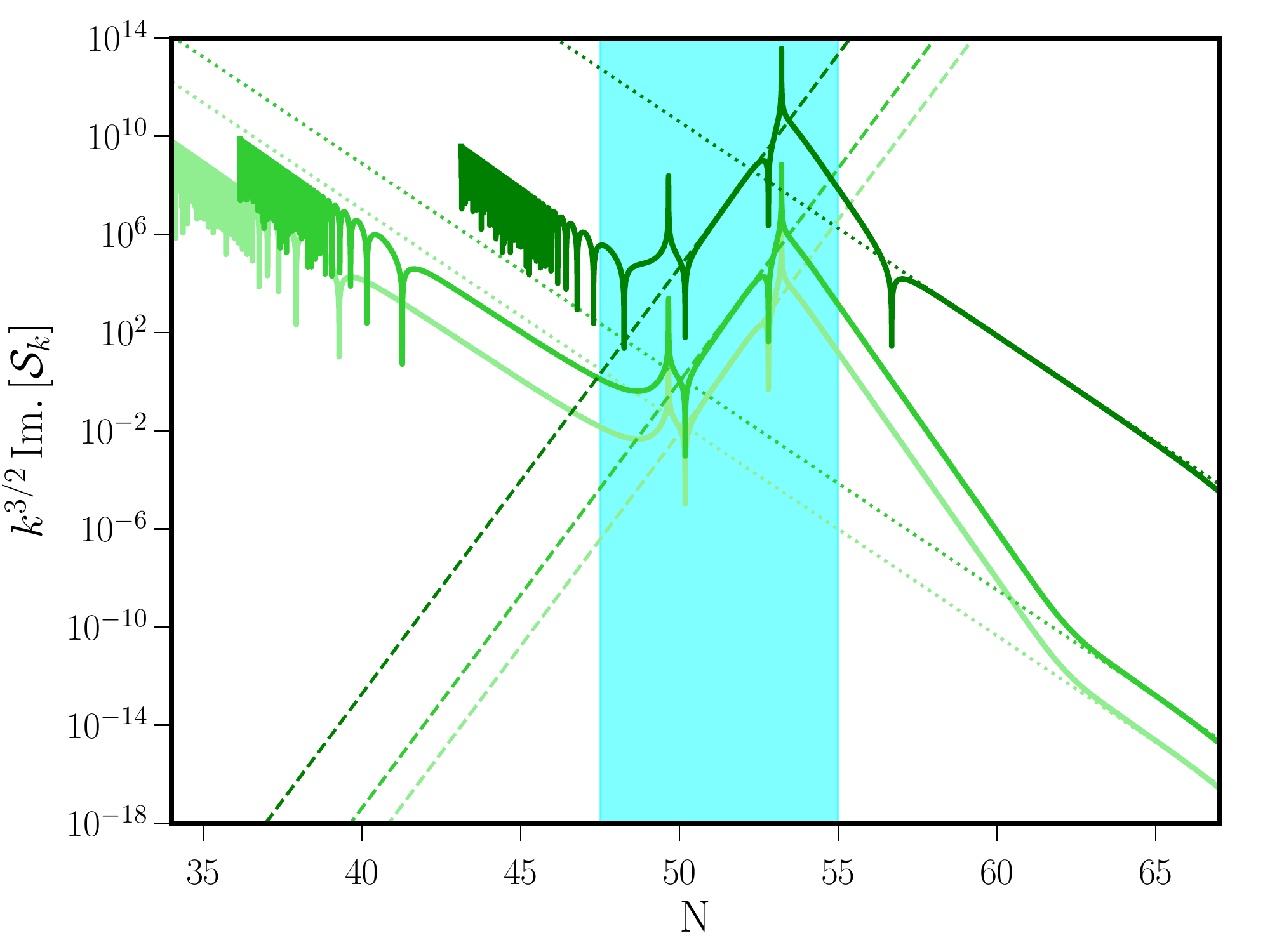}
\end{center}
\vskip -15pt
\caption{The evolution of the amplitudes of the imaginary parts of the 
curvature perturbation~$\cR_k$ (on the left) and the corresponding
intrinsic entropy perturbation~$\cS_k$ (on the right) have been 
plotted for the three wave numbers $k = 10^{10}\, \mathrm{Mpc}^{-1}$,
$10^{11}\,\mathrm{Mpc}^{-1}$ and $10^{14}\,\mathrm{Mpc}^{-1}$ (in 
light, lime and dark green, respectively) in the two models USR2 
(on top) and PI3 (at the bottom) as a function of e-folds.
We have also included the behavior of the first two slow roll parameters
$\epsilon_1$ and $\vert\epsilon_2\vert$ (in red and blue, respectively,
on the left) in these models to indicate the regime (demarcated by the
cyan band) over which the transition from slow roll to ultra slow roll 
occurs.
The first mode with the smallest wave number is already in the super-Hubble 
regime when the departure from slow roll sets in, and the amplitude of the 
corresponding curvature perturbation is hardly affected by the transition. 
The second mode is barely in the super-Hubble regime when the transition 
from slow roll begins.
The amplitude of its curvature perturbation is slightly attenuated as it 
emerges from the departure from slow roll.
Whereas, the amplitude of the curvature perturbation associated with the 
third mode, which leaves the Hubble radius right in the middle of the 
transition, exhibits a considerable enhancement due to the transition.
These changes in the curvature perturbations can be attributed to the 
rapid growth in the corresponding entropy perturbations (plotted on 
the right) during the transition.
We find that $\cS_k$ grows as either $\mathrm{e}^{3\,N}$ or 
$\mathrm{e}^{4\,N}$ (indicated as dashed lines) during the transition. 
We also find that the entropy perturbations eventually die down as 
$\mathrm{e}^{-2\,N}$ in the super-Hubble limit (indicated by dotted
lines) as expected.
It is these behaviors that lead to features in the inflationary scalar 
power spectra.}\label{fig:modes-usr2-pi3}
\end{figure}

To demonstrate these effects, in figure~\ref{fig:modes-usr2-pi3}, we have 
plotted the evolution of the curvature and the intrinsic entropy perturbations 
in the inflationary models USR2 and PI3. 
In order to highlight the differences in the behavior of the modes, we have 
plotted the evolution of the amplitudes for three modes which leave the Hubble 
radius just prior to the start of the departure from slow roll inflation, 
immediately after start of the period of transition, and during the middle 
of the transition.
We should point out that we have plotted the imaginary parts of~$\cR_k$
and~$\cS_k$ since they dominate at late times.
Moreover, they allow us to highlight the oscillations in the sub-Hubble regime.
The time when these oscillations cease is an indication that the modes have
crossed the Hubble radius.
Evidently, there is a sharp rise in the amplitude of the intrinsic entropy 
perturbation for all the modes during the departure from slow roll inflation.
We should add here that the corresponding real parts of $\cR_k$ and $\cS_k$ 
behave in a roughly similar manner.
It is the sharp rise in $\cS_k$ that is responsible for either an enhancement 
or a suppression in the asymptotic (i.e. late time) amplitude of the curvature 
perturbation, thereby leading to features in the power spectrum (for 
related discussions in this context, also see, for instance,  
refs.~\cite{Dalianis:2018frf,Cicoli:2018asa}).
In contrast, we find that there is relatively little effect of the deviation 
from slow roll on the evolution of the amplitude of the tensor perturbations.
Due to this reason, the tensor power spectrum exhibits far less sharper 
features than the scalar power spectrum.


\subsection{Scalar and tensor power spectra}

\begin{figure}[!t]
\begin{center}
\includegraphics[width=15cm]{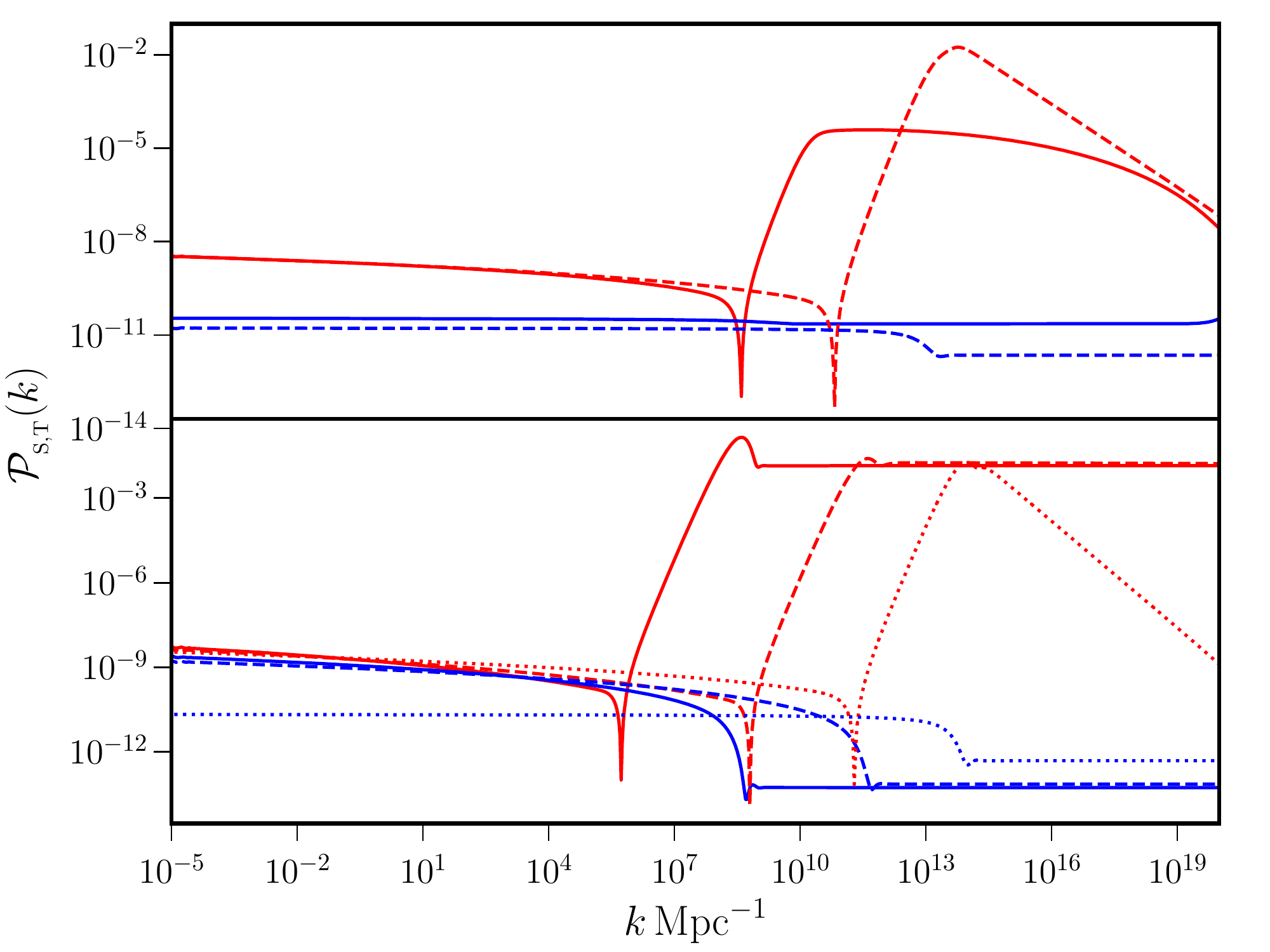}
\end{center}
\vskip -15pt
\caption{The scalar (in red) and tensor power spectra (in blue) have been 
plotted in the various ultra slow roll and punctuated inflationary models 
of our interest~---~USR1 and USR2 (as solid and dashed curves, on top) 
and PI1, PI2, and PI3 (as solid, dashed and dotted curves, at the
bottom)~---~over a wide range of scales.
Note that the enhancement of power on small scales is more in the case of 
USR2 than USR1.
Moreover, in the case of the punctuated inflationary models, the scalar 
power in PI1 and PI2 do not eventually come down at very small scales
due to the fact that inflation does not terminate in these models.
We should also point out that, in contrast to the scalar power spectra,  
the tensor power spectra have lower power at small scales when compared
to the large scales.}\label{fig:pps-usr-pi}
\end{figure}

We shall now turn to the scalar and tensor power spectra that arise in the 
ultra slow roll and punctuated inflationary scenarios we had discussed in 
the last section.
Barring the brief rise of $\epsilon_1$ above unity in the models of punctuated
inflation and the location of the deviations from slow roll inflation, we had 
seen that the behavior of the first three slow roll parameters were very similar 
in the different models of our interest (cf. figure~\ref{fig:epss-usr-pi}).
We can expect these features to be reflected in the corresponding power spectra.
In figure~\ref{fig:pps-usr-pi}, we have plotted the power spectra arising in
all the five models, viz. USR1, USR2, PI1, PI2 and PI3. 

\par

We shall first point out the features in the scalar power spectra that are 
common to all the models.
All the models exhibit a rise in scalar power on small scales corresponding 
to modes that leave the Hubble radius during the second stage of slow roll.
Moreover, the location of the rise in power is determined by the time when 
the deviation from slow roll occurs.
This is due to the fact that, as we discussed in the previous subsection,
it is the amplitude of the modes which exit the Hubble radius during the 
phase of departure from slow roll that are enhanced compared to the amplitudes 
of modes which leave during the initial phase of slow roll.
Further, the modes that exit the Hubble radius during the epoch of ultra 
slow roll carry the imprints of the extremely small values of the first 
slow roll parameter and hence exhibit higher amplitudes.

\par

Let us now consider the power spectra in the models USR1 and USR2.
The location of features in the spectra is determined by the finely 
tuned values of parameters of the potential and the time when the modes 
leave the Hubble radius.
Note that both USR1 and USR2 have a definite end of inflation.
Let us say that the pivot scale $k_\ast=0.05\,\mathrm{Mpc}^{-1}$ 
leaves the Hubble radius $N_\ast$ number of e-folds prior to the 
end of inflation.
For USR1 and USR2, to arrive at the power spectra
plotted in figure~\ref{fig:pps-usr-pi}, we have assumed that 
$N_\ast=(50.0,56.2)$.
The occurrence of a peak in the scalar power spectra 
at small scales in these models can be easily understood if we recall 
the behavior of the slow roll parameters in these cases.
Note that, in slow roll inflation, the scalar spectral index $\ns$ is 
given in terms of the first two slow roll parameters as $\ns = 1 - 2\, 
\epsilon_1 - \epsilon_2$. 
Though the regime of our interest does not strictly correspond to slow 
roll dynamics, we can utilize this relation to roughly understand the 
rise and fall of the scalar power spectra. 
We had earlier mentioned that, as $\epsilon_1$ decreases rapidly during
the epoch of ultra slow roll and eventually rises from its very small 
values, $\epsilon_2$ changes from relatively large negative values to 
positive values in USR1 and USR2.
Since $\epsilon_1$ is very small during the ultra slow roll regime, for
modes which leave around this epoch, the spectral index $\ns$ mimics the 
behavior of $-\epsilon_2$, changing from large positive values (corresponding 
to an initially blue spectrum) to negative values (corresponding to a 
red spectrum on smaller scales), leading to a peak in the power spectra.
Clearly, we also require that the power spectra at large scales 
are consistent with the current constraints on the scalar spectral 
index $\ns$ and the tensor-to-scalar ratio~$r$ from the CMB 
data~\cite{Ade:2015lrj,Akrami:2018odb}.
We find that the models USR1 and USR2 lead to $(\ns, r)= (0.945,0.015)$ 
and $(0.946,0.007)$ at the pivot scale.
We should add a word of caution in this regard.
The above values for $\ns$ and $r$ lie barely within the $2$-$\sigma$ 
limits on the respective parameters according to the latest constraints 
from Planck~\cite{Akrami:2018odb}.
Importantly, if one were to even slightly change the values of the model 
parameters, the features in the power spectra get considerably altered.
In other words, there is a severe fine tuning involved in arriving at the
desired power spectra, an aspect which is well known and has been 
highlighted earlier (in this regard, see, for instance, 
ref.~\cite{Germani:2017bcs}).

\par

Let us now turn to the power spectra arising in the punctuated inflationary
models.
Once again, we can understand the behavior of the spectra at small scales in 
these cases from the relation between the scalar spectral index and the slow 
roll parameters.
Recall that, while PI3 has a finite duration of inflation, there exists the 
problem of termination of inflation in the models PI1 and PI2.
Due to this reason, as should be evident from the power spectra plotted in
figure~\ref{fig:pps-usr-pi}, the power never comes down in PI1 and PI2 because
the eventual slow roll regime lasts for a long duration.
However, since the evolution of the slow roll parameters in PI3 mimic their
behavior in USR1 and USR2, the resulting scalar power spectrum exhibits a peak 
for the same reason that we discussed above, viz. the relatively large values
and the change in the sign of the second slow roll parameter~$\epsilon_2$.
For the three models of PI1, PI2 and PI3, we have set $N_\ast = (60.0, 60.0,
54.5)$ to arrive at their respective spectra presented 
in figure~\ref{fig:pps-usr-pi}.
We find that, for the choice of parameters that lead to COBE normalized
scalar amplitude on large scales, the scalar spectral index and the 
tensor-to-scalar ratio at the pivot scale prove to be 
$(\ns, r)= (0.885, 0.580)$, $(0.909, 0.461)$ and $(0.944, 0.009)$ in 
PI1, PI2 and PI3, respectively.
Evidently, PI1 and PI2 are ruled out due to the large tensor-to-scalar ratio 
(beyond the upper limits from Planck) generated on the CMB scales in these 
models. 
In contrast, PI3 leads to a rather small tensor-to-scalar ratio that is 
consistent with the bounds from the Planck data and also comes close to 
satisfying the constraints on~$\ns$~\cite{Ade:2015lrj,Akrami:2018odb}. 
As far as the extent of boosting the power on small scales and the tunability 
of the model parameters are concerned, PI3 seems to require the same extent of 
fine-tuning as USR1 and USR2.
In contrast to PI3, we find that it is easier to achieve sustained amplification 
of power over a wider range of scales in PI1 and PI2.
But, obviously, it is achieved at the high cost that inflation does not end within 
the desired duration, essentially making them unviable.
Nevertheless, we believe that there are lessons to be learnt from the simpler
models PI1 and PI2 and we will exploit the main features of these models to
reverse engineer desired potentials in the following section.

\par

Lastly, let us make a few remarks on the tensor power spectra that we obtain 
in the various models. 
Note that the tensor power spectra also exhibit a step-like feature in all the 
models, but the step is in the opposite direction as compared to the scalars,
with the amplitude of tensors at small scales being a few orders of magnitude
smaller than their amplitude over large scales~\cite{Jain:2008dw,Jain:2009pm,
Pi:2019ihn}.
This can be attributed to the fact that after the period of deviation from slow 
roll, the inflaton evolves over smaller values of the field and hence smaller 
values of the potential. 


\subsection{Challenges in constructing viable models}

With the experience of examining a handful of inflationary models, let us 
briefly summarize the challenges in constructing viable and well motivated 
models that lead to enhanced power on small scales.

\par

To begin with, we need to ensure that the scalar spectral index $\ns$ and 
the tensor-to-scalar ratio $r$ are consistent with the cosmological data 
over the CMB scales.
Moreover, in order to boost the extent of PBHs formed and the amplitude of 
the secondary GWs, we require enhanced power on small scales.
Simultaneously, we need to make sure that inflation ends in a reasonable 
number of (say, about $65$) e-folds.
It is found that, as one attempts to resolve one issue, say, reduce the level
of fine tuning or permit room to shift the location of the features in the 
scalar power spectrum, another difficulty, such as the prolonged duration of
inflation, creeps in.

\par

We should point out here that, a given potential which admits ultra slow 
roll inflation for a set of values of the parameters involved may permit 
punctuated inflation for another set (in this context, see 
appendix~\ref{app:d-usr-pi}).
For that reason, we should stress that the potentials themselves cannot 
always be classified as ultra slow roll or punctuated inflationary models.
Hence, the dichotomy of ultra slow roll and punctuated inflationary scenarios
that we have created may be considered somewhat artificial.
However, we find it intriguing that whenever a potential admits restoration
of inflation after a brief interruption, it seems to naturally result in a 
regime of ultra slow roll inflation.
We believe that this aspect ought to be exploited to construct well motivated 
and viable canonical, single field inflationary models that also lead to 
enhanced PBH formation and generate secondary GWs of significant amplitudes.

\par

With the eventual aim of overcoming these difficulties in single, canonical
scalar field models of inflation, we shall now attempt to reconstruct
potentials that possess the desired features.


\section{Reverse engineering potentials admitting ultra slow roll
and punctuated inflation}\label{sec:re}

In this section, we shall assume specific time-dependence for the first
slow roll parameter $\epsilon_1$ so that it leads to ultra slow roll or 
punctuated inflation.
With the functional form of $\epsilon_1(N)$ at hand, we shall reconstruct 
the potentials using the equations of motion for the background and evaluate
the resulting scalar and tensor power spectra that arise in the different 
scenarios~\cite{Hertzberg:2017dkh,Byrnes:2018txb,Motohashi:2019rhu}.


\subsection{Choices of $\epsilon_1(N)$}

We shall consider the following two forms for $\epsilon_1(N)$ which 
lead to ultra slow roll or punctuated inflation for suitable choice 
of the parameters involved:
\begin{subequations}
\label{eq:eps1-12}
\begin{eqnarray}
\epsilon_1^{\mathrm{I}}(N) 
&=& \l[{\epsilon_{1a}\,\l(1+\epsilon_{2a}\,N\r)}\r]\,
\l[1 - {\mathrm{tanh}}\l(\f{N - N_1}{\Delta N_1}\r)\r] 
+ \epsilon_{1b} + \mathrm{exp}\l(\f{N - N_2}{\Delta N_2}\r),\label{eq:eps1-1}\\
\epsilon_1^{\mathrm{II}}(N) 
&=&\epsilon_1^{\mathrm{I}}(N) 
+ {\mathrm{cosh}}^{-2}\l(\f{N - N_1}{\Delta N_1}\r).\label{eq:eps1-2}
\end{eqnarray}
\end{subequations}
We find that considering a parametrization of the first 
slow roll parameter rather than the quantity $z$ or the scale factor~$a$ 
proves to be much more convenient and easy to model the scenarios of our
interest (in this context, see the recent 
efforts~\cite{DAmico:2020euu,Tasinato:2020vdk}).
The approach we adopt also allows us to easily ensure that the CMB constraints 
on large scales are satisfied.
The above forms of $\epsilon_1(N)$ are supposed to represent the ultra 
slow roll and the punctuated inflationary scenarios we had discussed 
earlier.
For convenience, we shall hereafter refer to the reconstructed inflationary
scenarios arising from the forms of $\epsilon_1(N)$ in eqs.~\eqref{eq:eps1-1}
and~\eqref{eq:eps1-2} as RS1 and RS2, respectively.
We shall now highlight a few points concerning the above constructions before 
proceeding to calculate the resulting power spectra. 

\par

Consider RS1 described by $\epsilon_1(N)$ in eq.~\eqref{eq:eps1-1}.
Note that the functional form contains seven parameters, viz. $\epsilon_{1a}$, 
$\epsilon_{1b}$, $\epsilon_{2a}$, $N_1$, $N_2$, $\Delta N_1$ and $\Delta N_2$.
For suitable choices of these parameters, this form of $\epsilon_1(N)$ leads to 
a period of slow roll followed by an epoch of ultra slow roll, before inflation 
eventually ends, as encountered in the ultra slow models USR1 and USR2 we had
discussed in the last section.
While $\epsilon_{1a}$ and $\epsilon_{1b}$ determine the values of the first slow 
roll parameter during slow roll and ultra slow roll, the parameters~$N_1$ and~$N_2$ 
determine the duration of these two phases.  
Note that the first term in the functional form~\eqref{eq:eps1-1} is expressed
as a product of two parts.
The first part involving the parameter $\epsilon_{2a}$ induces a small time
dependence during the early stages.
Such a time dependence is necessary to achieve slow roll inflation which leads 
to scalar and tensor power spectra that are consistent with the CMB data.
Recall that, in slow roll inflation, the scalar spectral index and the 
tensor-to-scalar ratio are given by $\ns = 1-2\,\epsilon_1-\epsilon_2$
and $r = 16\,\epsilon_1$, with the slow roll parameters evaluated at 
the time when the modes cross the Hubble radius.
For suitable choices of $\epsilon_{1a}$ and $\epsilon_{2a}$, we find that we 
can arrive at spectra that are consistent with the constraints on $\ns$ 
and $r$ from CMB, viz. $\ns = 0.9649 \pm 0.0042$ and $r < 0.056$ at the pivot
scale~\cite{Ade:2015lrj,Akrami:2018odb}.
The second part of the first term containing the hyperbolic tangent function
aids in the transition from the slow roll to the ultra slow roll phase around 
the e-fold~$N_1$.
We need to set~$N_1$ so that all the large scale modes leave the Hubble radius
during the first slow roll phase.

\par 

The second term $\epsilon_{1b}$ in equation~\eqref{eq:eps1-1} essentially 
prevents the first slow parameter~$\epsilon_1$ from reducing to zero 
beyond~$N_1$. 
Since $\epsilon_{1b}$ defines the ultra slow roll phase of the model, 
we shall choose the parameter to be much smaller than $\epsilon_{1a}$.
The last term involving the exponential factor has been included to
essentially ensure that $\epsilon_1$ rapidly rises at later times, 
crossing unity at~$N_2$, resulting in the termination of inflation. 
Lastly, the rapidity of the transitions from slow roll to ultra slow roll 
and from ultra slow roll to the end of inflation are determined by the 
parameters~$\Delta N_1$ and~$\Delta N_2$, respectively.
In summary, since  $\epsilon_{1a}$ and  $\epsilon_{2a}$ are constrained
by the CMB data on large scales, we have five free parameters, viz.
$\epsilon_{1b}$, $N_1$, $N_2$, $\Delta N_1$ and $\Delta N_2$, to construct 
the features we desire in the scalar power spectra over small scales.

\par

Let us now turn to RS2 with $\epsilon_1(N)$ described by eq.~\eqref{eq:eps1-2}.
In this case, evidently, the term involving the hyperbolic cosine function 
has been added to the form of $\epsilon_1(N)$ in RS1. 
This additional terms leads to a brief interruption of inflation around the 
e-fold~$N_1$, as is encountered in the punctuated inflationary models PI1, 
PI2, and PI3 discussed earlier.

\par

Both the constructions of~$\epsilon_1$ above have been motivated to simplify
the study of models containing an epoch of ultra slow roll with or without 
punctuation and thus producing inflationary spectra with either extended or 
localized features on small scales. 
The advantage of these constructions is that the parameters are easy to tune,
which allows us to directly infer the corresponding effects on the background 
dynamics and importantly on the power spectra, unlike the specific inflationary
models examined earlier.
Of course, this has been possible due to the fact the reconstructions involve
more parameters than the potentials we have considered.


\subsection{Reconstructed potentials and the corresponding scalar and tensor 
power spectra}

Using the Friedmann equations and the equation of motion governing the
inflaton, it is straightforward to show that the time evolution of the scalar 
field~$\phi(N)$ and the Hubble parameter~$H(N)$ can be expressed in terms 
of the slow roll parameter $\epsilon_1(N)$ as follows:
\begin{subequations}
\begin{eqnarray}
\phi(N) &=& \phi_\mathrm{i} 
-\Mpl\, \int^N_{N_\mathrm{i}}\d N\,\sqrt{2\,\epsilon_1(N)},\\
H(N) &=& H_\mathrm{i}\;\mathrm{exp}\l[-\int^N_{N_\mathrm{i}}\d N\,
\epsilon_1(N)\r],
\end{eqnarray}
\end{subequations}
where $\phi_\mathrm{i}$ and $H_\mathrm{i}$ are the values of the scalar field
and the Hubble parameter at some initial e-fold~$N_\mathrm{i}$.
We can use the above relations to arrive at the required background quantities
given a functional form for $\epsilon_1(N)$.
These background quantities can then be utilized to evaluate the resulting scalar 
and tensor power spectra.
It is useful to note that the potential $V(N)$ can be expressed in terms of 
the Hubble parameter and the first slow roll parameter as
\begin{equation}
V(N) = \Mpl^2\,H^2(N)\,\l[3-\epsilon_1(N)\r].
\end{equation}
Having obtained $\phi(N)$ and $V(N)$, clearly, we can construct $V(\phi)$ 
parametrically.

\par

\begin{figure}[!t]
\begin{center}
\includegraphics[width=7.50cm]{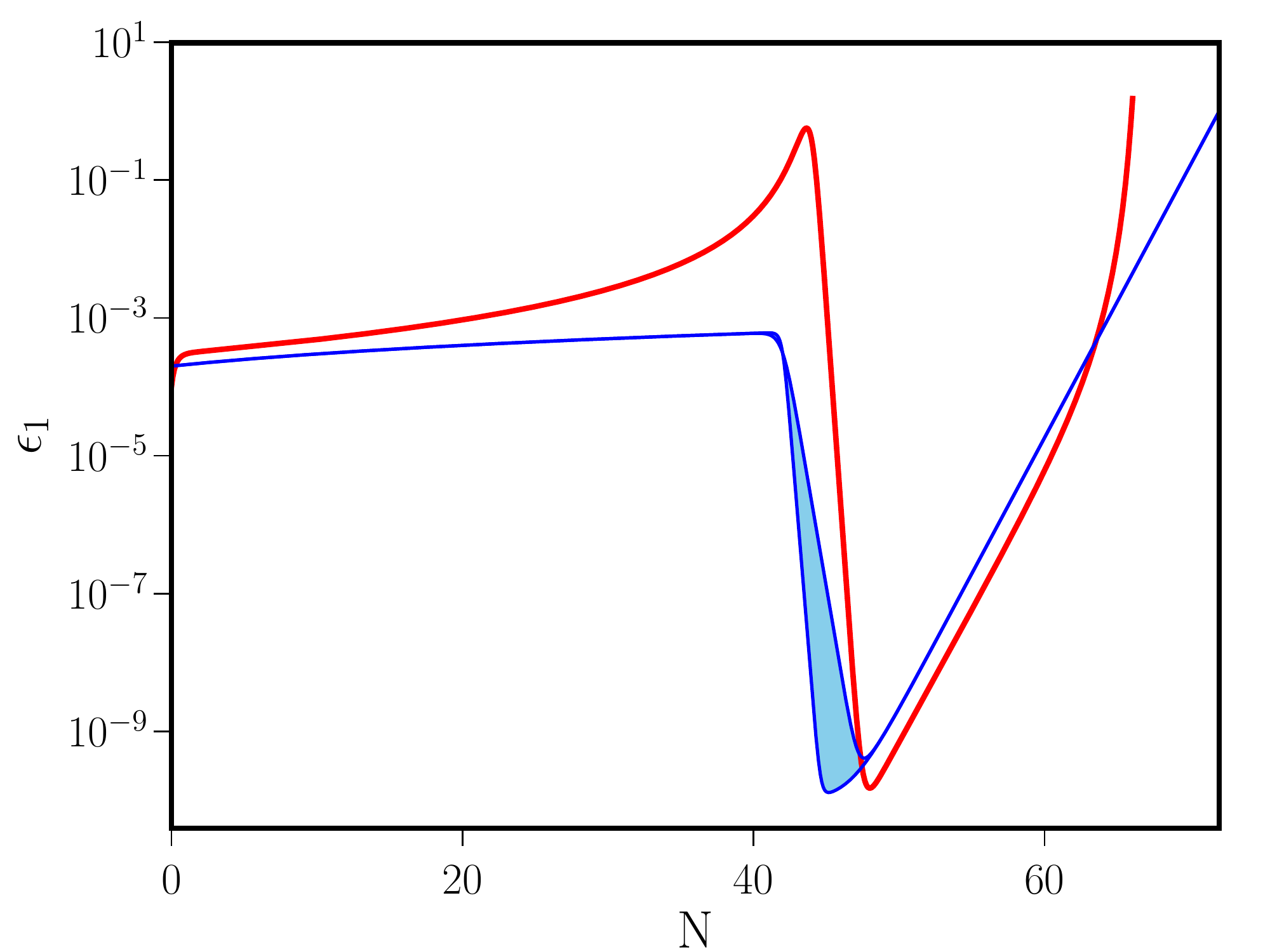}
\includegraphics[width=7.50cm]{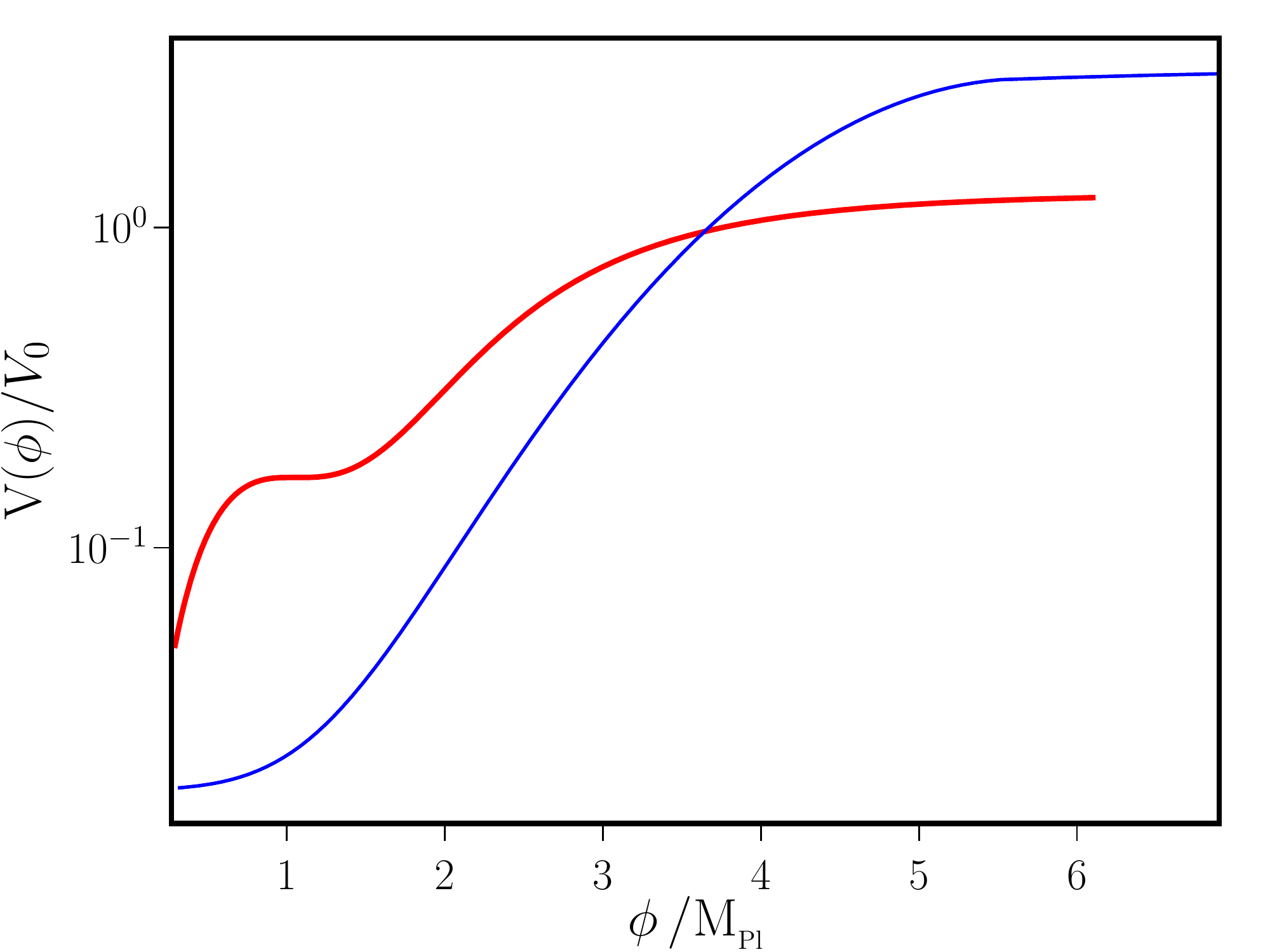}
\vskip 10pt
\includegraphics[width=7.50cm]{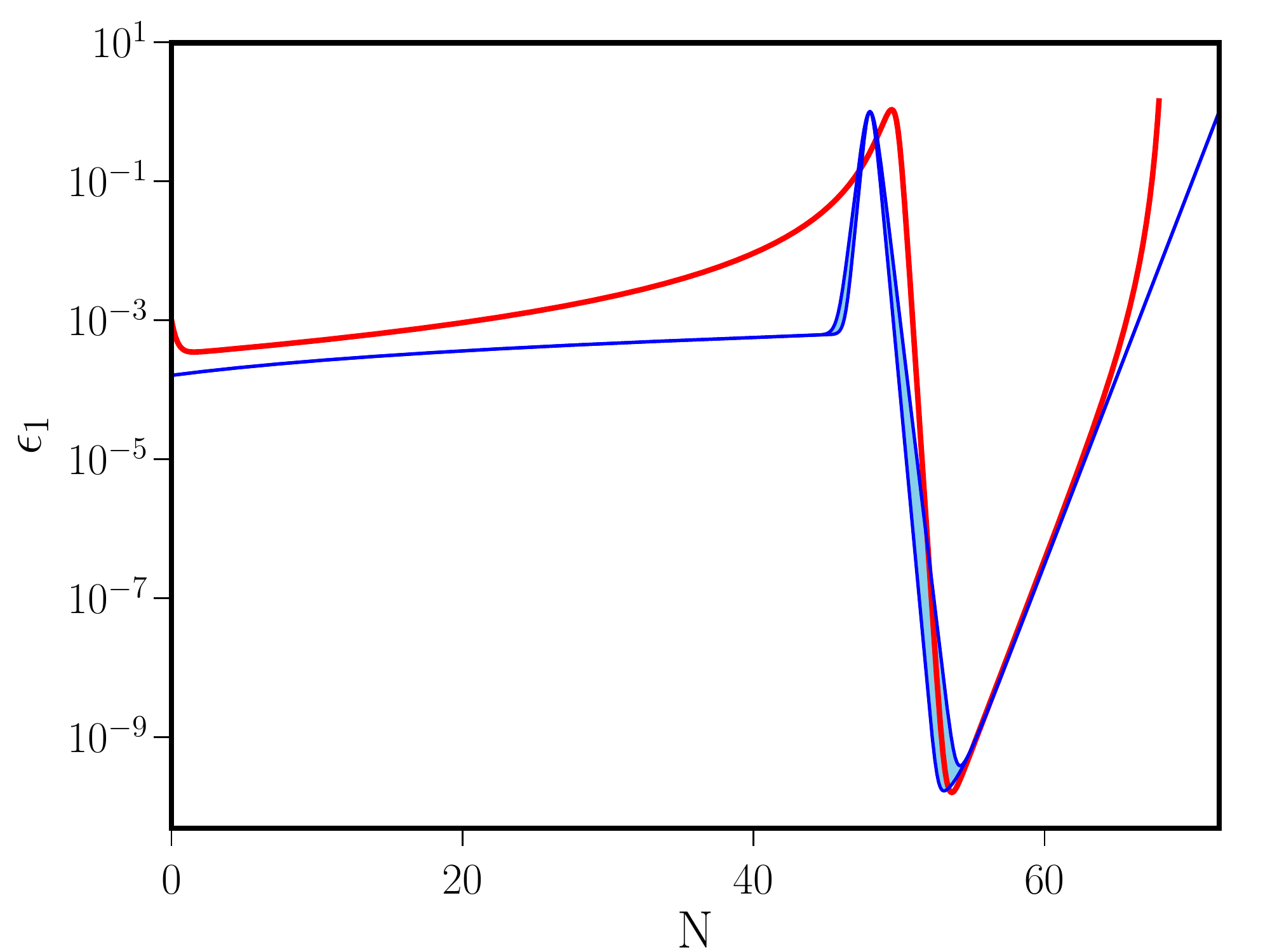}
\includegraphics[width=7.50cm]{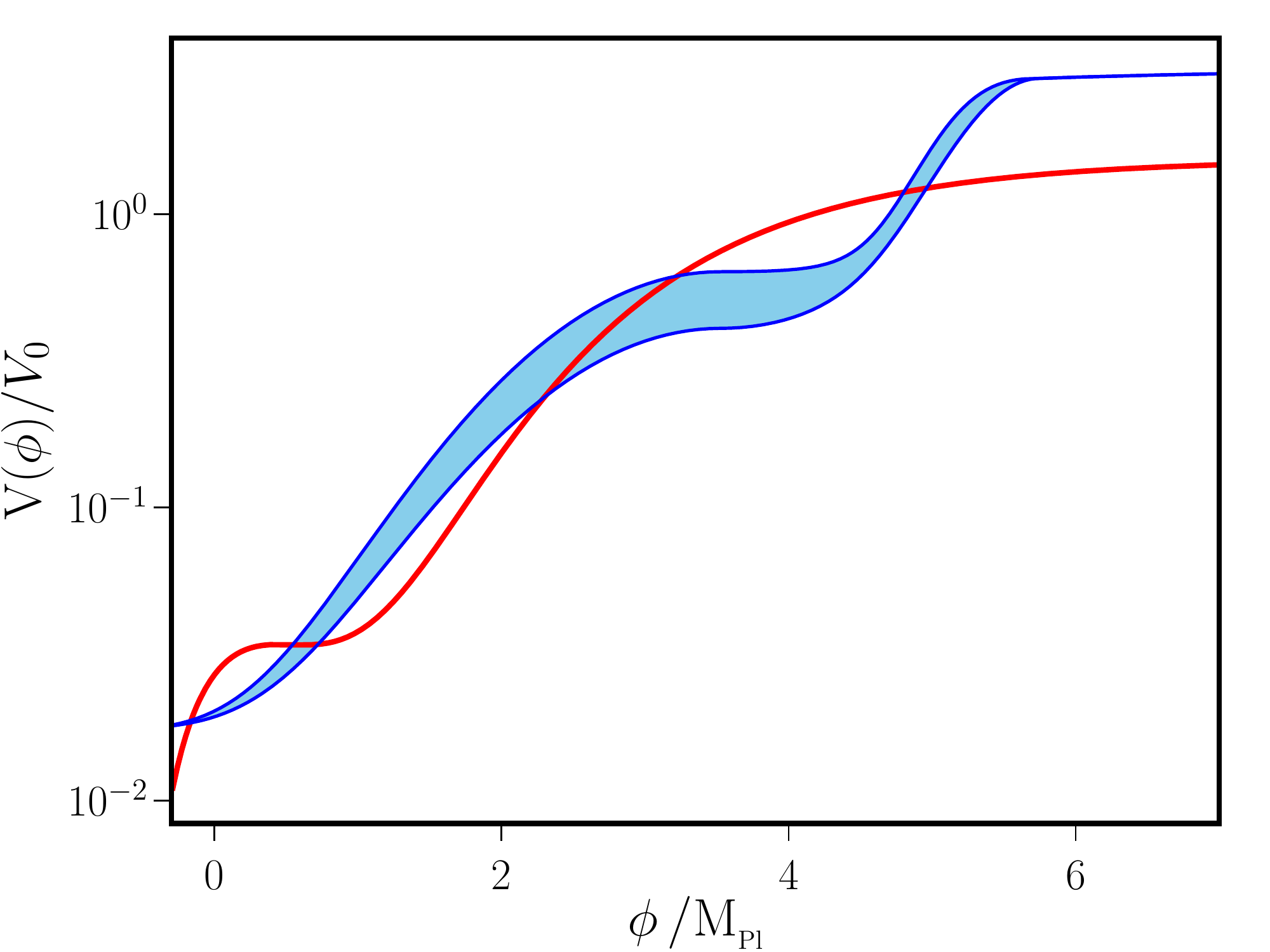}
\end{center}
\vskip -15pt
\caption{We have plotted the functional forms of $\epsilon_1(N)$ (in blue, on
the left) as well as the corresponding reconstructed potentials (in blue, on 
the right) in the cases of RS1 (on top) and RS2 (at the bottom) for suitable
values of the parameters involved. 
In fact, we have plotted the behavior in RS1 and RS2 as bands corresponding to 
a small range of the parameter $\Delta N_1$ which determines the duration of 
the transition from slow roll to ultra slow roll.
For comparison, we have also plotted the behavior of~$\epsilon_1$ (in red, on 
the left) and illustrated the potentials (in red, on the right) in the models 
USR2 (on top) and PI3 (at the bottom).
We have chosen the parameters in the cases of RS1 and RS2 so that they closely 
resemble the behavior of $\epsilon_1$ in the models USR2 and PI3.
Interestingly, we find that the reconstructed potentials always contain a point 
of inflection.
Note that, in the cases of RS1 and RS2, we have set $V_0=H_\mathrm{i}^2\,\Mpl^2$, 
which corresponds to $V_0=5.625\times 10^{-9}\,\Mpl^4$.}\label{fig:e1-V}
\end{figure}

In figure~\ref{fig:e1-V}, we have plotted the two choices~\eqref{eq:eps1-12}
for $\epsilon_1(N)$ and the corresponding potentials for a small range of
the parameter~$\Delta N_1$ that determines the duration of the transition 
from slow roll to ultra slow roll.
The parameters we have worked with in the case of the reconstructed 
scenario~RS1 are as follows: $\epsilon_{1a}=10^{-4}$, 
$\epsilon_{2a}=5\times 10^{-2}$,
$\epsilon_{1b} = 10^{-10}$, $N_1 = 42$, $N_2 = 72$ and $\Delta N_2 = 1.1$. 
We have varied the parameter $\Delta N_1$ over the range $(0.3345,0.7)$ 
to obtain the bands of $\epsilon_1$ and the corresponding potential in 
the figure.
Similarly, in the case of RS2, the parameters we have chosen to work
with are as follows: $\epsilon_{1a}=8\times 10^{-5}$, 
$\epsilon_{2a}=6.25\times 10^{-2}$,
$\epsilon_{1b}= 10^{-10}$, $N_1 = 48$, $N_2 = 72$
and $\Delta N_2 = 0.8$. 
The parameter $\Delta N_1$ has been varied over the range $(0.3847,0.5)$ 
to arrive at the bands of $\epsilon_1$ and the corresponding potential. 
We should note that the band describing the potential is more pronounced 
in the case of RS2 than in RS1. 
The choices for $\epsilon_{1a}$ and $\epsilon_{2a}$ have been made so that
the resulting power spectra are consistent with the Planck constraints on
the scalar spectral index~$\ns$ and the tensor-to-scalar ratio~$r$ at the 
pivot scale that we mentioned earlier.
For comparison, in the figure, we have also included the behavior 
of the first slow parameter as well as the form of the potential 
in the models USR2 and PI3.
It should be clear that, for suitable values of the parameters, our 
functional forms for $\epsilon_1(N)$ closely mimic the corresponding 
behavior in these models.
Moreover, from the parametric forms of $V(\phi)$ constructed numerically, 
we have been able to determine if the reconstructed potentials in the cases 
of RS1 and RS2 contain a point of inflection.
At an accuracy of $0.1\%$, we find that the reconstructed potentials 
indeed contain an inflection point.

\par

\begin{figure}[!t]
\begin{center}
\includegraphics[width=15cm]{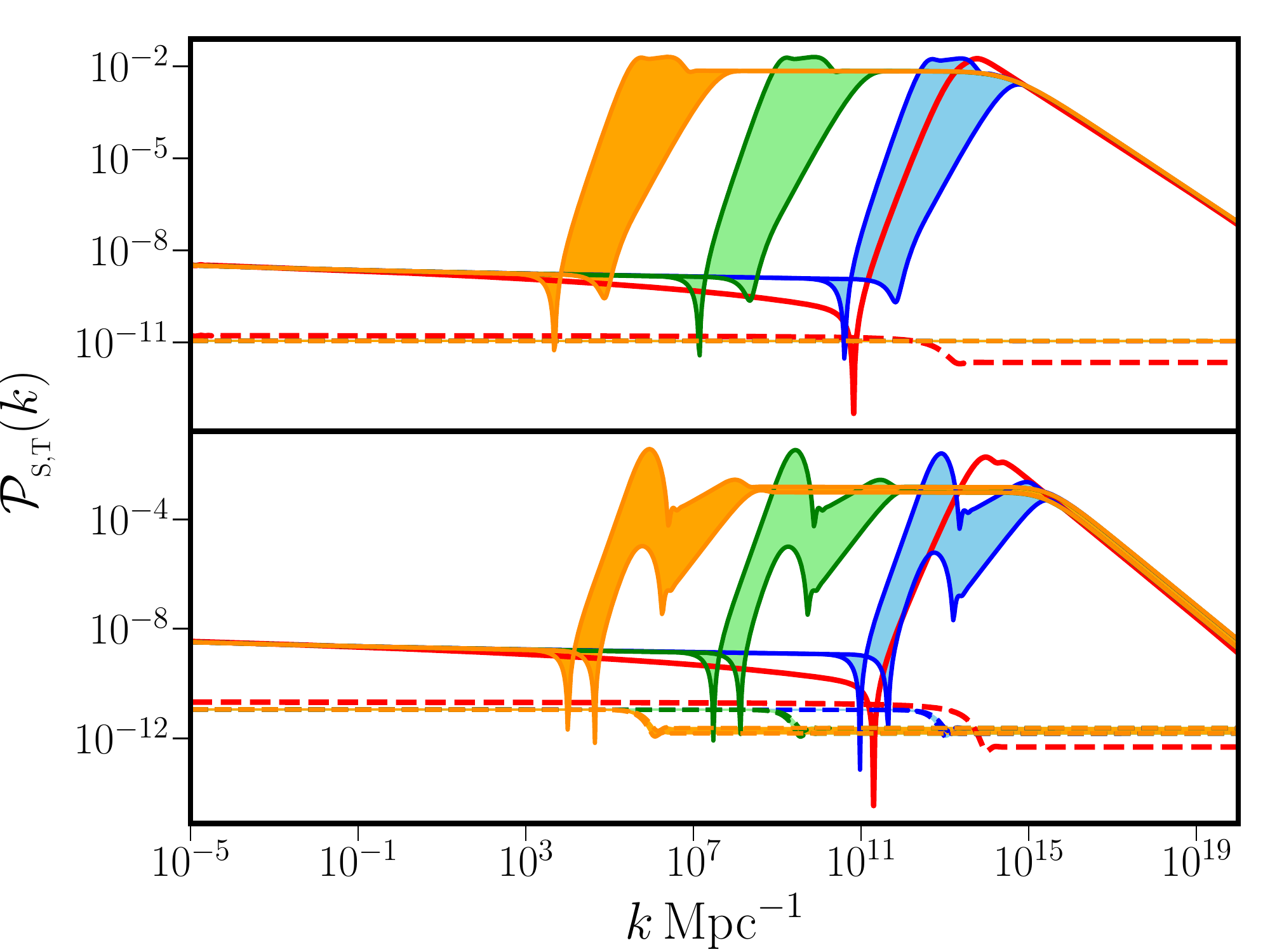}
\end{center}
\vskip -15pt
\caption{The scalar (in solid blue) and tensor power spectra 
(in dashed blue) resulting from the scenarios RS1 (on top) and RS2 (at the bottom)
have been plotted over a wide range of wave numbers.
For comparison, we have also plotted the scalar (in solid red) and tensor (in 
dashed red) power spectra that arise in the cases of USR2 (on top) and in PI3 
(at the bottom).
In the cases of RS1 and RS2~(plotted in blue), we have chosen the parameters 
so that the peak in the scalar power spectra roughly coincides with the peaks
in the models of USR2 and PI3 (plotted in red), respectively.
In addition, we have plotted the spectra arising in RS1 and RS2 for two other
values of the parameter~$N_1$ to produce peaks in the scalar power at smaller
wave numbers (in green and orange). 
Actually, we have plotted the spectra in RS1 and RS2 as bands (in blue, green
and orange) corresponding to a small range of the parameter~$\Delta N_1$ 
[cf. eqs.~\eqref{eq:eps1-12}].}\label{fig:pps-rcs}
\end{figure}

With the background quantities at hand, it is straightforward to compute the
power spectra by integrating the differential equations~\eqref{eq:de-p} for 
the curvature and the tensor perturbations.
In figure~\ref{fig:pps-rcs}, we have plotted the power spectra that arise 
in the scenarios RS1 and RS2.
We have also compared the power spectra in these cases with the spectra in 
USR2 and PI3.
It is clear that, while the scalar power spectra from the reconstructed 
potentials are indeed very similar to the power spectra from USR2 and PI3,
the corresponding tensor power spectra exhibit some differences. 
Since we shall be focusing on the observational imprints of the scalar 
perturbations generated during inflation, we shall ignore these differences 
for now.
We shall make a few clarifying remarks regarding this point in the 
concluding section.

\par

Earlier, we had emphasized the point that the models USR2
and PI3 are highly fine-tuned and that it is difficult to move the locations 
of the peaks in the scalar power spectra substantially without either 
considerably affecting the duration of inflation or the spectra over the 
CMB scales.
In contrast, because of the presence of the additional parameters, the scenarios
RS1 and RS2 are easier to tune and, as a result, we find that we can shift the
location of the peak as well as broaden its width. 
In figure~\ref{fig:pps-rcs}, apart from the spectra in RS1 and RS2 which
closely mimic the scalar spectra that arise in USR2 and PI3, we have 
plotted the power spectra for two other sets of parameters which lead to 
peaks at different locations and also exhibit a broader peak.
These spectra have been achieved by choosing different values for the 
parameter~$N_1$, while keeping the other parameters fixed at the values 
mentioned earlier. 
To arrive at the spectra with the broader peaks in figure~\ref{fig:pps-rcs},
we have set $N_1=34$ and $26$ in the case of RS1 and $N_1 = 40$ and $32$ in 
the case of RS2.
We should mention that a smaller choice of $N_1$ leads to a peak at a smaller
wave number.
Moreover, the bands associated with these two spectra correspond to the 
variation of the parameter $\Delta N_1$ over the domain we had mentioned 
before.

\par

In the next two sections, we shall study the imprints of the various power 
spectra on the formation of PBHs and the generation of secondary GWs.


\section{Formation of PBHs}\label{sec:pbhs}

Let us begin by recalling a few essentials.
Scales with wave numbers greater than $k\simeq 10^{-2}\, \mathrm{Mpc}^{-1}$ 
renter the Hubble radius during the radiation dominated epoch.
When these modes reenter the Hubble radius, the perturbations in the matter
density at the corresponding scales collapse to form structures.
We shall assume that the density contrast in matter characterized by the
quantity~$\delta$ is a Gaussian random variable described by the 
probability density
\begin{equation}
{\cal P}(\delta) 
= \f{1}{\sqrt{2\,\pi\,\sigma^2}}\; 
\mathrm{exp}{\l(-\f{\delta^2}{2\,\sigma^2}\r)},\label{eq:pd}
\end{equation}
where $\sigma^2$ is the variance of the spatial density fluctuations.
Let us assume that perturbations with a density contrast beyond a certain 
threshold, say, $\delta_\mathrm{c}$, are responsible for the formation 
of PBHs.
In such a case, the fraction, say, $\beta$, of the density fluctuations 
that collapse to form PBHs is described by the integral (in this context, 
see the reviews~\cite{Carr:2016drx,Carr:2018rid,Sasaki:2018dmp,Carr:2020xqk})
\begin{equation}
\beta = \int^{1}_{_{\delta_\mathrm{c}}} \d\delta\,
\cP(\delta)
\simeq \f{1}{2}\,\l[1-\mathrm{erf}\l(\f{\delta_\mathrm{c}}{\sqrt{2\,\sigma^2}}\r)\r],
\label{eq:b-pbh}
\end{equation}
where $\mathrm{erf}(z)$ denotes the error function.
Note that the lower limit of the above integral is the threshold value of 
the density contrast beyond which matter is expected to collapse to form PBHs.
We should clarify here that the value of $\delta_\mathrm{c}$ 
is not unique and it is expected to depend on the amplitude of the perturbation 
at a given scale (see refs.~\cite{Carr:1975qj,Green:2004wb};
in this context, also see the recent discussions~\cite{Sasaki:2018dmp,Germani:2018jgr,Germani:2019zez,
Escriva:2019nsa,Escriva:2019phb,Escriva:2020tak}).
The choice of $\delta_\mathrm{c}$ becomes important for the reason that the 
extent of PBHs formed is exponentially sensitive to its value.
In order to calculate the extent of PBHs formed, we shall work with the 
following values of $\delta_\mathrm{c}$: $1/3$, $0.35$ and $0.4$.

\par

During the radiation dominated epoch, the matter power spectrum~$P_\delta(k)$ 
and the inflationary scalar power spectrum $\ps(k)$ are related through the
expression 
\begin{equation}
P_\delta(k) 
= \f{16}{81}\,\l(\f{k}{aH}\r)^4\,\ps(k).\label{eq:pm-k}
\end{equation}
The variance in the spatial density fluctuations~$\sigma^2$, which determines
the fraction $\beta$ of PBHs formed [cf. eq.~\eqref{eq:b-pbh}], can be expressed
as an integral over the matter power spectrum~$P_\delta(k)$.
In order to introduce a length scale, say, $R$, the variance is smoothened over 
the scale with the aid of a window function~$W(k\,R)$.
The variance~$\sigma^2(R)$ can then be written as
\begin{equation}
\sigma^2(R) 
= \int_{0}^{\infty} \f{\d k}{k}\,P_\delta(k)\, W^2(k\,R),\label{eq:sigma2}
\end{equation}
and we shall work with a Gaussian window function of the form
$W(k\,R) = \mathrm{e}^{-(k^2\,R^2)/2}$.

\par

There remains the task of relating the scale~$R$ to the mass, say, $M$, 
of the PBHs formed.
Let~$M_{_\mathrm{H}}$ denote the mass within the Hubble radius~$H^{-1}$ 
at a given time.
It is reasonable to suppose that a certain fraction of the total mass 
within the Hubble radius, say, $M=\gamma\, M_\mathrm{H}$, goes on to
form PBHs when a mode with wave number~$k$ reenters the Hubble radius. 
The quantity~$\gamma$ that has been introduced reflects the efficiency 
of the collapse.
In the absence of any other scale, it seems natural to choose $k=R^{-1}$, 
and make use of the fact that $k=a\,H$ when the modes reenter the Hubble 
radius, to finally obtain the relation between~$R$ and~$M$.
One can show that $R$ and $M$ are related as follows:
\begin{equation}
R=\f{2^{1/4}}{\gamma^{1/2}}\,
\l(\f{g_{\ast,k}}{g_{\ast,\mathrm{eq}}}\r)^{1/12}\,
\l(\f{1}{k_\mathrm{eq}}\r)\,
\l(\f{M}{M_\mathrm{eq}}\r)^{1/2},\label{eq:R-M}
\end{equation}
where $k_\mathrm{eq}$ is the wave number that reenters the Hubble radius 
at the epoch of radiation-matter equality, and $M_\mathrm{eq}$ denotes 
the mass within the Hubble radius at equality.
Also, the quantities $g_{\ast,k}$ and $g_{\ast,\mathrm{eq}}$ represent
the number of relativistic degrees of freedom at the times of PBH 
formation and radiation-matter equality, respectively.
It can be easily determined that $M_\mathrm{eq} = 5.83\times 10^{47}\, \mathrm{kg}$, 
so that we can express the above relation between $R$ and $M$ in terms of the
solar mass $M_\odot$ as follows:
\begin{equation}
R=4.72\times10^{-7}\,\l(\f{\gamma}{0.2}\r)^{-1/2}\,
\l(\f{g_{\ast,k}}{g_{\ast,\mathrm{eq}}}\r)^{1/12}\,
\l(\f{M}{M_\odot}\r)^{1/2}\,\mathrm{Mpc}.\label{eq:R-Ms}
\end{equation}

\par 

On using the above arguments, we can arrive at the fraction of PBHs, say,
$\fpbh$, that contribute to the dark matter density today.
The quantity $\fpbh(M)$ can be expressed as  
\begin{equation}
\fpbh(M) = 2^{1/4}\;\gamma^{3/2}\,\beta(M)\,
\l(\f{\Omega_\mathrm{m}\,h^2}{\Omega_\mathrm{c}\,h^2}\r)\,
\l(\f{g_{\ast,k}}{g_{\ast,\mathrm{eq}}}\r)^{-1/4}\,
\l(\f{M}{M_\mathrm{eq}}\r)^{-1/2}, 
\end{equation}
where $\Omega_\mathrm{m}$ and $\Omega_\mathrm{c}$ are the dimensionless 
parameters describing the matter and cold matter densities, with the Hubble 
parameter, as usual, expressed as $H_0=100\,h\,\mathrm{km}\,\mathrm{sec}^{-1}\,
\mathrm{Mpc}^{-1}$.
In our calculations, we shall choose $\gamma = 0.2$, $g_{\ast,k} 
= 106.75$ and $g_{\ast,\mathrm{eq}} = 3.36$ and set $\Omega_\mathrm{m}\,h^2
= 0.14$, $\Omega_{\mathrm{c}}\,h^2 = 0.12$, with the last two being the best
fit values from the recent Planck data~\cite{Ade:2015xua,Aghanim:2018eyx}.
On substituting these values, one can arrive at the following expression 
for~$\fpbh(M)$:
\begin{equation}
\fpbh(M) 
=\l(\f{\gamma}{0.2}\r)^{3/2}\,
\l(\f{\beta(M)}{1.46\times 10^{-8}}\r)\, 
\l(\f{g_{\ast,k}}{g_{\ast,\mathrm{eq}}}\r)^{-1/4}\,
\l(\f{M}{M_\odot}\r)^{-1/2}.\label{eq:fpbh-f}
\end{equation}


\par

\begin{figure}[!t]
\begin{center}
\includegraphics[width=15cm]{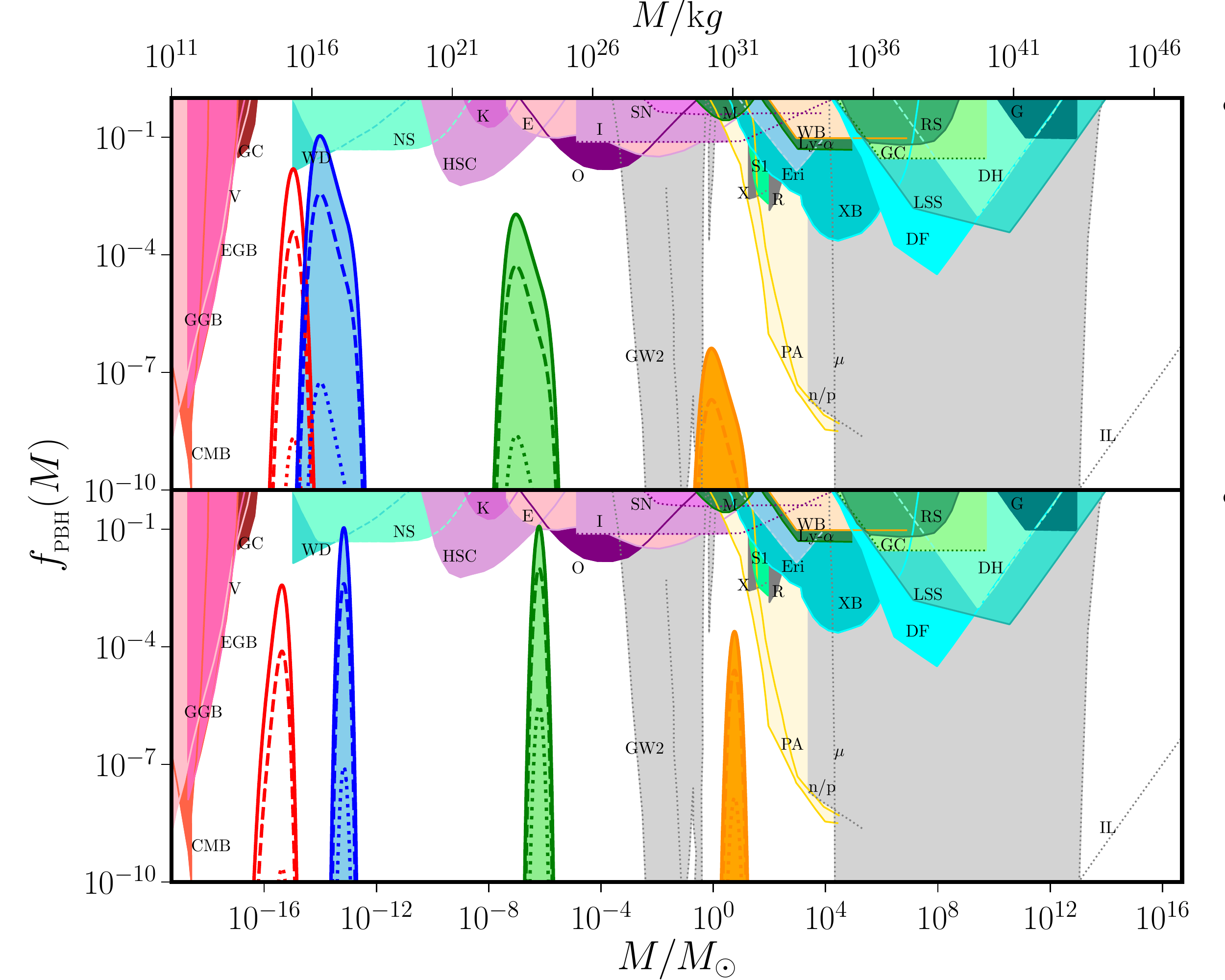}
\end{center}
\vskip -15pt
\caption{The fraction of PBHs contributing to the dark 
matter density today~$\fpbh$ has been plotted for the various models and 
scenarios of interest, viz. USR2 and RS1 (on top, in red and blue) 
and PI3 and RS2 (at the bottom, in red and blue).
We have plotted the quantity $\fpbh$ for the following three values of 
$\delta_\mathrm{c}$: $1/3$ (as solid curves) and $0.35$ (as dashed curves) 
and $0.4$ (as dotted curves).
In the cases of RS1 and RS2, apart from the original choices of parameters
that led to scalar spectra that closely matched the spectra in USR2 and PI3, 
we have plotted the quantity~$\fpbh$ for spectra which had exhibited broader
peaks starting at smaller wave numbers (cf. figure~\ref{fig:pps-rcs}).
As in the previous figure, in the cases of RS1 and RS2, we have plotted  
bands corresponding to a range of the parameter~$\Delta N_1$.
We have also indicated the latest direct (in different colors) and indirect
(in gray) constraints on~$\fpbh$ from a variety of observations.
We should mention here that the indirect constraints depend on additional 
assumptions.
Evidently, for the parameters of the potentials we have been working with, 
USR2 leads to a larger formation of PBHs than PI3.
Moreover, note that the existing observational constraints already limit 
the parameter $\Delta N_1$ in the reconstructions RS1 and RS2.}
\label{fig:fpbh}
\end{figure}

Given a primordial power spectrum $\ps(k)$, we can utilize the
relations~\eqref{eq:pm-k} and~\eqref{eq:sigma2} to arrive at 
the quantity~$\sigma^2(R)$.
Then, using the relation~\eqref{eq:R-M}, we can determine $\sigma^2$
as a function of~$M$ and utilize the result~\eqref{eq:b-pbh} to 
obtain~$\beta(M)$.
With $\beta(M)$ at hand, we can use the relation~\eqref{eq:fpbh-f} to
finally arrive at $\fpbh(M)$ for a given inflationary scalar power 
spectrum.
In figure~\ref{fig:fpbh}, we have plotted $\fpbh(M)$ for the models of 
USR2, PI3, RS1, and RS2.
In the figure, we have also indicated the constraints from the various 
observations such as constraints from gravitational 
lensing~\cite{Barnacka:2012bm,Katz:2018zrn}, 
constraints due to the limits on extragalactic background 
photons from PBH evaporation~\cite{Carr:2009jm}, constraints 
from microlensing searches by Kepler~\cite{Griest:2013esa},
MACHO~\cite{Allsman:2000kg}, EROS~\cite{Tisserand:2006zx} 
and OGLE~\cite{Wyrzykowski:2011tr}, 
constraints from the large scale structure~\cite{Carr:2009jm},
constraints from the CMB anisotropies due to accretion onto 
PBHs (FIRAS and WMAP3)~\cite{Ricotti:2007au} and, finally, constraints
from the dynamics of ultra-faint dwarf galaxies~\cite{Brandt:2016aco}. 
(For the latest and comprehensive list of these constraints 
and a detailed discussion, see refs.~\cite{Carr:2020gox,Green:2020jor}.
For related discussions in these contexts, also see 
refs.~\cite{Montero-Camacho:2019jte,Laha:2019ssq,Dasgupta:2019cae,Laha:2020ivk}.)
We find that, in the cases of USR2 and RS1, where the location of the peaks 
in the scalar power spectra approximately match, the maximum values of $\fpbh$
achieved are $1.5\times10^{-2}$ and $0.10$, respectively. 
For the models PI3 and RS2, when the peaks are located at roughly the same
wave number, we similarly obtain $\fpbh$ to be $3\times10^{-3}$ and $0.11$ at their 
respective maxima. 
In these cases, the maxmima in $\fpbh(M)$ are located over the domain $M
\simeq 10^{-16}$--$10^{-12}\, M_\odot$.
For peaks in the scalar power spectra that occur at smaller wave numbers in 
the cases of RS1 and RS2, as expected, the locations of the maxima in $\fpbh(M)$ 
shift towards larger masses of PBHs.
Interestingly, for the power spectra in RS1 and RS2 which exhibit a broad peak 
beginning at $k\simeq\,10^{6}\, \mathrm{Mpc}^{-1}$, there arise maxima in $\fpbh$ 
at tens of solar masses.
However, the corresponding maximum value of $\fpbh$ at $M\simeq 10\,M_\odot$ is a 
few orders of magnitude smaller than the maximum values we discussed above at 
smaller masses. 
This arises despite the fact the amplitude of the scalar power spectra at their
peak is the same in all these cases.
We believe that this result can be attributed to the dependence of $\fpbh$ on $M$
as $M^{-1/2}$ [cf. eq.~\eqref{eq:fpbh-f}].
We should point out here that the shaded bands corresponding to RS1 and RS2 
in figure~\ref{fig:fpbh} indicate the range of~$\fpbh$ that can be generated 
by varying the parameter~$\Delta N_1$ in the functional forms of $\epsilon_1(N)$ 
[cf. eqs.~\eqref{eq:eps1-12}].
The intersection of the shaded bands with the constraints readily translate 
to the limits on this parameter in our reconstructions RS1 and RS2. 
We find that a smaller~$\Delta N_1$ leads to a steeper growth of power and 
hence to a higher fraction of PBHs. 
Therefore, for a fixed set of values for the other parameters, the constraints
essentially restrict the rapidity of the transition of inflation from slow roll 
to ultra slow roll epoch in our reconstructions.


\section{Generation of secondary GWs}\label{sec:sgws}

In this section, we shall calculate the secondary power and bispectrum of 
GWs induced by the scalar perturbations at the second order.


\subsection{The secondary tensor power spectrum}

Earlier, we had described the scalar and tensor perturbations at first order 
in terms of the curvature perturbation~$\cR$ and the quantity~$\gamma_{ij}$ 
(cf. subsection~\ref{subsec:ps}).
It is well known that, at the linear order, the scalar and tensor perturbations 
evolve independently, with their evolution being governed by the corresponding
equations of motion, viz. eqs.~\eqref{eq:de-p}.
However, one finds that, at the second order, the tensor perturbations are 
sourced by quadratic terms involving the first order scalar perturbations 
(for early discussions in this context, see for instance, 
refs.~\cite{Ananda:2006af,Baumann:2007zm,Saito:2008jc,Saito:2009jt}).
These contributions due to the scalar perturbations become important particularly
when the amplitude of the scalar power spectrum is boosted over small scales such
as in the situations leading to enhanced formation of PBHs.
In this subsection, we shall calculate the dimensionless density parameter
associated with the GWs, say, $\ogw$, generated due to the scalar perturbations 
in the different models and scenarios of interest.

\par

Let us begin by outlining the primary steps towards the calculation of~$\ogw(f)$, 
where~$f$ is the frequency associated with the wave number $k$.
We shall start with the following perturbed metric:
\begin{equation}
\d s^2 = a^2(\eta)\, \l\{-\l(1+2\,\Phi\r)\, \d\eta^2
+ \l[\l(1-2\,\Psi\r)\,\delta_{ij}+\f{1}{2}\,h_{ij}\r]\, \d x^i \d x^j\r\},
\end{equation}
where $\Phi$ and $\Psi$ are the Bardeen potentials describing the scalar 
perturbations at the first order, while the quantity $h_{ij}$ represents 
the second order tensor perturbations.
We should clarify that we have denoted the second order tensor perturbation
as $h_{ij}$ in order to distinguish them from the first order tensor
perturbations~$\gamma_{ij}$ which we had introduced earlier.
The transverse and traceless nature of the tensor perturbations implies that 
$\pa^i\,h_{ij}=0$ and $h^i_i=0$.
In our discussion below, we shall assume that anisotropic stresses are absent
so that $\Phi=\Psi$.

\par

The tensor perturbations $h_{ij}$ can be decomposed in terms of the Fourier modes, 
say, $h_{\bm k}$, as 
\begin{equation}
h_{ij}(\eta,\vx)
=\int\f{\d^3\vk}{(2\,\pi)^{3/2}}\,
\l[e_{ij}^+(\vk)\, h_{\vk}^+(\eta)
+e_{ij}^\times(\vk)\, h_{\vk}^\times(\eta)\r]\, {\rm e}^{i\,\vk\cdot\vx},
\end{equation}
where $e_{ij}^+(\vk)$ and $e_{ij}^\times(\vk)$ denote the polarization tensors
which have non-zero components in the plane perpendicular to the direction of 
propagation, viz. $\hat{\vk}$.
The polarization tensors $e_{ij}^+(\vk)$ and $e_{ij}^\times(\vk)$ can be 
expressed in terms of the set of orthogonal unit vectors $(e(\vk),
{\bar e}(\vk),\hat{\vk})$ in the following manner (see, for instance, the
review~\cite{Maggiore:1999vm}):
\begin{subequations}
\begin{eqnarray}
e_{ij}^+(\vk)
&=&\f{1}{\sqrt{2}}\,\l[e_{i}(\vk)\,e_{j}(\vk)
-{\bar e}_{i}(\vk)\, {\bar e}_{j}(\vk)\r],\\
e_{ij}^\times(\vk)
&=&\f{1}{\sqrt{2}}\,\l[e_{i}(\vk)\,{\bar e}_{j}(\vk)
+{\bar e}_{i}(\vk)\, e_{j}(\vk)\r].
\end{eqnarray}
\end{subequations}
The orthonormal nature of the vectors $e(\vk)$ and ${\bar e}(\vk)$ lead to the 
normalization condition:~$e_{ij}^{\lambda}(\vk)\, e^{\lambda',ij}(\vk)
=\delta^{\lambda \lambda'}$, where $\lambda$ and $\lambda'$ can be 
either $+$ or $\times$.

\par

The equation of motion governing the Fourier modes $h_\vk$ can be arrived at using 
the second order Einstein equations describing the tensor perturbation~$h_{ij}$
and the Bardeen equation describing the scalar 
perturbation~$\Psi$ at the first order (see, for example, 
refs.~\cite{Ananda:2006af,Baumann:2007zm}; for recent discussions, see 
refs.~\cite{Bartolo:2016ami,Bartolo:2018evs,Bartolo:2018rku,Espinosa:2018eve}).
One finds that the equation governing $h_\vk$ can be written as
\begin{equation}
{h_\vk^\lambda}''+ 2\,\mathcal H\, {h_\vk^\lambda}'+ k^2\, h_\vk^\lambda
=S_\vk^\lambda\label{eq:sgw}
\end{equation}
with the source term $S_\vk^\lambda$ being given by
\begin{eqnarray}
S_\vk^\lambda(\eta)
&=& 4\, \int\frac{\d^3 \vp}{(2\,\pi)^{3/2}}\, e^\lambda(\vk,\vp)\,
\Biggl\{2\,\Psi_\vp(\eta)\,\Psi_{\vk-\vp}(\eta)\nn\\
& &+\,\f{4}{3\, (1+w)\,{\mathcal H}^2\,}\, 
\l[\Psi_\vp'(\eta)+{\cal H}\,\Psi_\vp(\eta)\r]\,
\l[\Psi_{\vk-\vp}'(\eta)+{\cal H}\,\Psi_{\vk-\vp}(\eta)\r]\,\Biggr\},
\end{eqnarray}
where, evidently, $\Psi_\vk$ represents the Fourier modes of the Bardeen potential,
while~${\cal H}$ and~$w$ denote the conformal Hubble parameter and the equation of
state parameter describing the universe at the conformal time~$\eta$.
Also, for convenience, we have defined the quantity $e^{\lambda}(\vk,\vp)=
e^{\lambda}_{ij}(\vk)\,p^i\,p^j$.
While discussing the formation of PBHs earlier, we had assumed that the scales of 
our interest reenter the Hubble radius during the epoch of radiation domination.
In such a case, we have $w=1/3$ and ${\cal H}=1/\eta$.
Moreover, during radiation domination, it is well known that we can express the 
Fourier modes $\Psi_\vk$ of the Bardeen potential in terms of the inflationary 
Fourier modes $\cR_\vk$ of the curvature perturbations generated during inflation
through the relation
\begin{equation}
\Psi_\vk(\eta)=\f{2}{3}\,\cT(k\,\eta)\, \cR_\vk,
\end{equation}
where $\cT(k\,\eta)$ is the transfer function given by
\begin{equation}
\cT(k\,\eta)=\f{9}{\l(k\,\eta\r)^2}\,
\l[\f{{\rm sin}\l(k\,\eta/\sqrt{3}\r)}{k\,\eta/\sqrt{3}}
-{\rm cos}\l(k\,\eta/\sqrt{3}\r)\r].
\end{equation}
Utilizing the Green's function corresponding to the tensor modes during 
radiation domination, we can express the inhomogeneous contribution to 
$h_\vk^\lambda$ as~\cite{Espinosa:2018eve}
\begin{eqnarray}
h_\vk^\lambda(\eta) 
&=& \frac{4}{9\,k^3\,\eta}\,
\int \f{\d^3 \vp}{(2\,\pi)^{3/2}}\, 
e^{\lambda}(\vk,\vp)\, \cR_\vk\,\cR_{\vk-\vp}\,
\l[\cI_c\l(\f{p}{k},\f{\vert\vk-\vp\vert}{k}\r)\,{\rm cos}\l(k\,\eta\r)
+\cI_s\l(\f{p}{k},\f{\vert\vk-\vp\vert}{k}\r)\,{\rm sin}\l(k\,\eta\r)\r],
\label{eq:hkl}
\end{eqnarray}
where the quantities $\cI_c(v,u)$ and $\cI_s(v,u)$ are described by the integrals
\begin{subequations}
\begin{eqnarray}
\cI_c(v,u)&=&-4\,\int_{0}^{\infty}\,\d \tau\,\tau\,{\rm sin}\,\tau\,
\biggl\{2\,\cT(v\,\tau)\,\cT(u\,\tau)
+\l[\cT(v\,\tau)+v\,\tau\,\cT_{v\tau}(v\,\tau)\r]\,
\l[\cT(u\,\tau)+u\,\tau\,\cT_{u\tau}(u\,\tau)\r]\biggr\},\\
\cI_s(v,u)&=& 4\,\int_{0}^{\infty}\,\d \tau\,\tau\,{\rm cos}\,\tau\,
\biggl\{2\,\cT(v\,\tau)\,\cT(u\,\tau)
+\l[\cT(v\,\tau)+v\,\tau\,\cT_{v\tau}(v\,\tau)\r]\,
\l[\cT(u\,\tau)+u\,\tau\,\cT_{u\tau}(u\,\tau)\r]\biggr\},
\end{eqnarray}
\end{subequations}
with $\cT_z=\d\cT/\d z$.
The above integrals can be carried out analytically and they are given by
\begin{subequations}\label{eq:cI}
\begin{eqnarray}
\cI_c(v,u) &=& -\f{27\,\pi}{4\,v^3\,u^3}\,
\Theta\l(v+u-\sqrt{3}\r)\, (v^2+u^2-3)^2,\\
\cI_s(v,u) &=& -\f{27}{4\,v^3\,u^3}\, (v^2+u^2-3)\,
\l[4\,v\,u+ (v^2+u^2-3)\;{\rm log}\,
\biggl\vert\f{3-(v-u)^2}{3-(v+u)^2}\biggr\vert\r],\qquad
\end{eqnarray}
\end{subequations}
where $\Theta(z)$ denotes the theta function.
It is useful to note that $\cI_{c,s}(v,u) =\cI_{c,s}(u,v)$.

\par

The power spectrum of the secondary GWs, say, $\ph(k,\eta)$, generated
due to the second order scalar perturbations can be defined as follows:
\begin{eqnarray}
\langle h_{\vk}^{\lambda}(\eta)\,h_{\vk'}^{\lambda'}(\eta)\rangle
=\f{2\,\pi^2}{k^3}\,\cP_h(k,\eta)\,\delta^{(3)}(\vk+\vk')\,\delta^{\lambda\lambda'}.
\end{eqnarray}
Note that $h_\vk^\lambda$ involves products of the Fourier modes $\cR_\vk$ 
and $\cR_{\vk-\vp}$ of the curvature perturbations generated during inflation 
[cf. eq.~\eqref{eq:hkl}].
Evidently, the power spectrum $\ph(k)$ of the secondary GWs will involve 
products of four such variables.
Since, the quantity $\cR_\vk$ is a Gaussian random variable, we can express
the four-point function in terms of the two-point functions or, equivalently,
the inflationary scalar power spectrum $\ps(k)$ [cf. eq.~\eqref{eq:sps-d}]
as
\begin{eqnarray}
\ph(k,\eta)
&=& \f{4}{81\,k^2\,\eta^2}
\int_{0}^{\infty}\d v\,\int_{\vert 1-v\vert}^{1+v}\d u\,
\l[\f{4\,v^2-(1+v^2-u^2)^2}{4\,u\,v}\r]^2\,\ps(k\,v)\,\ps(k\,u)\nn\\
& &\times\,
\l[\cI_c(u,v)\,{\rm cos}\l(k\,\eta\r)+\cI_s(u,v)\,{\rm sin}\l(k\,\eta\r)\r]^2.
\label{eq:ph}
\end{eqnarray}
We shall now choose to average $\ph(k,\eta)$ over small time scales so that 
the trigonometric functions in the above expressions are replaced by their
average over a time period.
In such a case, only the overall time dependence remains, leading 
to~\cite{Kohri:2018awv,Espinosa:2018eve}
\begin{eqnarray}
\overline{\ph(k,\eta)}
&=& \f{2}{81\,k^2\,\eta^2}
\int_{0}^{\infty}\d v\,\int_{\vert 1-v\vert}^{1+v}\d u\,
\l[\f{4\,v^2-(1+v^2-u^2)^2}{4\,u\,v}\r]^2\,\ps(k\,v)\,\ps(k\,u)\nn\\
& &\times\,
\l[\cI_c^2(u,v)+\cI_s^2(u,v)\r],\label{eq:phf}
\end{eqnarray}
where the line over $\ph(k,\eta)$ implies that we have averaged over 
small time scales.
The energy density of GWs associated with a Fourier mode corresponding to 
the wave number~$k$ at a time $\eta$ is given by~\cite{Maggiore:1999vm}
\begin{equation}
\rho_{_{\mathrm{GW}}}(k,\eta)
= \f{\Mpl^2}{8}\,\l(\frac{k}{a}\r)^2\,\overline{\ph(k,\eta)}.
\end{equation}
The corresponding dimensionless density parameter~$\ogw(k,\eta)$ can
be defined in terms of the critical density $\rho_{\mathrm{cr}}(\eta)$ 
as~\cite{Espinosa:2018eve}
\begin{equation}
\ogw(k,\eta)
=\f{\rho_{_{\mathrm{GW}}}(k,\eta)}{\rho_{\mathrm{cr}}(\eta)}
=\f{1}{24}\,\l(\f{k}{{\mathcal H}}\r)^2 \overline{\ph(k,\eta)}.\label{eq:ogw-rd}
\end{equation}

\par

Note that the dimensionless density parameter~$\ogw(k,\eta)$ above has been 
evaluated during the radiation dominated epoch.
Once the modes are inside the Hubble radius, the energy density of GWs 
decay just as the energy density of radiation does.
Upon utilizing this point, we can express $\ogw(k)$ today in terms of the 
above $\ogw(k,\eta)$ as follows:
\begin{eqnarray}
h^2\,\ogw(k)
&=& \l(\f{g_{\ast,k}}{g_{\ast,0}}\r)^{-1/3}\,\Omega_{\mathrm{r}}\,h^2\;
\ogw(k,\eta)\nn\\
& \simeq & 
1.38\times10^{-5}\, 
\l(\f{g_{\ast,k}}{106.75}\r)^{-1/3}\,
\l(\f{\Omega_{r}\,h^2}{4.16\times10^{-5}}\r)\,\ogw(k,\eta),\label{eq:ogw0}
\end{eqnarray}
where $\Omega_\mathrm{r}$ and $g_{\ast,0}$ denote the dimensionless energy 
density of radiation and the number of relativistic degrees of freedom today.
We should point out here that, since $\mathcal{H}\propto \eta^{-1}$ during 
radiation domination and $\ph(k,\eta)\propto \eta^{-2}$, the quantity 
$\ogw(k,\eta)$ in the expression~\eqref{eq:ogw-rd} is actually independent 
of time.
Moreover, the observable parameter today is usually expressed as a function 
of the frequency, say, $f$, which is related to the wave number~$k$ as
\begin{equation}
f = \frac{k}{2\,\pi} 
= 1.55\times10^{-15}\,\l(\f{k}{1\, \mathrm{Mpc}^{-1}}\r)\,\mathrm{Hz}.
\end{equation}

\par

\begin{figure}[!t]
\begin{center}
\includegraphics[width=15cm]{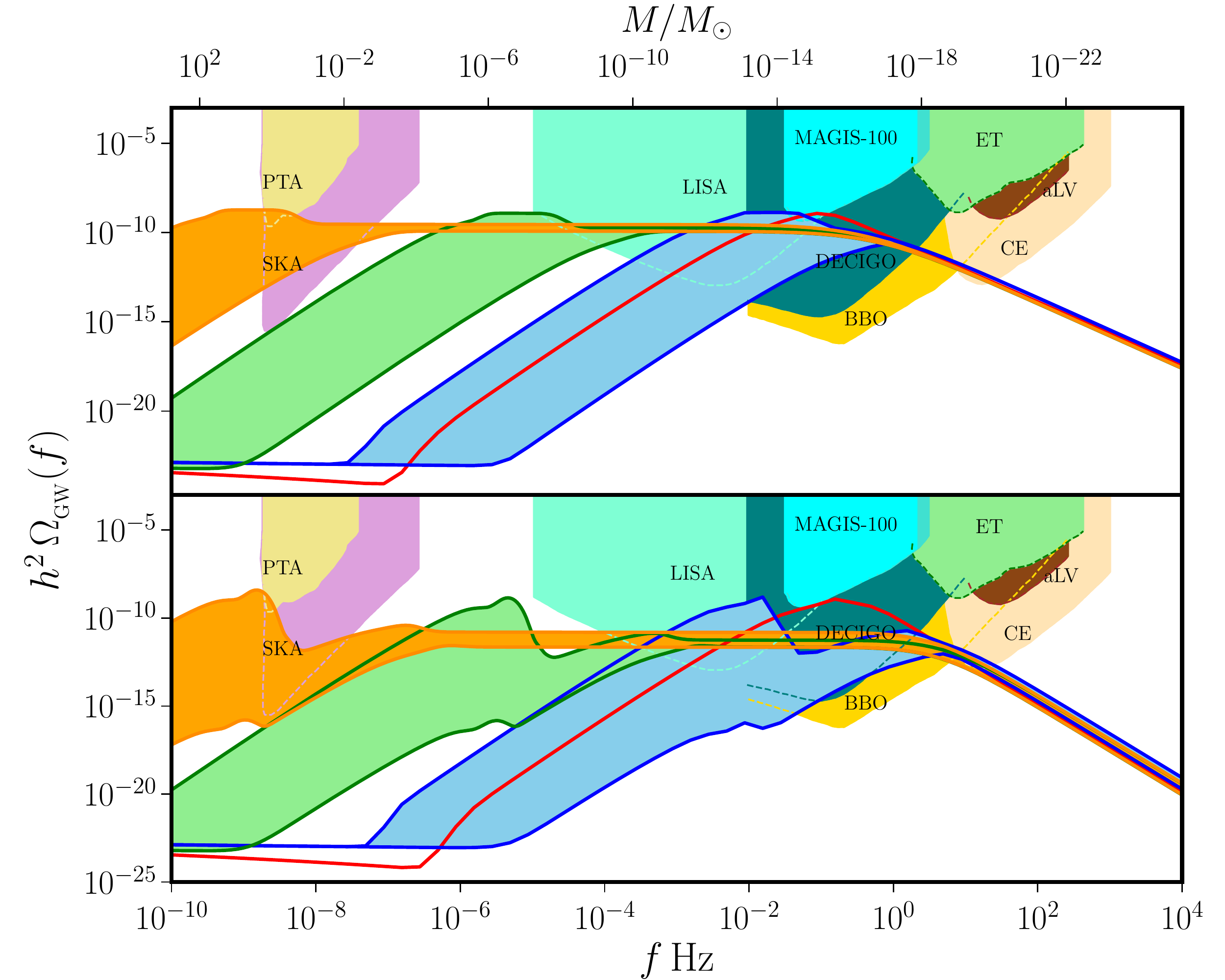}
\end{center}
\vskip -15pt
\caption{The dimensionless density parameter $\ogw$ associated with 
the secondary GWs generated in the models and reconstructed scenarios 
of USR2 and RS1 (in red and blue, on top) as well as PI3 and RS2 (in 
red and blue, at the bottom) have been plotted as a function of the 
frequency~$f$.
We have also plotted the $\ogw$ produced by the scenarios 
RS1 and RS2 with broader peaks beginning at smaller wave numbers (in green 
and orange).
The bands of spectra, as with the previous figures, correspond to variation
of the parameter $\Delta N_1$ for a given $N_1$.
Moreover, we have included the sensitivity curves of various existing and 
upcoming observational probes of GWs (as shaded regions, in the top part 
of the panels).
Clearly, it should be possible to detect the GWs generated in the models 
and scenarios of our interest by some of the forthcoming observatories.}
\label{fig:ogw}
\end{figure}

In figure~\ref{fig:ogw}, we have plotted the quantity $\ogw(f)$ arising
in the models USR2 and PI3 as well as the reconstructed scenarios RS1 
and RS2.
In the figure, we have also included the sensitivity curves associated  
with the various current and forthcoming observatories, viz. PTA and 
the Square Kilometre Array (SKA)~\cite{Moore:2014lga}, LISA~\cite{Bartolo:2016ami}, 
MAGIS-100~\cite{Coleman:2018ozp,Espinosa:2018eve}, 
BBO~\cite{Crowder:2005nr,Corbin:2005ny,Baker:2019pnp}, 
DECIGO~\cite{Kawamura:2011zz,Kawamura:2019jqt},
ET~\cite{Sathyaprakash:2012jk}, 
advanced LIGO $+$ Virgo~\cite{TheLIGOScientific:2016dpb,LIGOScientific:2019vic}
and CE~\cite{Evans:2016mbw}. 
(For a summary of the sensitivity curves and their updated versions, see
ref.~\cite{Moore:2014lga} and the associated web-page.)  
We should mention here that the estimated sensitivity curves have been arrived
at assuming a power law spectrum (the so-called `power-law integrated curves') 
over the bands of interest. 
These sensitivities are expected to be achieved by integrating over frequency 
in addition to integrating over time~\cite{Thrane:2013oya,Abbott:2009ws}.
It should be evident from the figure that the strength of the GWs generated 
in the models and scenarios we have examined here is significant enough to
be detectable by one or more of these observatories.
Recall that, spectra arising in the scenarios RS1 and RS2
with broad peaks starting from a wave number of about $10^{6}\, \mathrm{Mpc}^{-1}$ 
had led to PBHs with tens of solar masses.
It should be clear from figure~\ref{fig:ogw} that the constraints from PTA 
on $\ogw$ already rule out such spectra for certain values of $\Delta N_1$.


\subsection{The secondary tensor bispectrum}\label{sec:tbs}

In this section, we shall evaluate the secondary tensor bispectrum generated 
in the inflationary models and scenarios of our interest.
The secondary tensor bispectrum, say, 
$\cB^{\lambda_1\lambda_2\lambda_3}_h(\vka,\vkb,\vkc)$ is defined as
\begin{equation}
\l\langle h_{\vka}^{\lambda_1}(\eta)\, h_{\vkb}^{\lambda_2}(\eta)\,
h_{\vkc}^{\lambda_3}(\eta)\r\rangle 
=(2\,\pi)^3\, \cB_h^{\lambda_1\lambda_2\lambda_3}(\vka,\vkb,\vkc,\eta)\,
\delta^{(3)}(\vka+\vkb+\vkc).\label{eq:s-tbs-d}
\end{equation}
We can evaluate the above tensor bispectrum during the radiation dominated 
era by using the expression~\eqref{eq:hkl} for $h^{\lambda}_{\vk}(\eta)$.
As we had discussed, $h^{\lambda}_{\vk}(\eta)$ is quadratic in the 
Gaussian variables~$\cR_\vk$.
Therefore, obviously, the bispectrum 
$\cB_h^{\lambda_1\lambda_2\lambda_3}(\vka,\vkb,\vkc,\eta)$ will involve 
six of these variables.
Upon utilizing Wick's theorem applicable to Gaussian random variables, one 
can show that the tensor bispectrum consists of eight terms all of which lead 
to the same contribution~\cite{Bartolo:2018rku,Espinosa:2018eve}.
For convenience, we shall define 
$G_h^{\lambda_1\lambda_2\lambda_3}(\vka,\vkb,\vkc,\eta)
=(2\,\pi)^{-9/2}\;\cB_h^{\lambda_1\lambda_2\lambda_3}(\vka,\vkb,\vkc,\eta)$ 
and hereafter refer to $G_h^{\lambda_1\lambda_2\lambda_3}(\vka,\vkb,\vkc,\eta)$ 
as the secondary tensor bispectrum.
We find that the secondary tensor bispectrum can be expressed as
\begin{eqnarray}
G_h^{\lambda_1\lambda_2\lambda_3}(\vka,\vkb,\vkc,\eta)
&=& \l(\frac{8\,\pi}{9}\r)^3\,\f{1}{(k_1\,k_2\,k_3\,\eta)^3} \nn \\
& &\times\, \int \d^3 \vp_1\, e^{\lambda_1}(\vk_1,\vp_1)\,
e^{\lambda_2}(\vk_2,\vp_2)\,e^{\lambda_3}(\vk_3,\vp_3)\,\f{\ps(p_1)}{p^3_1}\,
\f{\ps(p_2)}{p^3_2}\,\f{\ps(p_3)}{p^3_3}\nn\\
& &\times\,J\l(\f{p_1}{k_1},\f{p_2}{k_1},\eta\r)\,
J\l(\f{p_2}{k_2},\f{p_3}{k_2},\eta\r)\,
J\l(\f{p_3}{k_3},\f{p_1}{k_3},\eta\r),\label{eq:Bh}
\end{eqnarray}
where $\vp_2 = \vp_1-\vka$, $\vp_3 = \vp_1 + \vkc$ and, for convenience, we 
have set
\begin{equation}
J\l(\f{p_1}{k_1},\f{p_2}{k_1},\eta\r)
=\cI_c\l(\f{p_1}{k_1},\f{p_2}{k_1}\r)\,{\rm cos}\l(k_1\,\eta\r)
+\cI_s\l(\f{p_1}{k_1},\f{p_2}{k_1}\r)\,{\rm sin}\l(k_1\,\eta\r),
\end{equation}
with $\cI_c(v,u)$ and $\cI_s(v,u)$ given by eqs.~\eqref{eq:cI}.
In a manner partly similar to the case of the secondary tensor 
power spectrum, we shall replace the trigonometric functions 
by their averages so that the function $J(x,y,\eta)$ is instead 
given by
\begin{equation}
\bar{J}(v,u) =\f{1}{\sqrt{2}}\, \l[\cI_c^2(v,u)+\cI_s^2(v,u)\r]^{1/2}.    
\end{equation}

\par

Our aim in this work is to understand the amplitude of the secondary tensor 
bispectrum generated due to the scalar perturbations for modes that reenter 
the Hubble radius during the radiation dominated era. 
For simplicity, we shall restrict our analysis to the equilateral limit of 
the bispectrum so that $k_1=k_2=k_3=k$.
In order to determine the integrals involved in the expression~\eqref{eq:Bh},
we shall choose a specific configuration for the vectors $\vka$, $\vkb$ and~$\vkc$.
We shall assume that the vectors lie in the $x$-$y$-plane with $\vkc$ oriented
along the negative $x$-direction.
In such a case, we find that the vectors $(\vka, \vkb, \vkc)$ in the equilateral 
limit are given by
\begin{equation}
\vka =\l(k/2, \sqrt{3}\,k/2, 0\r),\quad
\vkb = \l(k/2, -\sqrt{3}\,k/2, 0\r),\quad
\vkc = (-k, 0, 0).\label{eq:vk}
\end{equation}
We shall choose $\vp_1 =(p_{1x}, p_{1y}, p_{1z})$ so that, since 
$\vp_2 = \vp_1-\vka$ and $\vp_3 = \vp_1 + \vkc$, we have
\begin{equation}
\vp_2 = \l(p_{1x}-k/2, p_{1y}-\sqrt{3}\,k/2, p_{1z}\r),\quad
\vp_3 = (p_{1x}-k, p_{1y}, p_{1z}).\label{eq:vp}
\end{equation}
We find that such a choice of Cartesian coordinates
proves to be convenient to carry out the integrals involved than the 
cylindrical polar coordinates that have been adopted
earlier~\cite{Bartolo:2018rku,Espinosa:2018eve}.
Therefore, the tensor bispectrum in the equilateral limit 
$G_h^{\lambda_1\lambda_2\lambda_3}(k)$ can be written as
\begin{eqnarray}
k^6\,G_h^{\lambda_1\lambda_2\lambda_3}(k,\eta) 
&=& \l(\frac{8\,\pi}{9\,\sqrt{2}}\r)^3\,\f{1}{(k\,\eta)^3}\nn\\
& &\times\,\int_{-\infty}^{\infty} \d p_{1x}\, \int_{-\infty}^{\infty} \d p_{1y}\, 
\int_{-\infty}^{\infty} \d p_{1z}\,e^{\lambda_1}(\vk_1,\vp_1)\,
e^{\lambda_2}(\vk_2,\vp_2)\,e^{\lambda_3}(\vk_3,\vp_3)\nn\\ 
& &\times\,\f{\ps(p_1)}{p^3_1}\,\f{\ps(p_2)}{p^3_2}\,\frac{\ps(p_3)}{p^3_3}\,
\bar{J}\l(\frac{p_1}{k},\frac{p_2}{k}\r)\,
\bar{J}\l(\frac{p_2}{k},\frac{p_3}{k}\r)\,
\bar{J}\l(\frac{p_3}{k},\frac{p_1}{k}\r).\label{eq:Bhxyz}
\end{eqnarray}
The factors $e^{\lambda}(\vk,\vp)$ involving the polarization tensor can 
be readily evaluated for our configurations of $(\vka,\vkb,\vkc)$ and 
$(\vp_1,\vp_2,\vp_3)$ (for details, see appendix~\ref{app:pf}).
Since $\lambda$ can be~$+$ or~$\times$, clearly, the tensor bispectrum
$G_h^{\lambda_1\lambda_2\lambda_3}(k,\eta)$ has eight components.
However, we find that $e^{\times}(\vk,\vp)$ is odd in $p_{1z}$ 
[cf. eqs.~\eqref{eq:pfs}].
As a result, the tensor bispectrum proves to be non-zero only for the 
following combinations of $({\lambda_1\lambda_2\lambda_3})$: $(+++)$, 
$(+\times\times)$, $(\times+\times)$ and $(\times\times+)$.
Also, note that the integral above describing the tensor bispectrum in 
the equilateral limit is symmetric under the simultaneous interchange 
of $\lambda_1\leftrightarrow\lambda_2$, $\vka\leftrightarrow\vkb$ and
$\vp_1\leftrightarrow\vp_2$.
This implies that, in the equilateral limit of interest, the tensor 
bispectrum for the three components~$(+\times\times)$, $(\times+\times)$ 
and $(\times\times+)$ are equal.
Hence, we are left with only $G_h^{+++}(k,\eta)$ and, say, 
$G_h^{+\times\times}(k,\eta)$ to evaluate.

\par

We proceed to numerically evaluate $G_h^{+++}(k)$ and $G_h^{+\times\times}(k)$
in the situations of our interest, viz. namely USR2, PI3, RS1, and RS2. 
Because the scalar power spectra in these cases exhibit a localized maxima, 
we restrict our evaluation of the tensor spectrum to the range of wave numbers
around the peak.
We find that the integrand in eq.~\eqref{eq:Bhxyz} exhibits a maximum around
$\vert\vp_1\vert\simeq k$ and, beyond that, it quickly decreases in all the
three directions of integration.
In fact, the contributions to the integral prove to be negligible 
for $\vert \vp_{1}\vert \gtrsim 100\,k$.
So, we choose the limits for our integrals over $p_{1x}$, $p_{1y}$ 
and~$p_{1z}$ to be $(-10^3\,k, 10^3\,k)$.

\par

\begin{figure}[!t]
\begin{center}
\includegraphics[width=15cm]{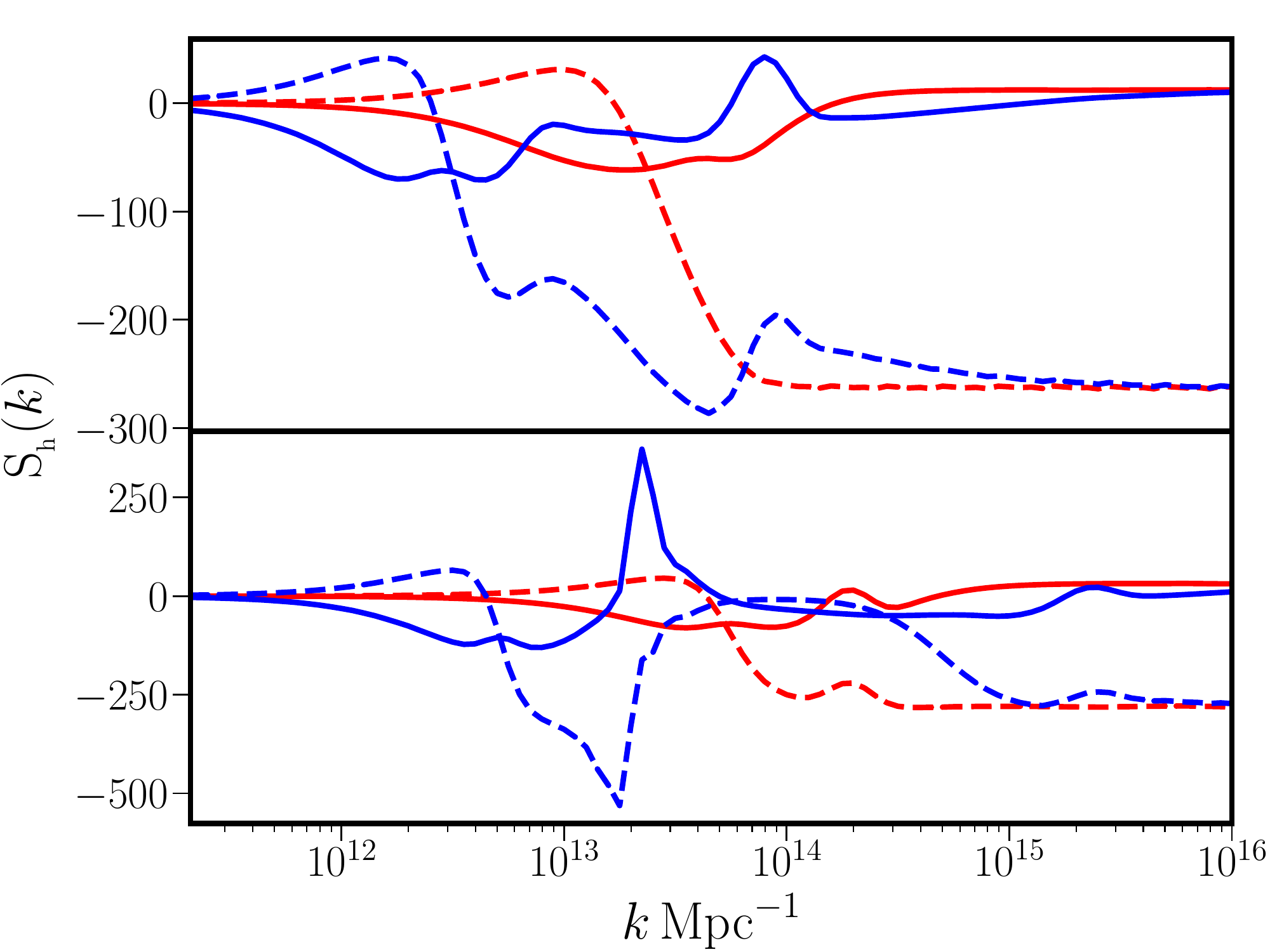}
\end{center}
\vskip -15pt
\caption{The dimensionless shape function~$S_h(k)$ characterizing the tensor 
bispectrum has been plotted in the equilateral limit for the models and scenarios 
of interest, viz. USR2 and RS1 (in red and blue, in the top panel) as well 
as PI3 and RS2 (in red and blue, in the bottom panel).
We have plotted both the non-zero components $S_h^{+++}(k)$ (as solid curves) 
and $S_h^{+\times\times}(k)$ (as dashed curves) for all the cases.
In plotting the results for RS1 and RS2, we have set $N_1=42$ 
and $48$ and chosen $\Delta N_1$ to be the lowest value within our windows, viz.
$0.3345$ and $0.3847$.
We find that, at large wave numbers [when compared to the location of the peak 
in the scalar power spectra (cf. figures~\ref{fig:pps-usr-pi}
and~\ref{fig:pps-rcs})], the amplitudes of $S_h^{+++}(k)$ and
$S_h^{+\times\times}(k)$ settle down to around $10$ and $-250$, respectively. 
Also, at wave numbers smaller than the location of the peak, the amplitudes of 
both the components prove to be of order unity or less in all the cases.}
\label{fig:Sh}
\end{figure}

In order to understand the behavior of the tensor bispectrum, we shall calculate 
the dimensionless quantity referred to the shape function, say, $\cS_h(k)$, which 
is defined as~\cite{Bartolo:2018rku,Espinosa:2018eve} 
\begin{equation}
S_h^{\lambda_1\lambda_2\lambda_3}(k) 
= \f{k^6\,G^{\lambda_1\lambda_2\lambda_3}_h(k,\eta)}{\sqrt{\cP_h^3(k,\eta)}}.
\end{equation}
Note that, in this expression, both the quantities 
$k^6\,G^{\lambda_1\lambda_2\lambda_3}_h(k)$ and $\cP_h^3(k)$ are 
dimensionless.
Moreover, the overall dependence on time cancels leading to a shape function that 
is time-independent.
In figure~\ref{fig:Sh}, we have plotted the shape functions $S_h^{+++}(k)$ and
$S_h^{+\times\times}(k)$ for the four cases of interest, viz. USR2, PI3, RS1
and RS2.
We find that the amplitude of $S_h(k)$ for a given model or scenario is maximum 
around the wave number where the scalar power spectrum exhibits a peak.
This is true for both the cases of  $S_h^{+++}(k)$ and $S_h^{+\times\times}(k)$
though there is a certain asymmetry in the behavior of the functions about the peak.
Note that the amplitude of $S_h(k)$ remains large over large wave numbers, while 
it quickly reduces to small values at smaller wave numbers.
In fact, this behavior should not come as a surprise since such a behavior was also
encountered in the case of $\ogw(f)$ (cf. figure~\ref{fig:ogw}).
It is interesting to note that $S_h^{+++}(k)$ and $S_h^{+\times\times}(k)$ 
settle down to about $10$ and $-250$, respectively, at large wave numbers.
Recall that the secondary tensor bispectra and hence the shape functions we have 
illustrated in figure~\ref{fig:Sh} have been evaluated during the radiation 
dominated epoch, when the modes are well inside the Hubble radius. 
They will have to be evolved until today to examine the
corresponding observational imprints which may possibly be detected by 
upcoming missions such as, say, LISA and PTA (in this context, see 
ref.~\cite{Bartolo:2018rku}; also see
refs.~\cite{Tsuneto:2018tif,Powell:2019kid,Iacconi:2020yxn}).


\section{Contributions to PBH formation and secondary GWs from 
scalar non-Gaussianities}\label{sec:sgws-dng}

Until now, we have focused on the imprints of the scalar power spectrum 
on the extent of PBHs formed and the generation of secondary GWs.
Clearly, if the scalar non-Gaussianities prove to be large in a given 
inflationary model, it seems plausible that they would significantly alter
the observables~$\fpbh$, $\ogw$ and $S_h$~\cite{Chongchitnan:2006wx,
Seery:2006wk,Hidalgo:2007vk,Motohashi:2017kbs,Atal:2018neu,Franciolini:2018vbk,
Kehagias:2019eil,Atal:2019erb,DeLuca:2019qsy,Passaglia:2018ixg,
Ezquiaga:2019ftu,Cai:2018dig,Unal:2018yaa}.
To understand the possible effects of non-Gaussianities on~$\fpbh$, $\ogw$ 
as well as $S_h$, in this section, we shall first calculate the scalar 
bispectrum and thereby the corresponding non-Gaussianity parameter~$\fnl$ 
in the two inflationary models USR2 and PI3 and the reconstructed scenarios 
RS1 and RS2.
We shall then discuss the corresponding contributions from the scalar 
bispectrum to $\fpbh$, $\ogw$ and~$S_h$.


\subsection{Evaluating the scalar bispectrum}

The scalar bispectrum is the three point function of the curvature 
perturbation in Fourier space, and it is defined in terms of the
operator~$\hat{\cR}_\vk$ that we had introduced earlier as 
follows~\cite{Ade:2015ava,Akrami:2019izv}:
\begin{equation}
\langle \hat{\cR}_{\vka}(\ee)\, 
\hat{\cR}_{\vkb}(\ee)\, \hat{\cR}_{\vkc}(\ee)\rangle 
=(2\,\pi)^3\, \cB_{_{\rm S}}(\vka,\vkb,\vkc)\,
\delta^{(3)}(\vka+\vkb+\vkc).\label{eq:bi-s}
\end{equation}
Recall that, $\ee$ is a time close to the end of inflation and, in this expression,
the expectation value on the left hand side is to evaluated in the perturbative 
vacuum~\cite{Maldacena:2002vr,Seery:2005wm,Chen:2010xka}.
Note that the three wave vectors $(\vka,\vkb,\vkc)$ form the edges of a triangle.
For convenience, we shall hereafter set
\begin{equation}
\cB_{_{\rm S}}(\vka,\vkb,\vkc)=
(2\,\pi)^{-9/2}\, G(\vka,\vkb,\vkc)
\end{equation}
and refer to $G(\vka,\vkb,\vkc)$ as the scalar bispectrum.

\par

The so-called Maldacena formalism is the most complete approach to 
evaluate the scalar bispectrum in a given 
inflationary model~\cite{Maldacena:2002vr,Seery:2005wm,Chen:2010xka}.
In this approach, one first obtains the third order action governing 
the curvature perturbation.
With the third order action at hand, the scalar bispectrum is
evaluated using the standard rules of perturbative quantum field 
theory.
For the case of inflation driven by a single, canonical scalar field,
the third order action is found to consist of six bulk terms terms,
apart from the boundary terms~\cite{Arroja:2011yj}.
One can show that the scalar bispectrum $G(\vka,\vkb,\vkc)$ generated
by such an action can be expressed as follows (see, for instance, 
refs.~\cite{Martin:2011sn,Hazra:2012yn}; in this context, also see
ref.~\cite{Ragavendra:2020old}):
\begin{eqnarray}
G(\vka,\vkb,\vkc) 
&=& \sum_{C=1}^{7}\; G_{_{C}}(\vka,\vkb,\vkc)\nn\\
&=& \Mp^2\; \sum_{C=1}^{6}\; 
\Biggl[f_{k_1}(\ee)\, f_{k_2}(\ee)\,f_{k_3}(\ee)\, 
\cG_{_{C}}(\vka,\vkb,\vkc)+{\mathrm{complex\;conjugate}}\Biggr]
+ G_{7}(\vka,\vkb,\vkc),\label{eq:sbs}
\end{eqnarray}
where, as we discussed earlier, $f_k$ are the positive frequency Fourier
modes of the curvature perturbation.
Amongst the seven terms in the above expression for the scalar bispectrum, 
the first six correspond to the bulk terms in the third order action, whereas
the seventh arises due to a boundary term, and it is usually absorbed through
a field redefinition~\cite{Arroja:2011yj}.
The quantities~$\cG_{_{C}}(\vka,\vkb,\vkc)$, with $C=(1,6)$, are integrals  
associated with the bulk terms in the action and, as one can expect, apart 
from the background quantities, they involve the modes~$f_k$ and its 
derivative~$f_k'$.
(We have listed these integrals explicitly in appendix~\ref{app:cG}.)
The seventh term $G_{7}(\vka,\vkb,\vkc)$ that arises due to the contribution 
from a boundary term can be expressed as~\cite{Arroja:2011yj,Ragavendra:2020old}
\begin{eqnarray}
G_{7}(\vka,\vkb,\vkc)
&=& -i\,\Mpl^2\,\l[f_{k_1}(\ee)\,f_{k_2}(\ee)\,f_{k_3}(\ee)\r]\nn\\
& &\times\, \biggl[a^2\epsilon_1\epsilon_{2}\,
f_{k_1}^{\ast}(\eta)\,f_{k_2}^{\ast}(\eta)\,f_{k_3}'^{\ast}(\eta) 
+\mathrm{two~permutations} \biggr]_{\eta_i}^{\ee} 
+~\mathrm{complex~conjugate},\label{eq:G7}
\end{eqnarray} 
where $\ei$ is the time when the initial conditions are imposed on the
scalar perturbations.
We should mention that the remaining boundary terms do not contribute 
in the scenarios of our interest.

\par

As in the case of the scalar power spectrum,  due to the deviation
from slow roll, it proves to be difficult to evaluate the scalar 
bispectrum analytically in the inflationary models of interest.
Therefore, we resort to numerics.
There now exists a standard procedure to numerically compute the scalar 
bispectrum in inflationary models involving a single, canonical 
scalar field~\cite{Chen:2008wn,Hazra:2012yn}.
Recall that, in the case of the power spectrum, it is adequate to 
impose the Bunch-Davies initial conditions on the modes when they 
are sufficiently inside the Hubble radius.
Apart some special situations wherein the boundary conditions 
may need to be imposed deeper inside the Hubble, one often imposes 
the conditions when $k/(a\,H)\simeq 10^2$.
Since the amplitude of the scalar as well as tensor perturbations 
freeze when they are adequately outside the Hubble radius, say, 
when $k/(a\,H)\simeq 10^{-5}$, one can evaluate the power spectra
at such a time for the different modes.
Note that, in order to arrive at the bispectrum we need to carry
out integrals which involve the background quantities, the scalar 
modes~$f_k$ and its time derivative~$f_k'$ 
[cf eqs.~\eqref{eq:sbs}~and~\eqref{eq:cG}].
These integrals need to be carried out from a time~$\ei$ when the 
initial conditions are imposed on the modes until the late 
time~$\ee$ towards the end of inflation.
We had mentioned that the amplitudes of the modes freeze soon after
they leave the Hubble radius.
Due to this reason, one finds that, the super-Hubble contributions
to the scalar bispectrum prove to be negligible~\cite{Hazra:2012yn}.
Therefore, one can carry out the integrals from the time when 
$k/(a\,H)\simeq 10^2$ to the time when $k/(a\,H)\simeq 10^{-5}$.
However, since the bispectrum involves three modes, in general, one 
needs to integrate from the time when the smallest of the three wave 
numbers is well inside the Hubble radius to the time until when the 
largest of the wave numbers is sufficiently outside.
Moreover, in order to choose the correct perturbative vacuum, one
has to impose a cut-off in the sub-Hubble regime~\cite{Seery:2005wm}.
We impose a democratic (in wave number) cut-off of the 
form $\mathrm{exp}-\l[\kappa\,(k_1+k_2+k_3)/(3\,a\,H)\r]$, 
where~$\kappa$ is a positive definite and small
quantity~\cite{Chen:2008wn,Hazra:2012yn,Ragavendra:2020old}.
In fact, such a cut-off aids in the efficient numerical computation 
of the integrals involved.
One can choose a suitable value of~$\kappa$ depending on how deep
from inside the Hubble radius the integrals are to be carried out.


\subsection{Amplitude and shape of $\fnl$}

The non-Gaussianity parameter, say, $\fnl(\vka,\vkb,\vkc)$, corresponding 
to the scalar bispectrum is defined as (see, for 
instance, Refs.~\cite{Martin:2011sn,Hazra:2012yn})
\begin{eqnarray}
\fnl(\vka,\vkb,\vkc)
& =&-\frac{10}{3}\,\frac{1}{\l(2\,\pi\r)^4}\;k_1^3\, k_2^3\, k_3^3\;
G(\vka,\vkb,\vkc)\, \biggl[k_1^3\,\ps(k_2)\,\ps(k_3) 
+ {\mathrm{two~permutations}}\biggr]^{-1},\label{eq:fnl}
\end{eqnarray}
where $\ps(k)$ denotes the scalar power spectrum [cf. eq.~\eqref{eq:sps}].
With the scalar power and bispectra at hand, evidently, it is straightforward 
to arrive the non-Gaussianity parameter~$\fnl$ for a given model.

\par

Based on prior experience, we would like to emphasize a few points 
concerning the expected shape and amplitude of the scalar 
bispectrum before we go on to present the results for~$\fnl$ in the 
different models and scenarios we have introduced earlier.
As is well known, in slow roll inflationary models involving a single,
canonical scalar field, the scalar non-Gaussianity parameter~$\fnl$ 
proves to be of the order of the first slow roll 
parameter~$\epsilon_1$~\cite{Maldacena:2002vr,Seery:2005wm,Chen:2010xka}.
In other words, the parameter~$\fnl$ is typically of the order of $10^{-2}$ 
or smaller in such situations.
Moreover, the bispectrum is found to have an equilateral shape, with 
the $\fnl$ parameter slightly peaking when $k_1=k_2=k_3$ (in this 
context, see, for instance, ref.~\cite{Hazra:2012yn}).
However, when departures from slow roll occur, the non-Gaussianity
parameter~$\fnl$ can be expected to be of the order of unity or larger,
depending on the details of the background dynamics.
Further, in contrast to the slow roll case, wherein there is only a 
weak dependence of the parameter $\fnl$ on scale, when departures from
slow roll occur, the parameter turns out to be strongly scale dependent.
Needless to say, we can expect that the non-Gaussianity parameter~$\fnl$
to be relatively large as well as strongly scale dependent in the
situations of our interest.

\par

\begin{figure}[!t]
\begin{center}
\includegraphics[width=15cm]{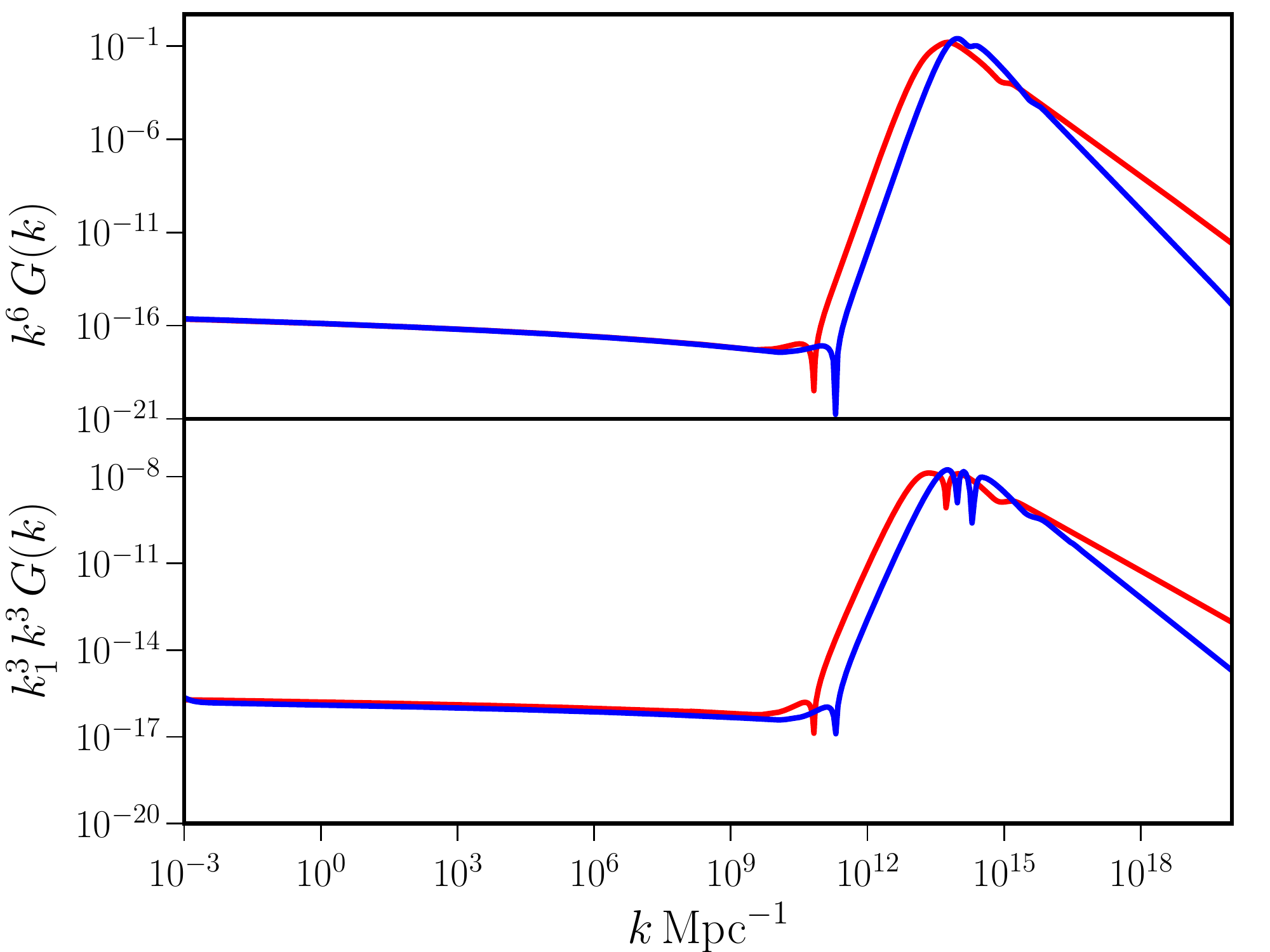}
\end{center}
\vskip -15pt
\caption{The amplitude of the dimensionless scalar bispectra has been plotted 
in the equilateral (on top) and squeezed limits (at the bottom) for the models 
USR2 (in red) and PI3 (in blue).
Clearly, the bispectra have approximately the same shape as the corresponding 
power spectra (cf. figure~\ref{fig:pps-usr-pi}).
Note that, at small scales, the dimensionless bispectra have considerably lower 
amplitudes in the squeezed limit when compared to their values in the equilateral 
limit, whereas they have roughly the same amplitude over the CMB scales.}
\label{fig:k6G}
\end{figure}

Let us now discuss the results we obtain in the different models we 
have introduced.
In order to illustrate the complete shape of the bispectrum, the 
non-Gaussianity parameter~$\fnl$ is usually presented as a density plot in, 
say, the $(k_3/k_1)$-$(k_2/k_1)$-plane~\cite{Hazra:2012yn,Komatsu:2010hc}.
It proves to be a bit of a numerical challenge to compute the complete 
shape of the bispectrum across the wide range of wave numbers over which 
we have evaluated the power spectra. 
As a result, we shall focus on the amplitude of $\fnl$ in the equilateral 
and the squeezed limits, i.e. when $k_1=k_2=k_3=k$ and when $k_1\to 0$, 
$k_2\simeq k_3=k$, respectively.
It is easier to calculate the scalar bispectrum in the equilateral
limit as we just need to follow the evolution of one mode at a time.
To arrive at the scalar bispectrum in the squeezed limit, we shall
set $k_2=k_3=k$ and choose $k_1=10^{-3}\,k$.
We have confirmed that our results are robust against choosing a smaller
value of $k_1$.
Before we go to illustrate the amplitude and shape of the non-Gaussianity 
parameter~$\fnl$, let us understand the behavior of the scalar 
bispectrum~$G(\vka,\vkb,\vkc)$ itself.
In figure~\ref{fig:k6G}, we have plotted the scalar bispectra that arise
in the equilateral and squeezed limits in the models of USR2 and PI3.
We would like to highlight a few aspects regarding the amplitude and shape of 
the bispectra.
Note that the scalar bispectra have roughly the same shape in the equilateral
and squeezed limits.
Also, they closely resemble the corresponding scalar power spectra and, in 
particular, they exhibit a dip and a peak around the same locations
(cf. figure~\ref{fig:pps-usr-pi}).
Moreover, at small scales, the scalar bispectra have a larger amplitude in the
equilateral limit than in the squeezed limit.
Further, in the equilateral limit, the scalar bispectra have almost the same
amplitude as the power spectra near the peak.

\par

\begin{figure}[!t]
\begin{center}
\includegraphics[width=15cm]{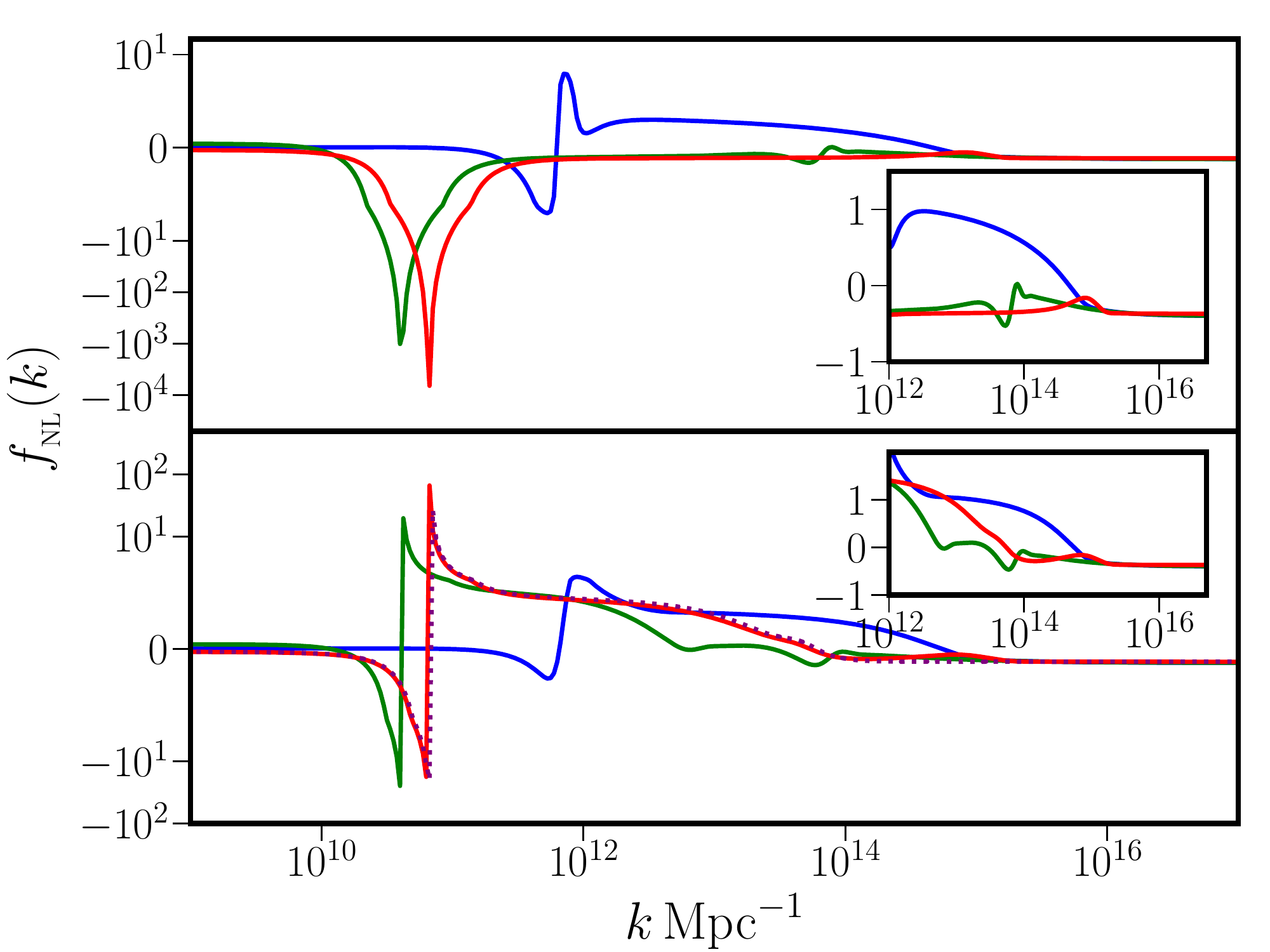}
\end{center}
\vskip -15pt
\caption{The scalar non-Gaussianity parameter $\fnl$ has been plotted in the 
equilateral (on top) and the squeezed (at the bottom) limits for the model of 
USR2 (in red) and the reconstructed scenario RS1 (in blue and green).
Note that, in the case of RS1, we have worked with our original choice of 
$N_1=42$ and plotted the lower (in blue) and the upper (in green) bounds 
of~$\fnl$ corresponding to the range over which the parameter $\Delta N_1$ 
is varied.
In the case of USR2, we have also plotted the consistency condition
$\fnl^{_\mathrm{CR}}(k)=(5/12)\, [\ns(k)-1]$ (as purple dots) along 
with the results in the squeezed limit.
Despite the deviations from slow roll leading to strong features in the scalar
power and bispectra, we find that the consistency condition is always satisfied.
The insets highlight the $\fnl$ around the wave numbers where the scalar power 
spectra exhibit their peaks.
It is clear that the parameter $\fnl$ attains larger values in the equilateral 
(where $\fnl\simeq 10^1$--$10^4$ at its maximum) than the squeezed (where 
$\fnl\simeq 1$--$10$) limit.
Importantly, we find that $\fnl$ is at most of order unity near the peaks of
the scalar power spectra.}\label{fig:fnl-usr2-rs1}
\end{figure}

\begin{figure}[!h]
\begin{center}
\includegraphics[width=15cm]{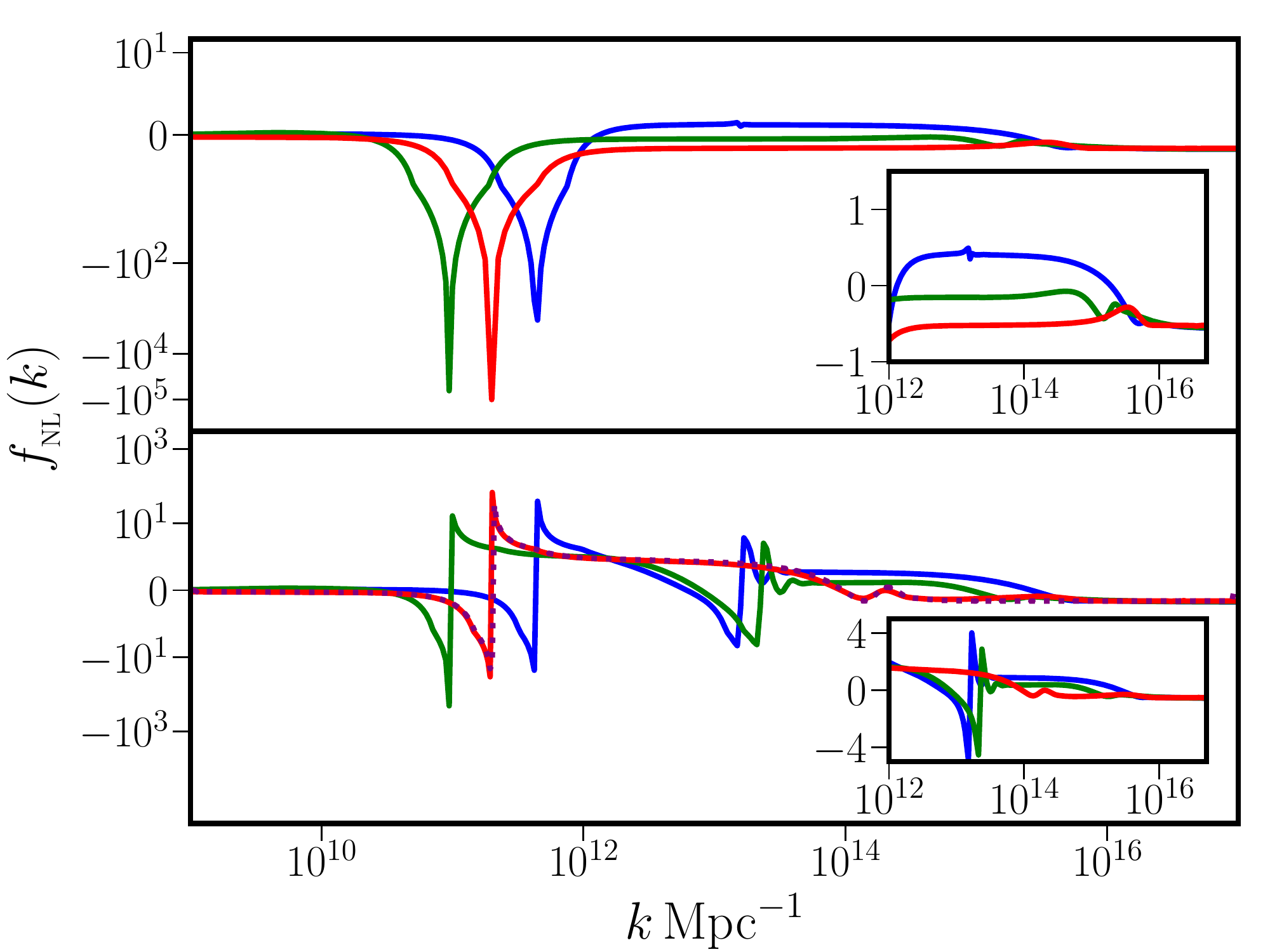}
\end{center}
\vskip -15pt
\caption{The scalar non-Gaussianity parameter $\fnl$ has been plotted in the 
equilateral and the squeezed limits for the model PI3 and the reconstructed 
scenario RS2 in the same manner (and the same choices of colors) as in the 
cases of USR2 and RS1 in the previous figure.
In the case of RS2, we have worked with our initial choice of $N_1=48$ and 
plotted the lower (in blue) and the upper (in green) bounds of $\fnl$ 
corresponding to the range over which the parameter $\Delta N_1$ is varied.
It should be evident that our earlier comments regarding the results for USR2 
and RS1 apply to the cases of PI3 and RS2 as well.}\label{fig:fnl-pi3-rs2}
\end{figure}

Let us now understand the behavior of the non-Gaussianity parameter~$\fnl$.
In figures~\ref{fig:fnl-usr2-rs1} and~\ref{fig:fnl-pi3-rs2}, we have plotted 
the behavior of the~$\fnl$ parameter in the equilateral and squeezed limits 
over a wide range of wave numbers in the models USR2 and PI3 as well as the 
scenarios RS1 and RS2.
The following points are evident from the two figures.
Firstly, in the equilateral limit, the non-Gaussianity parameter $\fnl$ proves 
to be fairly large (of the order of $10^1$--$10^4$) over a small range of wave 
numbers.
In fact, the $\fnl$ exhibit an upward spike in their amplitude around exactly 
the same wave numbers wherein the scalar power spectra exhibit a 
downward spike (cf. figures~\ref{fig:pps-usr-pi} and~\ref{fig:pps-rcs}).
Since the definition of the parameter $\fnl$ [cf. eq.~\eqref{eq:fnl}] contains
the scalar power spectrum in the denominator, the upward spike can be
partly attributed to the downward spike in the power spectrum.
If we ignore the large spike, we find that $\fnl\simeq 1$--$10$ around these 
wave numbers.
It is worth noting that these wave numbers correspond to those modes which
leave the Hubble radius just prior to or during the transition from the 
slow roll to the ultra slow roll regime.
In contrast, the non-Gaussianity parameter~$\fnl$ proves to be relatively 
small (at most of order unity) over wave numbers where the scalar power 
spectra exhibit their peak.
However, we should clarify that, though the value of $\fnl$ is smaller than unity 
around this domain, it is considerably larger than its typical value in slow roll
inflation (of about $10^{-2}$, such as over the CMB scales in our models).
For instance, in USR2 and PI3, we find that, in the equilateral limit, $\fnl$ is
about $-0.37$ and $-0.44$, respectively, near the locations of the peak in the 
power spectra.
This can be attributed to the large value of $\epsilon_2$ during the ultra slow 
roll regime.
Secondly, in the squeezed limit, the scalar bispectrum is expected to 
satisfy the so called consistency condition wherein it can be completely
expressed in terms of the scalar power spectrum~\cite{Maldacena:2002vr,
Creminelli:2004yq}.
This translates to the condition $\fnl^{_\mathrm{CR}}(k)=(5/12)\, [\ns(k)
-1]$ in the squeezed limit, where $\ns(k)-1=\d\, \mathrm{ln}\,\ps(k)/\d\, 
\mathrm{ln}\, k$ is the scalar spectral index.
In figures~\ref{fig:fnl-usr2-rs1} and~\ref{fig:fnl-pi3-rs2}, apart from 
plotting $\fnl$ in the squeezed limit, we have also plotted the quantity
$\fnl^{_\mathrm{CR}}$ obtained from the scalar spectral index.
We should add that we have also examined the validity of
the consistency relation more closely by working with a smaller~$k_1$.
We find that the consistency condition is indeed satisfied even when there 
arise strong features in the scalar power spectrum in all the scenarios of 
our interest (in this context, however, see appendix~\ref{app:cr}).
Therefore, in the squeezed limit, we find that $\fnl$ is at most of order
unity around the peaks of the scalar power spectra.

\par 

It seems important that we clarify a point regarding the 
validity of the consistency condition at this stage of our discussion. 
One may be concerned if the period of ultra slow roll, with its large value 
of $\epsilon_2$, could lead to a violation of the consistency condition over 
wave numbers that leave the Hubble radius during this epoch (in this context, 
see refs.~\cite{Namjoo:2012aa,Martin:2012pe,Motohashi:2014ppa}).
Recall that the amplitude of scalar modes over a certain range of wave numbers 
are modified to some extent during the transition from slow roll to ultra 
slow roll (cf. figure~\ref{fig:modes-usr2-pi3}).
However, since, in the cases of our interest, the epoch of ultra slow roll
ends leading to the eventual termination of inflation, the amplitude of the
scalar modes asymptotically freeze at sufficiently late times (for further 
details, see appendix~\ref{app:modes-USR}; in this context, also see 
refs.~\cite{Sreenath:2014nca,Passaglia:2018ixg}).
Due to this asymptotic behavior of the scalar modes, it should not come as a 
surprise that the consistency condition is satisfied in the models and scenarios 
of our interest despite the phase of ultra slow roll 
(for very recent 
discussions in this context, see Refs.~\cite{Bravo:2020hde,Pajer:2020wxk}).


\subsection{Imprints of $\fnl$ on $\fpbh$ and $\ogw$}

Recall that the observationally relevant dimensionless, scalar non-Gaussianity
parameter~$\fnl$ is usually introduced through the following relation (see
ref.~\cite{Komatsu:2001rj}; also see refs.~\cite{Martin:2011sn,
Hazra:2012yn}):
\begin{equation}
\cR(\eta, \vx)=\cR^{\mathrm{G}}(\eta, \vx)
-\f{3}{5}\,\fnl\, \l[\cR^{_\mathrm{G}}(\eta, \vx)\r]^2
\label{eq:i-fnl}
\end{equation}
where $\cR^{\mathrm{G}}$ denotes the Gaussian contribution. 
In Fourier space, this relation can be written as (see, for instance,
ref.~\cite{Martin:2011sn})
\begin{equation}
\cR_\vk
=\cR^{\mathrm{G}}_\vk
-\f{3}{5}\,\fnl\, \int \f{\d^{3}{\vp}}{(2\,\pi)^{3/2}}\, \cR^{\mathrm{G}}_{\vp}\;
\cR^{\mathrm{G}}_{\vk-\vp}.
\end{equation}
If one uses this expression for~$\cR_\vk$ and evaluates the corresponding
two-point correlation function in Fourier space, one obtains 
that~\cite{Cai:2018dig,Unal:2018yaa} 
\begin{equation}
\langle \hat{\cR}_\vk\, \hat{\cR}_{\vk'}\rangle
=\f{2\,\pi^2}{k^3}\,\delta^{(3)}(\vk +\vk')\,
\l[\ps(k)+\l(\f{3}{5}\r)^2\,\f{k^3}{2\,\pi}\,\fnl^2\,
\int {\rm d}^{3}{\vp}\,
\f{\ps(p)}{p^3}\,\f{\ps\l(\vert \vk-\vp\vert\r)}{\vert\vk-\vp\vert^3}\r],
\end{equation}
where $\ps(k)$ is the original scalar power spectrum defined in the Gaussian 
limit [cf. eq.~\eqref{eq:sps-d}], while the second term represents the 
leading non-Gaussian correction.
We find that we can write the non-Gaussian correction to the scalar power 
spectrum, say, $\pc(k)$, as follows:
\begin{eqnarray}
\pc(k)&=&\l(\f{3}{5}\r)^2\,\fnl^2\,
\int_{0}^{\infty}\d v\int_{\vert 1-v\vert}^{1+v}\f{\d u}{v^2\,u^2}\,
\ps(k\,v)\, \ps(k\,u)\nn\\
&=&\l(\f{12}{5}\r)^2\,\fnl^2\,
\int_{0}^{\infty}\d s\int_{0}^{1}\f{\d d}{(s^2-d^2)^2}\,
\ps[k\,(s+d)/2]\, \ps[k\,(s-d)/2].\label{eq:pc-k}
\end{eqnarray}

\par

\begin{figure}[!t]
\begin{center}
\includegraphics[width=15cm]{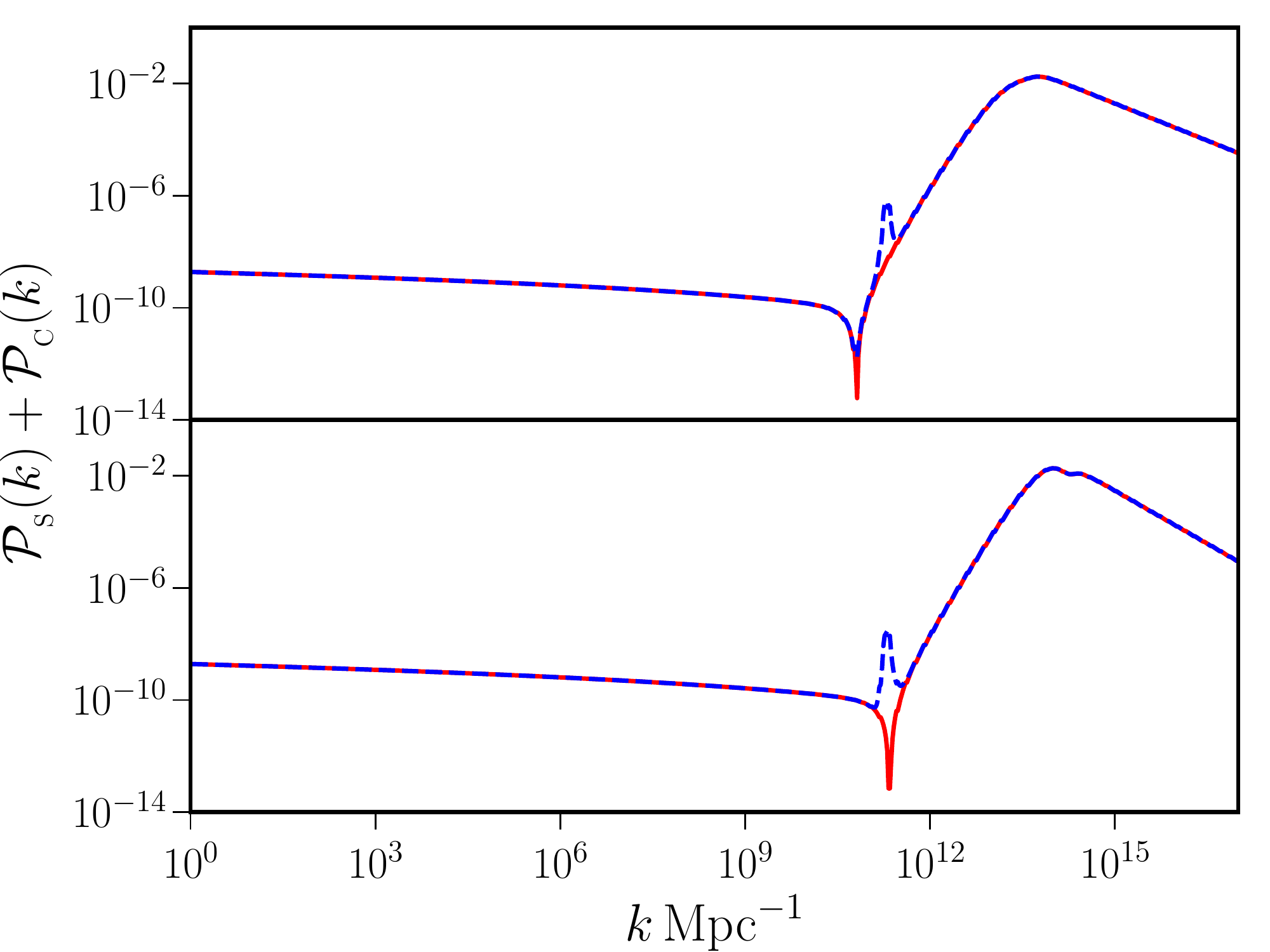}
\end{center}
\vskip -15pt
\caption{The original scalar power spectrum $\ps(k)$ (in solid red) and the 
modified spectrum $\ps(k)+\pc(k)$  (in dashed blue) arrived at upon including 
the non-Gaussian corrections, have been plotted for the models of USR2 (on 
top) and PI3 (at the bottom). 
In these models, the non-Gaussianity parameter~$\fnl$ had exhibited sharp spikes 
in its amplitude around wave numbers where the Gaussian scalar power spectrum 
had contained downward spikes.
We should clarify here that, in order to arrive at the modified power spectra, 
we have regulated the spikes in the $\fnl$ parameter so that its maximum value 
around these wave numbers is~$10^2$.
Clearly, the modifications to the scalar spectra, particularly at their peak, 
is hardly significant.}\label{fig:ps-m}
\end{figure}

Since we have evaluated the scalar non-Gaussianity parameter in the inflationary
models of our interest, we can now calculate the non-Gaussian corrections~$\pc(k)$ 
to the scalar power spectrum and the corresponding modifications to~$\fpbh$, $\ogw$
and $S_h$.
However, before we do so, we need to clarify an important point.
In introducing the scalar non-Gaussianity parameter through the
relation~\eqref{eq:i-fnl}, it has been assumed that $\fnl$ is local, i.e. 
it is independent of the wave number~\cite{Komatsu:2001rj}.
In contrast, the parameter $\fnl$ proves to be strongly scale dependent in all 
the situations we have considered.
In order to be consistent with the fact that the $\fnl$ in eq.~\eqref{eq:i-fnl}
is local, we shall consider the squeezed limit of the parameter (in this 
context, also see the discussions in ref.~\cite{Motohashi:2017kbs}).
Moreover, in the expression~\eqref{eq:pc-k} for $\pc(k)$, we shall assume that 
$\fnl$ is dependent on the wave number~$k$, with $k_2=k_3\simeq k$ and $k_1\ll k$ 
to be consistent with the squeezed limit. 
In figure~\ref{fig:ps-m}, we have plotted the original Gaussian power spectrum
as well the modified power spectrum including the non-Gaussian corrections~$\pc(k)$.
Recall that the non-Gaussianity parameter $\fnl$ had contained sharp spikes 
around the wave numbers where the Gaussian scalar power spectra had
exhibited a downward spike (cf. figures~\ref{fig:fnl-usr2-rs1} 
and~\ref{fig:fnl-pi3-rs2}). 
While evaluating the modified power spectra, we have regulated the maximum value
of these spikes to be $\vert \fnl \vert \simeq 100$.
Evidently, the non-Gaussian corrections to the scalar power spectrum are 
insignificant.
This can be attributed to the fact that the peaks in the original power 
spectrum~$\ps(k)$ and the non-Gaussianity parameter $\fnl$ are located 
at different wave numbers.
Therefore, we find the corresponding modifications to~$\fpbh$, $\ogw$ and 
$S_h$ are insignificant as well.
This conclusion can also be understood from the fact the amplitude of the
dimensionless bispectrum in the squeezed limit is considerably smaller 
than the amplitude of the scalar power spectrum around its peak 
(cf. figure~\ref{fig:k6G}).

We should clarify a particular point regarding the non-Gaussian corrections 
we have calculated in this section.
Note that we have calculated the cubic order non-Gaussian corrections to the 
power spectrum.
This method proves to be adequate to examine the imprints of non-Gaussianities 
on the dimensionless energy density~$\ogw$ describing the secondary GWs. 
However, the approach does not completely account for the effects of 
non-Gaussianities on the fraction~$\fpbh$ of PBHs produced (for an early
discussion on the topic, see ref.~\cite{Byrnes:2012yx}; for recent 
discussions, see refs.~\cite{Taoso:2021uvl,Riccardi:2021rlf}).
In the context of PBHs, the non-Gaussianities also change the shape of the 
probability distribution characterizing the over-densities at the time of 
their formation, which we have assumed to be a Gaussian [cf. eq.~\eqref{eq:pd}]
These effects due to the non-Gaussianities are expected to be larger (than the 
corrections to the power spectrum we have calculated), and they need to be taken 
into account to arrive at the modified~$\fpbh$~\cite{Taoso:2021uvl}.


\section{Conclusions}\label{sec:c}

In this work, we had considered models involving a single, canonical scalar 
field that lead to ultra slow roll or punctuated inflation.
All these models had contained a point of inflection, which seems essential 
to achieve the epoch of ultra slow roll required to enhance scalar power on 
small scales.
We had also examined the extent of PBHs formed and the secondary GWs generated 
in these models and had compared them with the constraints on the corresponding
observables~$\fpbh$ and~$\ogw$.
These models require a considerable extent of fine tuning in order to 
lead to the desirable duration of inflation (of say, $60$--$70$ e-folds), 
be consistent with the constraints from the CMB on large scales, and 
simultaneously exhibit higher scalar power on small scales.

\par

In order to explore the possibilities in single field models further, 
we had also considered scenarios wherein the functional forms for the 
first slow roll parameter closely mimic the typical behavior in ultra 
slow roll and punctuated inflation.
We had reconstructed the potentials associated with these scenarios,
evaluated the resulting scalar and tensor power spectra as well as
the corresponding imprints on~$\fpbh$, $\ogw$ and $S_h$.
The presence of extra parameters in the choices for $\epsilon_1(N)$
had allowed us to construct the required scenarios rather easily.
Interestingly, we had found that the reconstructed potentials too 
contain a point of inflection as the original models do.
This lends further credence to the notion that a point of inflection 
is essential to achieve ultra slow roll or punctuated inflation.
However, we should add a note of caution that, while we were able to 
broadly capture the expected shape of the scalar power spectra in the 
reconstructed scenarios, there were some differences in the tensor 
power spectra in these scenarios and the original models.
Moreover, we find that these reconstructed scenarios allow us to easily
examine the rate of growth of the scalar power from the CMB scales to 
small scales 
(for a discussion in this context, 
see refs.~\cite{Byrnes:2018txb,Ozsoy:2019lyy}).
While the steepest growth possible in the reconstructed scenario RS1 has 
$\ns-1\simeq 4$, we find that the growth is non-uniform but faster in 
RS2 with $\ns-1$ between $4$ and $6$ over the relevant range of wave
numbers (for details, see appendix~\ref{app:sg}).
Further, though we have been able to reconstruct the potentials numerically
in the scenarios RS1 and RS2, it would be worthwhile to arrive at analytical
forms of these potentials~\cite{Hertzberg:2017dkh,Byrnes:2018txb,
Motohashi:2019rhu}.

\par

We had also computed the scalar bispectrum and the associated non-Gaussianity
parameter $\fnl$ is these models and scenarios.
We had found that the parameter $\fnl$ is strongly scale dependent in all
the cases.  
Also, the non-Gaussianities had turned out to be fairly large (with, say, 
$\fnl>10$ over a range of wave numbers) in the equilateral limit.
Moreover, we had found that the consistency condition governing the 
non-Gaussianity parameter is always satisfied, despite the period of
sharp departure from slow roll, implying that the non-Gaussianity 
parameter in the squeezed limit is at most of order unity around the 
domain where the scalar power spectra exhibit their peak.
Due to this reason, we had found that the non-Gaussian corrections to 
power spectra were negligible leading to insignificant modifications to
the observables $\fpbh$, $\ogw$ and $S_h$ on small scales.
However, we should point out that the effects of non-Gaussianities on 
$\fpbh$ and~$\ogw$ have been included in a simple fashion and a more
detailed approach seems required to account for the complicated scale 
dependence of~$\fnl$~\cite{Atal:2018neu,Franciolini:2018vbk,
Kehagias:2019eil,Atal:2018neu,DeLuca:2019qsy,Passaglia:2018ixg}.
It has recently been argued that, in the squeezed limit
of the bispectrum, the part satisfying the consistency relation should 
be subtracted away as it cannot be observed (in this context, see 
refs.~\cite{Tada:2016pmk,Suyama:2020akr}; however also
see Ref.~\cite{Matarrese:2020why}).
If this is indeed so, since the scalar bispectrum satisfies the consistency 
condition in the squeezed limit in the models and scenarios we have examined, 
the cubic order non-Gaussian corrections to the power spectrum would then 
identically vanish.

\par

Moreover, we had calculated the secondary tensor bispectrum 
generated in the different inflationary models of interest during 
the radiation dominated epoch.
Interestingly, we had found that the shape function characterizing the 
tensor bispectrum has an amplitude of about $10$--$250$ at small wave 
numbers in all the models and scenarios of interest.
It seems important to evolve the shape function until today and examine 
the possibility of observing its imprints in ongoing efforts such as
PTA~\cite{Tsuneto:2018tif} and forthcoming missions such as 
LISA~\cite{Bartolo:2018rku,Powell:2019kid,Iacconi:2020yxn}.
We are currently investigating these issues
in a variety of single and two field models of inflation~\cite{Mishra:2019pzq,
Cai:2019amo,Cai:2019bmk,Ashoorioon:2019xqc,Lin:2020goi,Yi:2020kmq,Palma:2020ejf,
Fumagalli:2020adf,Braglia:2020eai}.


\acknowledgments

The authors wish to thank Dhiraj Hazra, Rajeev Jain and Subodh Patil for
discussions and detailed comments on the manuscript. 
HVR and LS also wish to thank Arindam Chatterjee, Arul Lakshminarayan and 
J\'{e}r\^{o}me Martin for related discussions.
HVR and PS would like to thank the Indian Institute of Technology Madras~(IIT
Madras), Chennai, India, for support through the Half-Time Research
Assistantship and the Institute Postdoctoral Fellowship, respectively.
The authors wish to acknowledge use of the cluster computing facilities at 
IIT Madras, where some of the numerical computations were carried out.
LS also wishes to acknowledge support from the Science and Engineering Research  
Board, Department of Science and Technology, Government of India, through the 
Core Research Grant CRG/2018/002200.


\appendix

\section{The dichotomy of ultra slow roll and punctuated 
inflation}\label{app:d-usr-pi}

\begin{figure}[!t]
\begin{center}
\includegraphics[width=10.00cm]{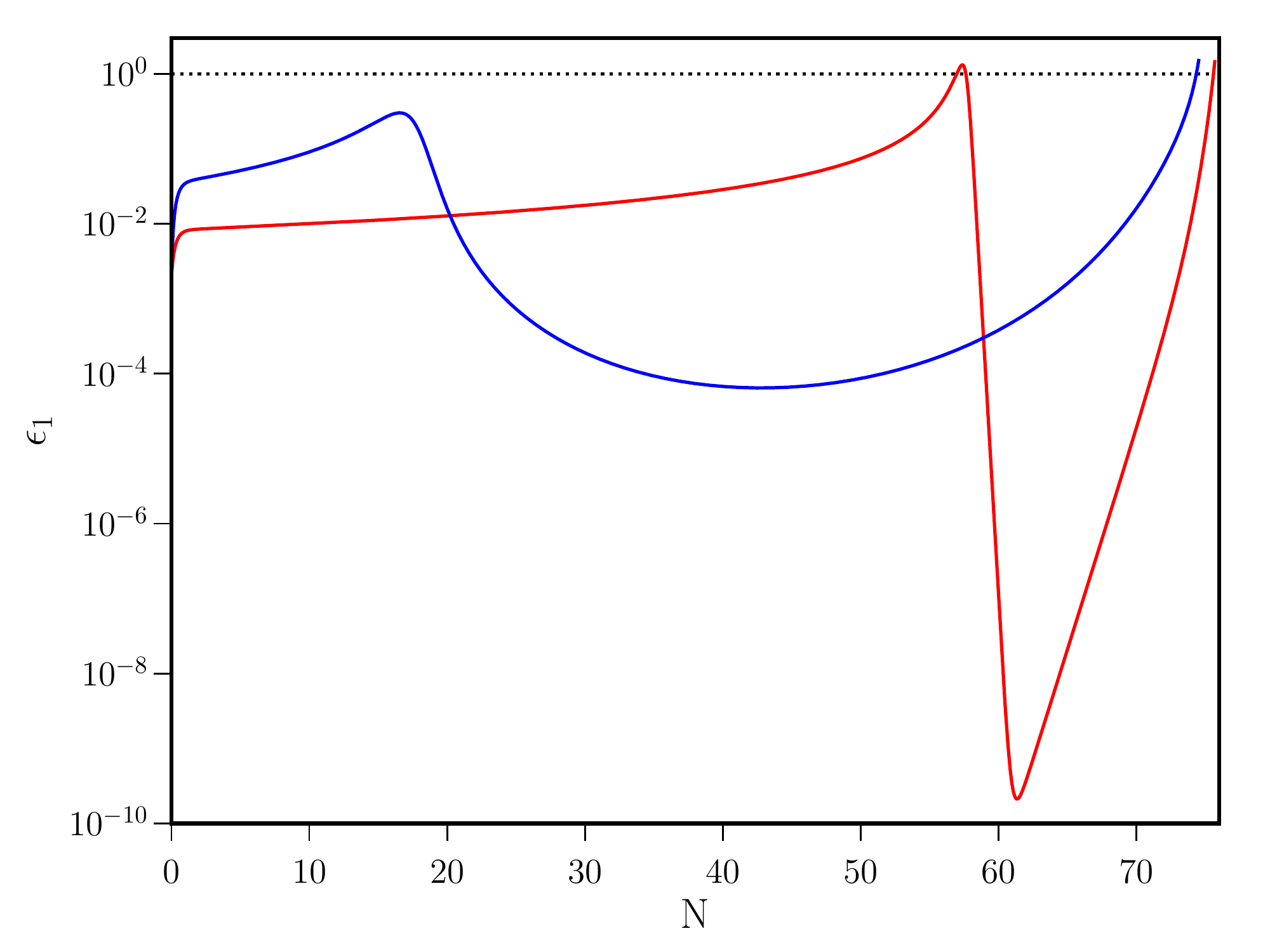}
\end{center}
\vskip -15pt
\caption{The behavior of the first slow roll parameter~$\epsilon_1$ has been
plotted for two sets of parameters describing the potential~\eqref{eq:phi6} 
and suitable initial conditions that lead to about $75$~e-folds of inflation. 
Note that the first set of values for the parameters leads to punctuated
inflation with~$\epsilon_1$ (plotted in red) crossing unity (indicated as a 
dotted horizontal line) twice, once prior to the regime of ultra slow roll 
and eventually when inflation terminates.
The second set of parameters leads to an extended period of ultra slow roll 
(plotted in blue) without any interruption of inflation until the very end.}
\label{fig:eps1-dichotomy}
\end{figure}

With the help of an example, in this appendix, we shall illustrate that a 
given inflationary potential can permit ultra slow roll as well as punctuated 
inflation for different sets of parameters.
The potential that we shall consider, when expressed in terms of the 
quantity $x = \phi/v$ that we had introduced in the context of USR1, is
given by~\cite{Bhaumik:2019tvl}
\begin{equation}
V(\phi) = V_0\,\frac{\alpha\,x^2-\beta\,x^4+\gamma\,x^6}{(1+\delta\,
x^2)^2}.\label{eq:phi6}
\end{equation}
In figure~\ref{fig:eps1-dichotomy}, we have plotted the evolution of the 
first slow roll parameter $\epsilon_1$ in the above potential for the 
following two sets of parameters:
$V_0/\Mpl^4=1.3253\times10^{-9}$, $\gamma=1$, $\delta= 1.5092$ and 
$(v/\Mpl,\alpha,\beta)=(4.3411,8.522\times 10^{-2},0.469)$
and $(10,8.53\times 10^{-2}, 0.458)$.
We obtain about $75$ e-folds of inflation in these cases for
$\phi_\mathrm{i}=17.245\,\Mpl$ and $\phi_\mathrm{i}=13.4\,\Mpl$.
It is clear from the figure that, while the first set of parameters lead to 
punctuated inflation, the second set does not permit an interruption of 
inflation until the very end.
This example illustrates the point that a potential itself cannot be classified
as an ultra slow roll or a punctuated inflationary model.


\section{The functional forms of the polarization factors}\label{app:pf}

Recall that, $e^\lambda(\vk,\vp)=e_{ij}^\lambda(\vk)\,p^i\,p^j$.
For our choice of $(\vka,\vkb,\vkc)$ and $(\vp_1,\vp_2,\vp_3)$ [cf.
eqs. \eqref{eq:vk} and \eqref{eq:vp}], we find that 
$e^{\lambda}(\vk,\vp)$ can be evaluated to be
\begin{subequations}
\label{eq:pfs}
\begin{eqnarray}
e^{+}(\vka,\vp_1) 
&=&\f{1}{4\,\sqrt{2}}\, \l(3\,p_{1x}^2 + p_{1y}^2 
- 2\,\sqrt{3}\,p_{1x}\,p_{1y} - 4\,p_{1z}^2\r),\\
e^{+}(\vkb,\vp_2) 
&=&\f{1}{4\,\sqrt{2}}\,\biggl(3\,p_{1x}^2 + 3\,k^2 + p_{1y}^2
+ 2\,\sqrt{3}\,p_{1x}\,p_{1y}- 6\,k\,p_{1x}
- 2\,\sqrt{3}\,k\,p_{1y} - 4\,p_{1z}^2\biggr),\\
e^{+}(\vkc,\vp_3) 
&=& \f{1}{\sqrt{2}}\,\l(p_{1y}^2-p_{1z}^2\r),\\
e^{\times}(\vka,\vp_1) 
&=& -\f{1}{\sqrt{2}}\,\l(\sqrt{3}\,p_{1x}-p_{1y}\r)\,p_{1z},\\
e^{\times}(\vkb,\vp_2) 
&=& \f{1}{\sqrt{2}}\,\l[\sqrt{3}\,(p_{1x} - k) + p_{1y}\r]\,p_{1z},\\
e^{\times}(\vkc,\vp_3) &=& -\sqrt{2}\,p_{1y}\,p_{1z}.
\end{eqnarray}
\end{subequations}


\section{Integrals determining the scalar bispectrum}\label{app:cG}

The quantities $\cG_{_{C}}(\vka,\vkb,\vkc)$ appearing in the 
expression~\eqref{eq:sbs} for the scalar bispectrum represent 
six integrals that involve the scale factor, the slow roll 
parameters, the modes~$f_k$ and their time derivatives~$f_k'$.
They correspond to the six bulk terms appearing in the cubic 
order action governing the curvature perturbation, and they are 
described by the following 
expressions~\cite{Martin:2011sn,Hazra:2012yn,Ragavendra:2020old}: 
\begin{subequations}\label{eq:cG}
\begin{eqnarray}
\cG_1(\vka,\vkb,\vkc)
&=& 2\,i\,\int_{\ei}^{\ee} \d\eta\; a^2\, 
\epsilon_{1}^2\, \l(f_{k_1}^{\ast}\,f_{k_2}'^{\ast}\,
f_{k_3}'^{\ast} + {\mathrm{two~permutations}}\r),\label{eq:cG1}\\
\cG_2(\vka,\vkb,\vkc)
&=&-2\,i\;\l(\vka\cdot \vkb + {\mathrm{two~permutations}}\r)\, 
\int_{\ei}^{\ee} \d\eta\; a^2\, 
\epsilon_{1}^2\, f_{k_1}^{\ast}\,f_{k_2}^{\ast}\,
f_{k_3}^{\ast},\label{eq:cG2}\\
\cG_3(\vka,\vkb,\vkc)
&=&-2\,i\,\int_{\ei}^{\ee} \d\eta\; a^2\,\epsilon_{1}^2\,
\l(\f{\vka\cdot\vkb}{k_2^2}\,
f_{k_1}^{\ast}\,f_{k_2}'^{\ast}\, f_{k_3}'^{\ast}
+ {\mathrm{five~permutations}}\r),\label{eq:cG3}\\
\cG_4(\vka,\vkb,\vkc)
&=& i\,\int_{\ei}^{\ee} \d\eta\; a^2\,\epsilon_{1}\,\epsilon_{2}'\, 
\l(f_{k_1}^{\ast}\,f_{k_2}^{\ast}\,f_{k_3}'^{\ast}
+{\mathrm{two~permutations}}\r),\label{eq:cG4}\\
\cG_5(\vka,\vkb,\vkc)
&=&\frac{i}{2}\,\int_{\ei}^{\ee} \d\eta\; 
a^2\, \epsilon_{1}^{3}\;\l(\f{\vka\cdot\vkb}{k_2^2}\,
f_{k_1}^{\ast}\,f_{k_2}'^{\ast}\, f_{k_3}'^{\ast} 
+ {\mathrm{five~permutations}}\r),\label{eq:cG5}\\
\cG_6(\vka,\vkb,\vkc) 
&=&\frac{i}{2}\,\int_{\ei}^{\ee}\d\eta\, a^2\, \epsilon_{1}^{3}\,
\l(\f{k_1^2\,\l(\vkb\cdot\vkc\r)}{k_2^2\,k_3^2}\, 
f_{k_1}^{\ast}\, f_{k_2}'^{\ast}\, f_{k_3}'^{\ast} 
+ {\mathrm{two~permutations}}\r).\qquad\label{eq:cG6}
\end{eqnarray}
\end{subequations}
These integrals are to be evaluated from a sufficiently early time, say, $\ei$, 
when all the modes are well inside the Hubble radius, until suitably late times,
which can be conveniently chosen to be a time close to the end of 
inflation, say, $\ee$.


\section{A closer examination of the consistency relation}\label{app:cr}

\begin{figure}[!t]
\begin{center}
\includegraphics[width=5.75cm]{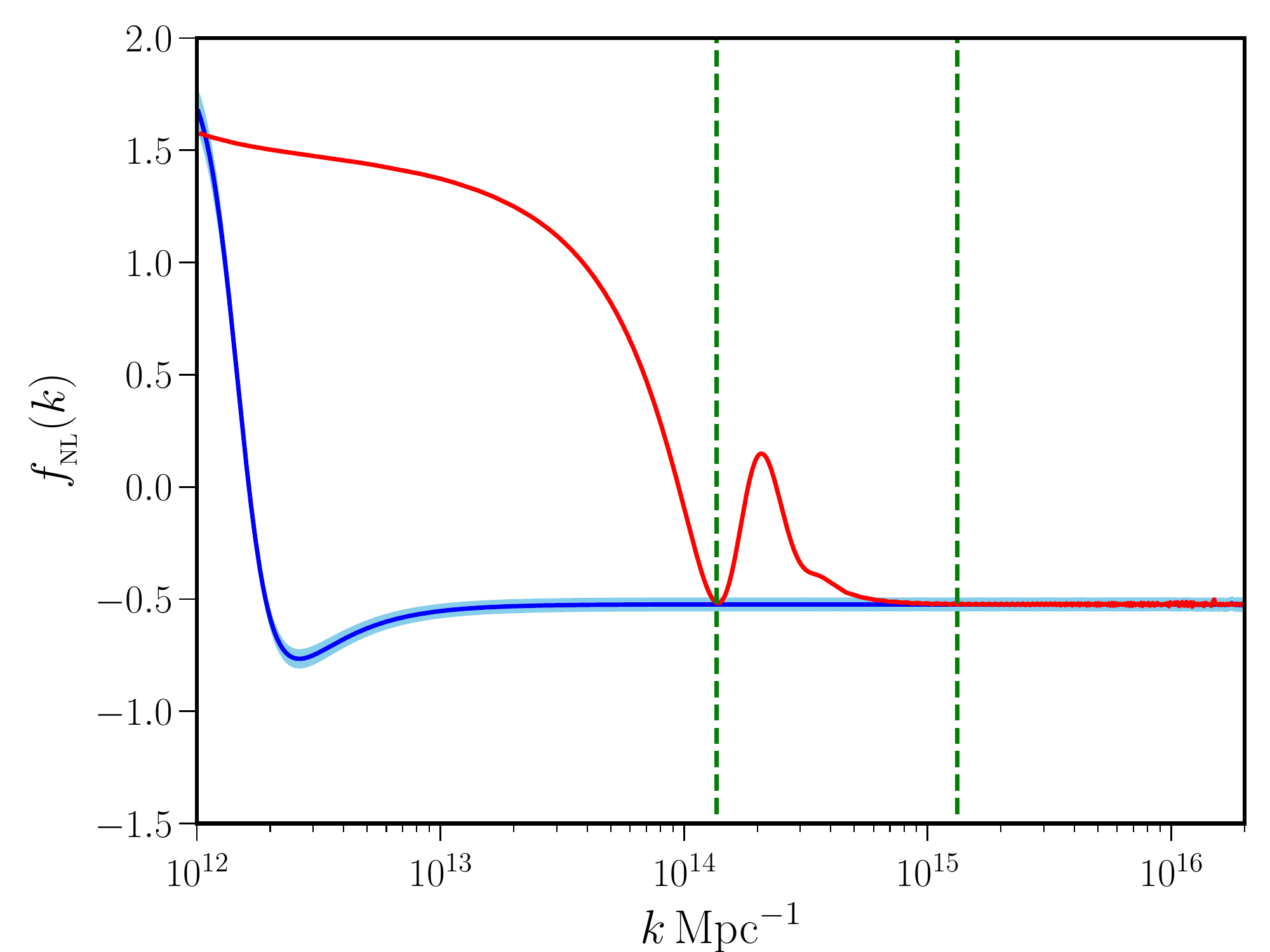}
\includegraphics[width=5.75cm]{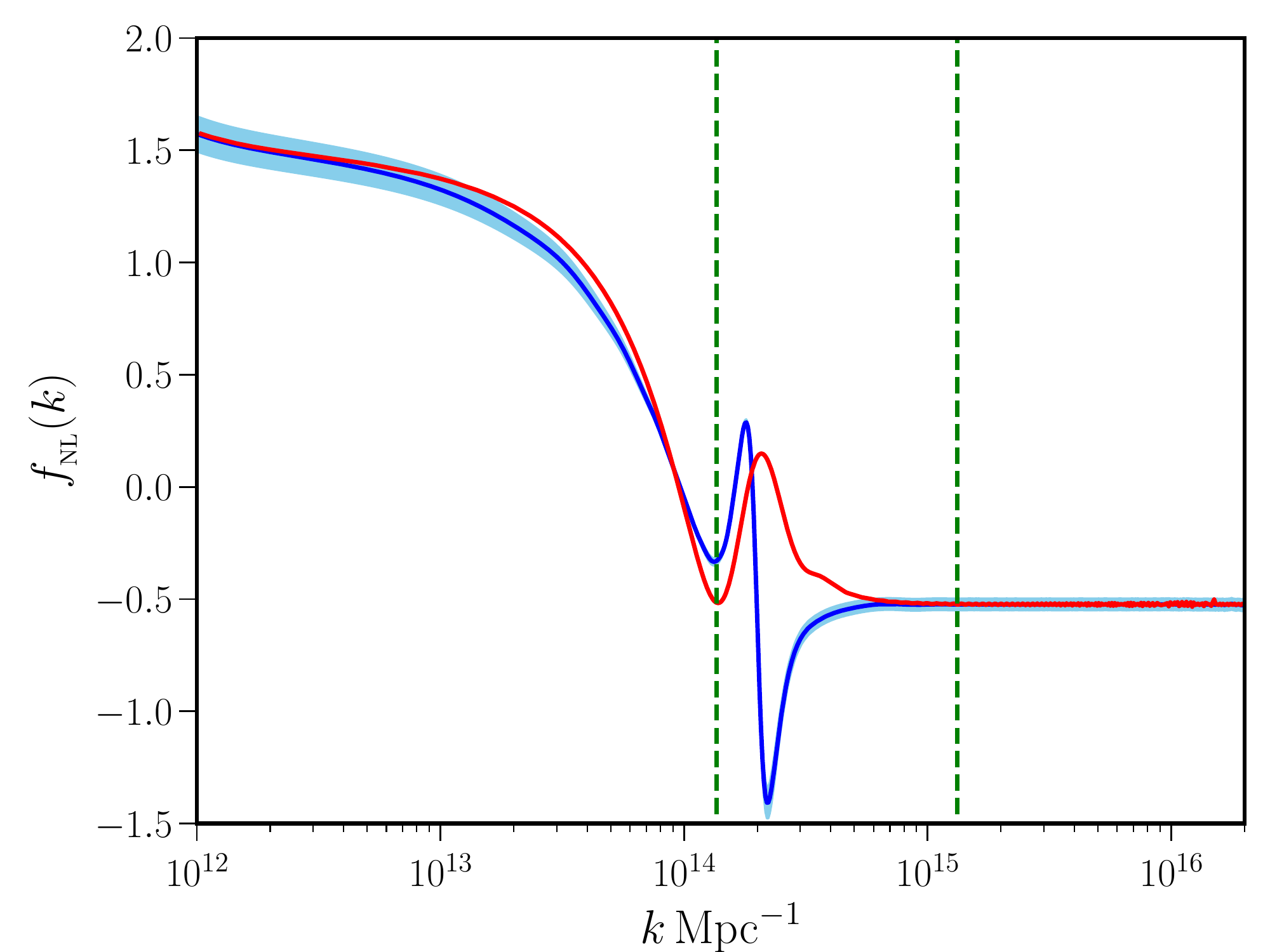}
\includegraphics[width=5.75cm]{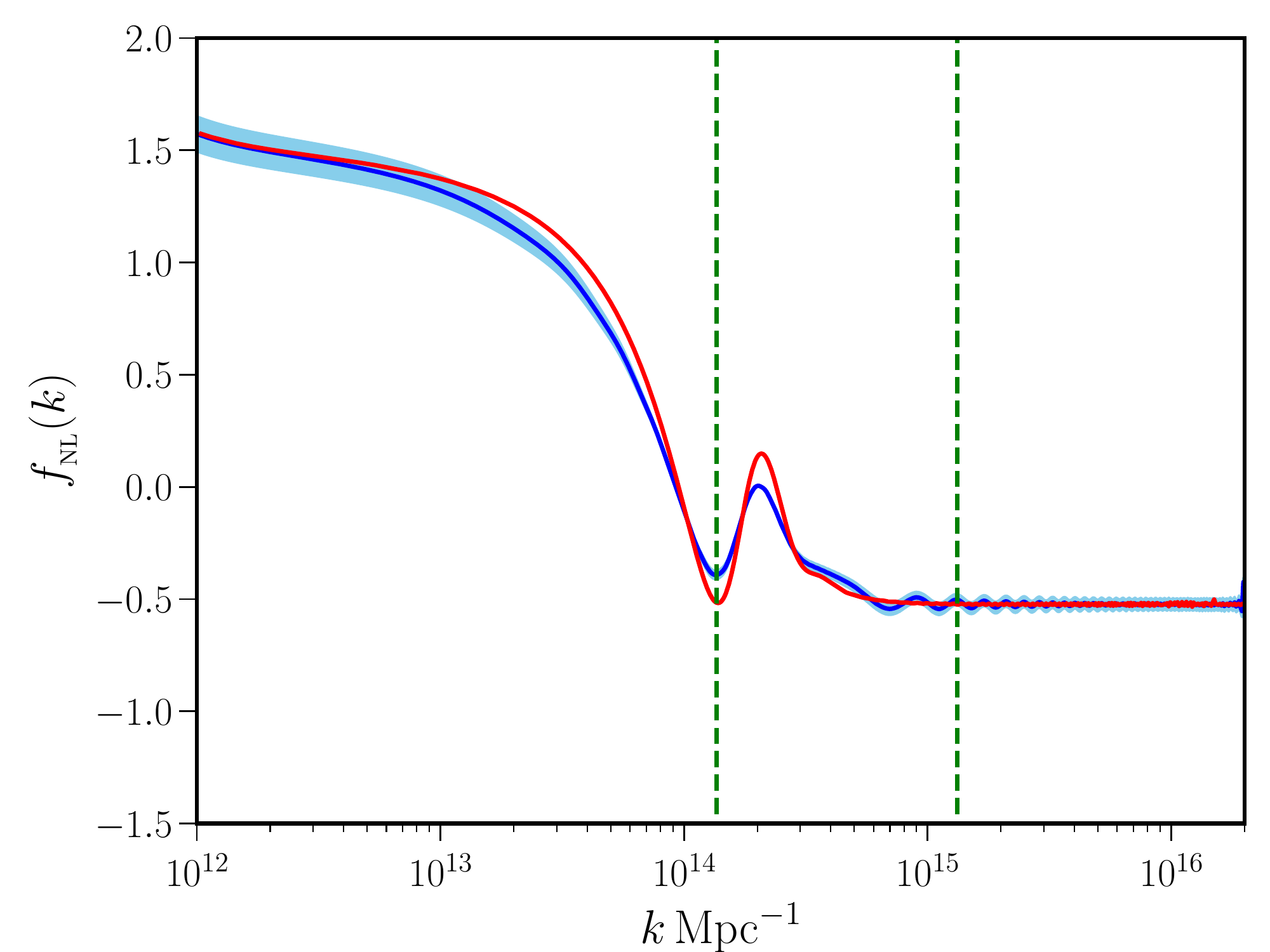}
\end{center}
\vskip -15pt
\caption{The non-Gaussianity parameter $\fnl$ in the squeezed limit (in
blue) and the consistency condition $\fnl^{_\mathrm{CR}}$ (in red) have 
been plotted for the model PI3 over wave numbers around the peak in the 
scalar power spectrum.
We have set the squeezed mode to be $k_1 = 10^{-1}\,k$ (on the left), 
$k_1 = 10^{-3}\,k$ (in the middle) and $k_1 = 10^{-5}\,k$ (on the right) 
in plotting these figures.
We have also indicated the $5\%$ uncertainty in our numerical estimate
as bands~(in blue).
Moreover, we have demarcated the range of modes (by vertical, dashed, green 
lines) that leave the Hubble radius during the epoch of ultra slow roll in
the model.
Obviously, the choice of $k_1=10^{-1}\,k$ is insufficient for $k_1$ to be
considered a squeezed mode. 
Such a choice has been made to illustrate the point that the value of $\fnl$ 
proves to be of order unity even when we confine to modes that leave the 
Hubble radius during the period of ultra slow roll.
Evidently, there is an improvement in the extent to which the consistency 
condition is satisfied when we choose to work with smaller and smaller values 
of~$k_1$.
Though the match improves as we work with a smaller $k_1$, we still seem
to notice some deviation.
This is possibly an artefact arising due to the reason that, numerically,
we are unable to work with an adequately small value of $k_1$.}
\label{fig:fnl-cr}
\color{black}
\end{figure}

We had pointed out that, in the squeezed limit, i.e. when $k_2\simeq k_3
=k$ and $k_1\to 0$, the non-Gaussianity parameter $\fnl$ is expected to satisfy 
the consistency condition $\fnl^{_\mathrm{CR}}(k)=(5/12)\, [\ns(k)-1]$, 
where $\ns(k)-1=\d\, \mathrm{ln}\,\ps(k)/\d\, \mathrm{ln}\, k$ is the scalar 
spectral index.
In the results presented earlier (in figures~\ref{fig:fnl-usr2-rs1} and
\ref{fig:fnl-pi3-rs2}), we had worked with $k_1=10^{-3}\, k$ to arrive
at $\fnl$ in the squeezed limit.
While we find that the consistency condition is satisfied to better than
$5\%$ over a wide range of scales, we notice that there is some departure 
around wave numbers corresponding to the peak in the scalar power spectrum.
To investigate this point more closely, in figure~\ref{fig:fnl-cr}, we have 
plotted the numerical results around the peak in the scalar power spectrum
for the original choice of $k_1$ as well as for $k_1 = 10^{-1} \,k$ and 
$k_1 = 10^{-5}\,k$ in the case of the model PI3.
We have considered the case of $k_1=10^{-1}\,k$ since we find that roughly a 
decade of modes exit the Hubble radius during the ultra slow roll phase.
Evidently, such a value of $k_1$ would be insufficient for it to be considered
a squeezed mode.
We find that the value of $\fnl$ remains of order unity even when we confine 
to modes which leave the Hubble radius during the period of ultra slow roll.
Also, as one would expect, we find that the consistency condition is satisfied 
better and better as we work with a smaller value of $k_1$.
We should clarify that adequate care needs to be taken while evaluating the 
integrals involved in the calculation of the bispectrum during the ultra 
slow roll regime. 
Since there occur rapid changes in the slow roll parameters during this epoch,
we should regulate the integrals with an appropriate choice for the cut-off 
parameter~$\kappa$, especially for the dominant contribution~$G_4(\vka,\vkb,\vkc)$ 
[cf.~appendix~\ref{app:cG}].
With an appropriate cut-off and with smaller values for the squeezed mode~$k_1$, 
we find that the match between $\fnl$ and $\fnl^{_\mathrm{CR}}$ indeed improves.
Nevertheless, even with a smaller of choice of $k_1$, we still notice some 
difference near the peak in the power spectrum.
We feel that this is an artefact and we believe that the difference can be 
overcome with a further smaller value for~$k_1$.
However, working with a very small $k_1$ poses certain numerical challenges, 
and we will leave it for future investigation.
We should mention that this an independent issue and stress that it does not 
affect our main conclusions related to PBHs and GWs. 


\section{Asymptotic behavior of the curvature perturbations}\label{app:modes-USR}

As we mentioned, it has been shown that an indefinite ultra slow 
roll regime of inflation leads to the violation of the consistency 
condition~\cite{Namjoo:2012aa,Martin:2012pe}.
Since all the models of our interest contain an ultra slow roll 
phase, one may wonder if a violation of the consistency condition 
would occur in these cases.
As we have seen, the consistency condition is satisfied in all the
cases we have considered.
This is primarily due to the fact that the ultra slow roll phase lasts 
only for a finite duration in our models, permitting the eventual 
freezing of the amplitude of the curvature perturbations.

\par

\begin{figure}[!t]
\begin{center}
\includegraphics[width=15cm]{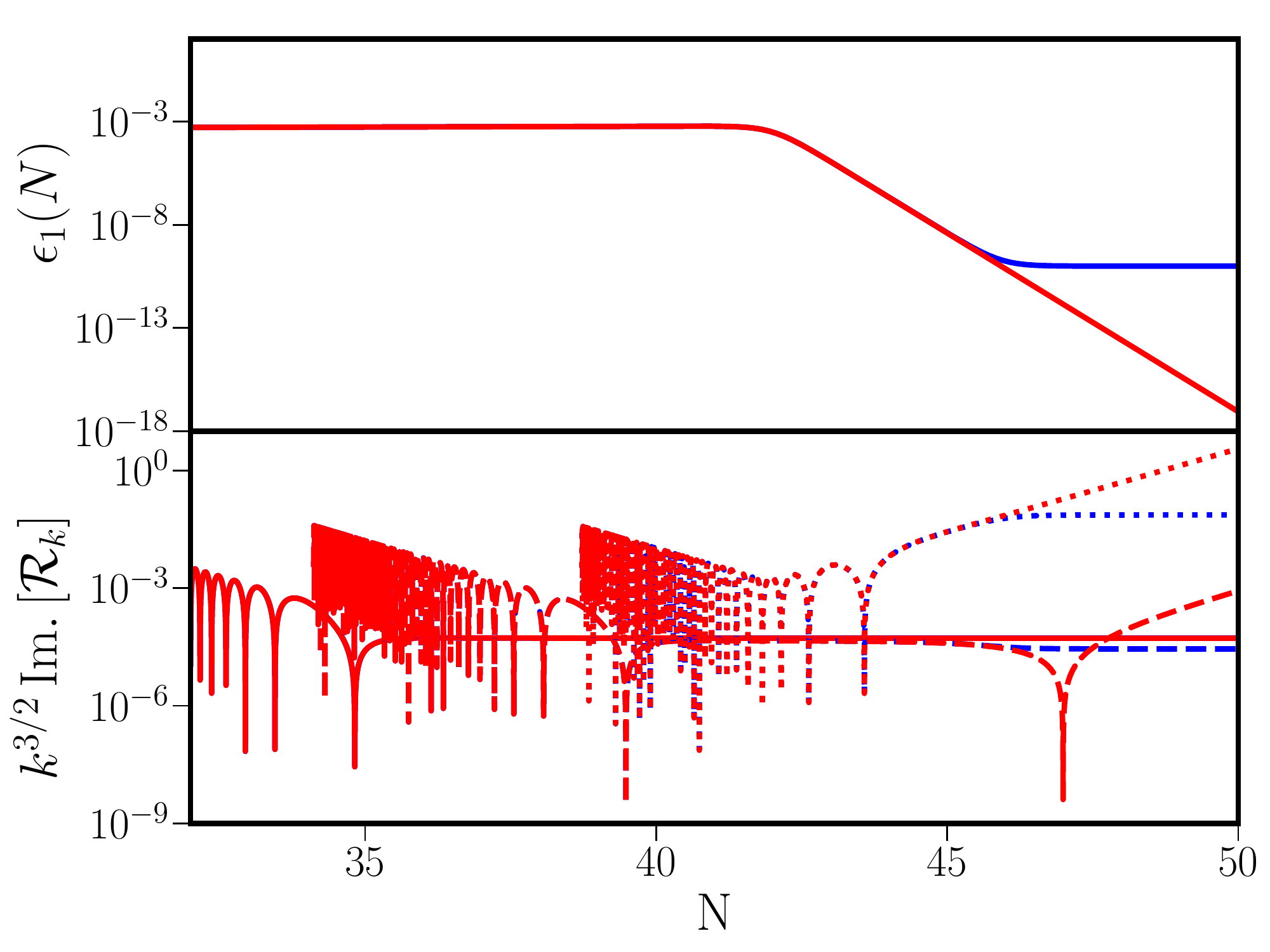}
\end{center}
\vskip -15pt
\caption{The functional forms $\epsilon_1^{\mathrm{III}}(N)$ (in red)
and $\epsilon_1^{\mathrm{IV}}(N)$ (in blue) for the first slow roll 
parameter have been plotted as function of e-folds (on top).
We have also illustrated the evolution of the dominant, imaginary part,
of the curvature perturbation $\cR_k$ for three representative modes 
in these two scenarios (as solid, dashed and dotted curves, in red and 
blue, respectively, at the bottom).
It is easy to see that (upon comparison of, say, the dotted red and blue
curves) that the end of the ultra slow phase ensures that the amplitude
of the curvature perturbations eventually freeze.}\label{fig:modes-USR}
\end{figure}

In this appendix, we shall illustrate this point with the aid of a 
truncated version of the scenario RS1.
We shall consider the following two functional forms for $\epsilon_1(N)$:
\begin{eqnarray}
\epsilon_1^{\mathrm{III}}(N) 
&=& \l[{\epsilon_{1a}\,\l(1+\epsilon_{2a}\,N\r)}\r]\,
\l[1 - {\mathrm{tanh}}\l(\f{N - N_1}{\Delta N_1}\r)\r] ,\label{eq:eps1-usr-1}\\
\epsilon_1^{\mathrm{IV}}(N) 
&=& \l[{\epsilon_{1a}\,\l(1+\epsilon_{2a}\,N\r)}\r]\,
\l[1 - {\mathrm{tanh}}\l(\f{N - N_1}{\Delta N_1}\r)\r] + \epsilon_{1b}.
\label{eq:eps1-usr-2}
\end{eqnarray}
Evidently, while the first choice lead to an indefinite period of ultra slow 
roll beyond the e-fold~$N_1$, the second choice restores slow roll 
when $\epsilon_1(N)$ attains the value of $\epsilon_{1b}$.
In figure~\ref{fig:modes-USR}, we have plotted the behavior of these slow roll
parameters as well as the evolution of the curvature perturbation for three
modes which leave the Hubble radius just prior to and after the onset of 
the ultra slow roll phase.
We have worked with the following values for parameters involved in plotting
the figure: $\epsilon_{1a} = 10^{-4}$, $\epsilon_{2a} = 0.05$, $N_1=42$, 
$\Delta N_1 = 0.5$ and $\epsilon_{1b} = 10^{-10}$.
It should be clear that, while the amplitude of the curvature perturbation 
grows indefinitely when the ultra slow roll continues, the amplitude 
freezes when slow roll inflation is restored.

\section{The steepest growth of the scalar power spectrum}\label{app:sg}

\begin{figure}[!t]
\begin{center}
\includegraphics[width=7.50cm]{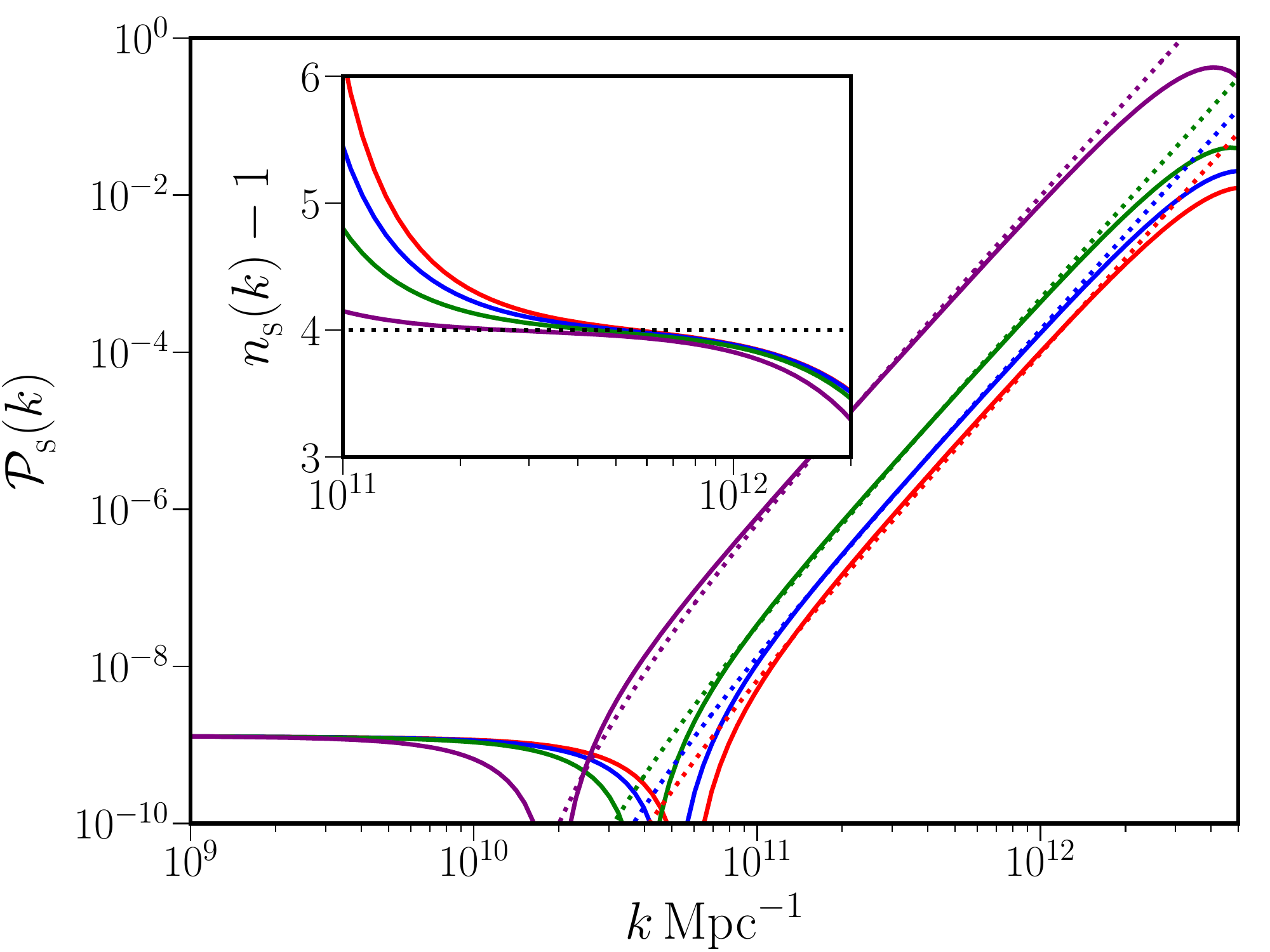}
\includegraphics[width=7.50cm]{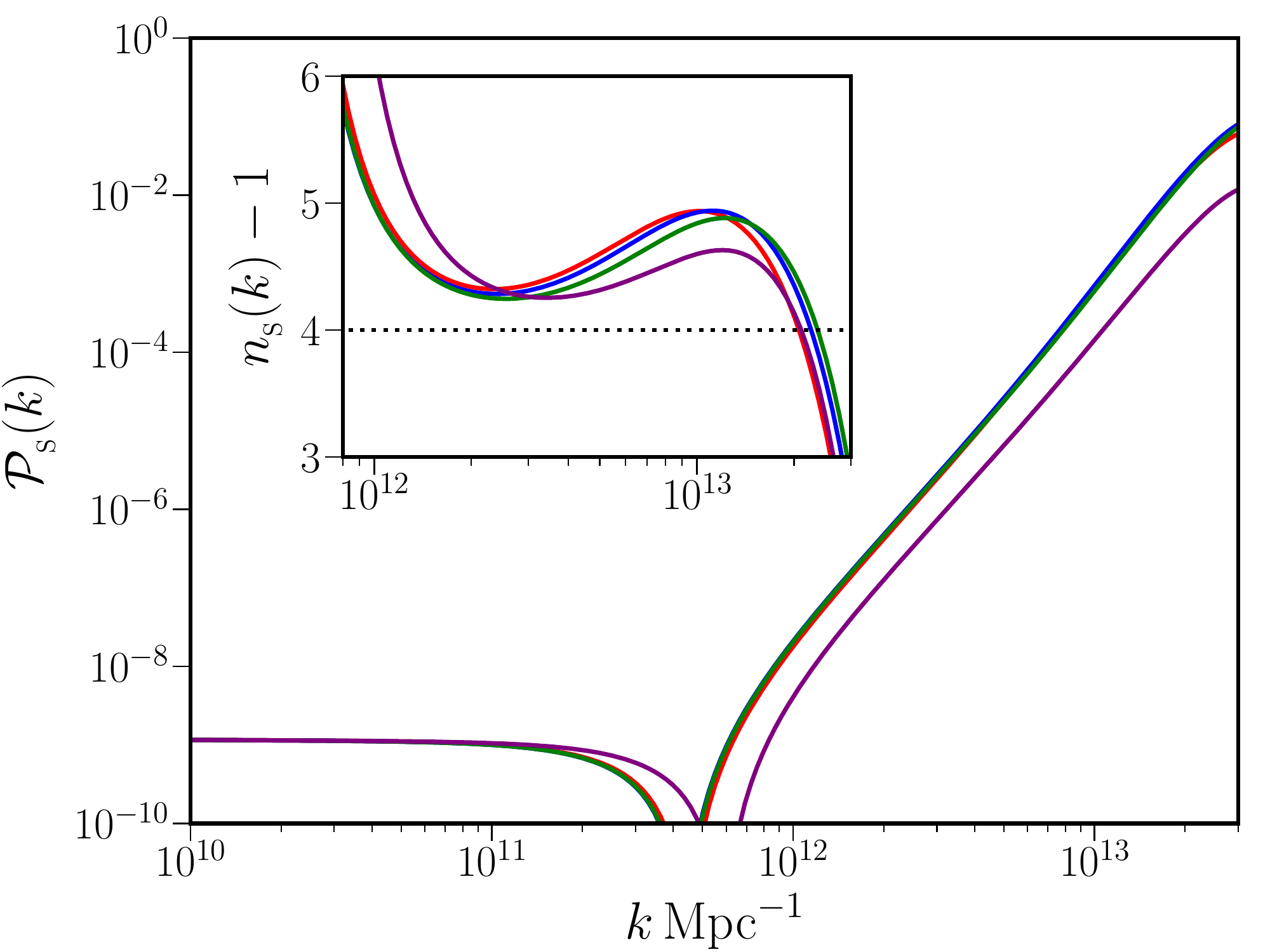}
\end{center}
\vskip -15pt
\caption{The scalar power spectra around the region where they exhibit 
the sharpest growth have been plotted in the cases of RS1 (on the left) 
and RS2 (on the right) for a set of values of~$\Delta N_1$. 
We have plotted the spectra for the following four values of $\Delta N_1$:
$(0.1, 0.08, 0.05, 0.01)$ (in red, blue, green and purple, respectively).
The insets illustrate the corresponding spectral indices $\ns - 1$. 
We have also indicated the $k^4$ behavior in the case of RS1 (as dotted
lines of corresponding colors on the left) to show how well it matches 
the spectra during the growth.
It should be evident that, while RS1 leads to a growth corresponding to 
$\ns-1 \simeq 4$, RS2 permits a steeper but non-uniform growth with $\ns-1$ 
varying between $4$ and $6$ over the relevant wave numbers.}\label{fig:ps-steep}
\end{figure}

In models of ultra slow roll and punctuated inflation, we have seen that 
the scalar power grows rapidly from its COBE normalized values on the CMB 
scales to higher values at smaller scales over wave numbers that leave 
the Hubble radius during the transition from slow roll to ultra slow roll. 
An interesting issue that is worth understanding is the steepest such growth 
that is possible in models of inflation driven by a single, canonical scalar 
field.
It has been argued that the fastest growth will
have $\ns-1\simeq 4$ over this range of wave numbers (in this 
context, see ref.~\cite{Byrnes:2018txb}; also 
see ref.~\cite{Ozsoy:2019lyy}).
We find that the reconstructed scenarios RS1 and RS2 easily permit us to 
examine this issue.  
Recall that, in these scenarios, the parameter $\Delta N_1$ determines the 
rapidity of the transition from the slow roll to the ultra slow roll regime
[cf. eqs.~\eqref{eq:eps1-12}].
We find that it is this parameter that dictates the steepness of the growth
in the corresponding scalar power spectra, with smaller $\Delta N_1$ producing 
a faster rise.
We have examined the rate of growth in the cases of RS1 and RS2 by varying 
$\Delta N_1$ over a certain range, while keeping the other parameters fixed.
In figure~\ref{fig:ps-steep}, we have illustrated the spectra for four values 
of~$\Delta N_1$ which are relatively smaller than those we had used for the 
reconstructions discussed earlier.
It should be clear from the figure that, in the case of RS1, the rise is 
fairly steady as the value of $\Delta N_1$ is made smaller, with $\ns-1 
\simeq 4$ over the growing regime. 
In the case of RS2, we find that $\ns-1$ varies between $4$ and $6$ over 
the growing regime and therefore corresponds to a steeper but non-uniform 
growth of the spectra.

\bibliographystyle{apsrev4-2}
\bibliography{pbh-sgw}
\end{document}